\documentclass[a4paper,11pt]{article}
\pdfoutput=1

\usepackage{jheppub}

\usepackage[T1]{fontenc}
\usepackage{amssymb,amsmath}
\usepackage{braket,bbm,bm}
\usepackage{mathtools}
\usepackage[usenames,dvipsnames,svgnames,table]{xcolor}
\usepackage[utf8]{inputenc}
\usepackage{soul, color}
\usepackage{subcaption}
\usepackage{lmodern}
\usepackage{footnote}
\usepackage[normalem]{ulem}
\usepackage{glossaries-extra}
\setabbreviationstyle[acronym]{long-short}
\glssetcategoryattribute{acronym}{nohyperfirst}{true}

\usepackage{slashed}
\usepackage{multirow}
\usepackage{here}
\usepackage{hyperref}
\usepackage{verbatim}
\usepackage[justification=justified,singlelinecheck=false]{caption}
\usepackage{cleveref}
\usepackage{listings}
\usepackage{fancyvrb}
\usepackage{dsfont}
\usepackage{nicefrac,xfrac}
\usepackage{verbatim}

\Crefname{equation}{eq.}{eqs.}
\Crefname{section}{section}{sections}
\Crefname{figure}{figure}{figures}
\Crefname{appendix}{appendix}{appendices}

\newcommand{\cA}{\mathcal{A}}
\newcommand{\cB}{\mathcal{B}}
\newcommand{\cC}{\mathcal{C}}
\newcommand{\cD}[0]{\mathcal D}

\newcommand{\cI}[0]{\mathcal I}

\newcommand{\cK}[0]{\mathcal K}
\newcommand{\cL}[0]{\mathcal L}
\newcommand{\cM}[0]{\mathcal M}
\newcommand{\cO}[0]{\mathcal O}
\newcommand{\cR}[0]{\mathcal R}

\newcommand{\cT}[0]{\mathcal T}
\newcommand{\cY}[0]{\mathcal Y}

\renewcommand{\k}{\bm{k}}
\newcommand{\p}{\bm{p}}
\newcommand{\q}{\bm{q}}

\renewcommand{\P}{\bm{P}}
\newcommand{\mone}{\mathbbm{1}}

\newcommand{\wt}[0]{\widetilde}

\newcommand{\kdf}{\mathcal{K}_{\text{df},3} }

\newcommand{\YL}[3]{\boldsymbol{\mathcal Y}^{[kab] \dagger}_{(#1#2#3)}}
\newcommand{\YR}[3]{\boldsymbol{\mathcal Y}^{[kab]}_{(#1#2#3)}}

\newcommand{\ThreeBody}[0]{%
Briceno:2012rv,
Polejaeva:2012ut,
Hansen:2014eka,
Hansen:2015zga,
Hammer:2017uqm,
Hammer:2017kms,
Mai:2017bge,
Jackura:2019bmu,
Hansen:2020zhy,
Pang:2020pkl,
Blanton:2020gmf,
Blanton:2020jnm,
Blanton:2021mih,
Muller:2021uur,
Blanton:2021eyf,
Draper:2023xvu,
Briceno:2024txg,
Xiao:2024dyw,
Feng:2024wyg}
\usepackage{mfirstuc} % to uppercase the name
\newcommand{\addReviewer}[2]{
  \expandafter\newcommand\csname #1\endcsname[1]{{\bf \color{#2} \capitalisewords{#1}:\,##1}}
  \expandafter\newcommand\csname #1cor\endcsname[2]{{\color{#2} \capitalisewords{#1}:\,\st{##1}{\bf ##2}}}
  \expandafter\newcommand\csname #1color\endcsname{#2}
}

\definecolor{cardinal}{rgb}{0.77, 0.12, 0.23}

\addReviewer{sebastian}{cardinal}
\addReviewer{fernando}{orange}
\addReviewer{steve}{Maroon}

\newcommand{\HSQCa}[0]{Hansen:2014eka}
\newcommand{\HSQCb}[0]{Hansen:2015zga}

\newcommand{\implement}[0]{Blanton:2021eyf}
\newcommand{\tetraquark}[0]{Hansen:2024ffk}

\newcommand{\BSnondegen}[0]{Blanton:2020gmf}
\newcommand{\BStwoplusone}[0]{Blanton:2021mih}
\newcommand{\IntegralEquations}[0]{%
Jackura:2020bsk,
Hansen:2020otl,
Mai:2021nul,
Dawid:2021fxd,
Dawid:2023jrj,
Jackura:2023qtp}

\newcommand{\cU}[0]{\mathcal U}

\preprint{{\small MIT-CTP/5774\\}}

\title{\boldmath Finite- and infinite-volume study of
$DD\pi$ scattering}

\author[a]{Sebastian M. Dawid}
\author[b,c]{, Fernando Romero-L\'opez}
\author[a]{, and Stephen R. Sharpe}

\affiliation[a]{Physics Department, University of Washington, Seattle, WA 98195-1560, USA}
\affiliation[b]{Albert Einstein Center, Institute for Theoretical Physics, University of Bern, 3012 Bern, Switzerland}
\affiliation[c]{Center for Theoretical Physics, Massachusetts Institute of Technology, Cambridge, MA 02139, USA}

\emailAdd{dawids@uw.edu}
\emailAdd{fernando.romero-lopez@unibe.ch}
\emailAdd{srsharpe@uw.edu}

\abstract{
We develop a comprehensive framework for extracting the pole position and properties of the doubly-charmed tetraquark $T_{\rm cc}^+(3875)$ from lattice QCD data using the relativistic three-particle formalism. This approach incorporates the effect of the one-pion exchange diagram in $DD\pi$ and $DD^*$ scattering, making it applicable at energies coinciding with the left-hand cut in the partial-wave projected $DD^*$ amplitude. We present an example application of this framework to existing lattice QCD data at $m_\pi = 280$ MeV. We solve the integral equations describing the $DD\pi$ reaction, use LSZ reduction to determine the corresponding $DD^*$ amplitude, and find the values of the infinite-volume two- and three-body $K$ matrices that lead to agreement with lattice $DD^*$ phase shifts within their uncertainties. Using these $K$ matrices in the three-particle quantization condition, we describe the finite-volume $DD^*$ spectrum and find good agreement with the lattice QCD energies. Our results suggest that, at this pion mass, the tetraquark appears as a pair of subthreshold complex poles whose precise location strongly depends on the value of the $DD\pi$ three-particle $K$ matrix.
}
\allowdisplaybreaks

\begin{document} 
\maketitle
\clearpage
\flushbottom

%%%%%%%%%%%%%%%%%%%%%%%%%%%%%%%%%%%%%%%%%
% SECTION
%%%%%%%%%%%%%%%%%%%%%%%%%%%%%%%%%%%%%%%%%
\section{Introduction}
\label{sec:intro}

The doubly-charmed tetraquark, $T_{\rm cc}^+(3875)$, has been observed by the LHCb collaboration as a resonance between the rest energies of $DD^*$ and $DD\pi$~\cite{LHCb:2021vvq, LHCb:2021auc}. Various phenomenological models have been able to predict a ${(I)J^P=(0)1^+}$ state in this mass window~\cite{Heller:1986bt, Zouzou:1986qh, Lipkin:1986dw, Carlson:1987hh, Janc:2004qn}. As for many other heavy-flavor multi-quark exotics, its proximity to the relevant threshold makes it a promising meson-meson molecule candidate; however, parallel descriptions as a compact diquark-antidiquark state or a mixture of different configurations were also successful. A recent review can be found in ref.~\cite{Chen:2022asf}. A fully controlled determination of this state from QCD has not been yet achieved; however, lattice computations are already exploring the $T_{\rm cc}^+$ at heavier-than-physical (and equal) light quark masses~\cite{Padmanath:2022cvl, Lyu:2023xro, Chen:2022vpo, Ortiz-Pacheco:2023ble, Collins:2024sfi, Whyte:2024ihh}.

The physical tetraquark decays exclusively into the $DD\pi$ state, requiring the inclusion of three-body thresholds and physical one-pion exchanges (OPEs) in phenomenological analyses~\cite{Albaladejo:2021vln, Du:2021zzh, Achasov:2022onn, Wang:2023iaz, Zhang:2024dth, Abolnikov:2024key}. However, when the pion is sufficiently heavy, the $D^*$ meson becomes a $D\pi$ bound state, and the $DD^*$ and $DD\pi$ thresholds are inverted. Rather than a three-body unstable resonance, the tetraquark may appear as a shallow two-body bound or a virtual state pole in the $DD^*$ scattering amplitude for a considerable range of quark masses~\cite{Collins:2024sfi, Abolnikov:2024key}. Although threshold inversion could, in principle, allow one to effectively describe $T_{cc}^+$ as a two-body system, it has become clear that a rigorous determination of the $DD^*$ amplitude must incorporate three-body effects~\cite{Meng:2023bmz, Hansen:2024ffk, Abolnikov:2024key}.

The main reason is that, at heavy pion masses, one-pion exchanges become virtual and induce a non-analyticity (left-hand cut) in the partial-wave projected $DD^*$ scattering amplitude near the expected $T_{cc}^+$ energy. A near-threshold branch point invalidates the application of commonly used finite- and infinite-volume two-body techniques, such as the L\"uscher formalism~\cite{Luscher:1986n2} or na\"ive effective-range expansions (ERE)~\cite{Du:2023hlu, Du:2024snq}. It complicates the extraction of the tetraquark pole position using lattice QCD and has motivated several authors to resolve this issue using different methods~\cite{Meng:2023bmz, Raposo:2023oru, Bubna:2024izx, Hansen:2024ffk}. 

Among these proposals, ref.~\cite{Hansen:2024ffk} advocates using the three-body finite-volume formalism~\cite{\ThreeBody}, specifically that in the generic relativistic field theory (RFT) approach of refs.~\cite{\HSQCa,\HSQCb}, to describe the $DD^*$ and $DD\pi$ spectrum, since it naturally includes the effects from the OPE diagrams. The first step in this approach is to apply the three-body quantization condition to the finite-volume $DD^*$ and $DD\pi$ energy levels to determine the properties of $DD\pi$ scattering at generic isospin in isosymmetric QCD (including $I=0$ for the $T_{\rm cc}^+$). The resulting infinite-volume information takes the form of two- and three-meson $K$ matrices. The former describe interactions of particles in the two-particle subchannels, $DD$ and $D\pi$, and incorporate the $D^*$ meson as a resonance or bound-state pole in the $I=1/2$, $p$-wave $D\pi$ scattering amplitude. The three-body $K$ matrix parametrizes short-range three-particle interactions.

In this approach, there is then a second step, in which
two- and three-meson $K$ matrices are translated into the elastic $DD\pi$ scattering amplitude by using relativistic integral equations~\cite{Hansen:2015zga, Briceno:2018aml, Briceno:2018mlh, Jackura:2020bsk, Jackura:2022gib, \tetraquark}. The solutions of the integral equations provide a three-body amplitude consistent with $S$ matrix unitarity, which can be analytically continued to energies near the OPE branch points and $D^*$ pole. By applying the Lehmann-Symanzik-Zimmermann (LSZ) reduction formula to the $DD\pi$ solution of these equations, one can extract the $DD^*$ amplitude at energies below the left-hand branch point and reliably determine the pole position of $T_{cc}^+$~\cite{Dawid:2023jrj, Hansen:2024ffk}. One of the advantages of this approach is that it is applicable at any pion mass, allowing one to analyze and connect results obtained from lattice ensembles characterized by both stable and unstable $D^*$.

In this work, we present a complete implementation of the strategy sketched above, in which we carry out all the steps needed to connect lattice QCD energies to the $T_{cc}^+$. We numerically implement the proposal of ref.~\cite{Hansen:2024ffk} in both finite and infinite volume and show how to extract the pole position of the tetraquark. To make contact with existing lattice QCD results, we focus on the case of a bound $D^*$ meson, such that the tetraquark appears in the $DD^*$ scattering amplitude. For concreteness, we tune the scattering parameters to resemble the lattice QCD setting of Refs.~\cite{Padmanath:2022cvl, Collins:2024sfi}.

Our work consists of three main parts. In the first, we solve the relativistic three-body integral equations in the $J^P=0^-$ and $J^P=1^+$ channel corresponding to the $T_{\rm cc}^+$. This requires generalization of existing tools~\cite{Hansen:2020otl, Jackura:2020bsk, Dawid:2023jrj, Dawid:2023kxu, Garofalo:2022pux} to a system with non-zero partial waves, non-degenerate masses, and general three-body couplings.\footnote{While this work was in preparation, ref.~\cite{Briceno:2024ehy} appeared, which also addresses non-zero partial waves.} We constrain the two-body subchannel interactions using the available $D\pi$ lattice data~\citep{Mohler:2012na, Becirevic:2012pf, Moir:2016srx, Gayer:2021xzv, Yan:2024yuq} and heavy-light meson Chiral Perturbation Theory (ChPT)~\cite{Georgi:1990um, Wise:1992hn, Wise:1993wa}, and tune the three-body $K$ matrix such that the results are in reasonable agreement with the $s$-wave $DD^*$ scattering phase shift of ref.~\cite{Padmanath:2022cvl}. 
The resulting amplitude is continued to the unphysical Riemann sheet and evaluated below the $DD^*$ threshold where we identify the $T_{cc}^+$ as a sub-threshold complex pole, in agreement with other analyses ~\citep{Du:2023hlu, Meng:2023bmz, Collins:2024sfi, Abolnikov:2024key,Gil-Dominguez:2024zmr}.

In the next part, we solve the three-particle quantization condition using the same $K$ matrices that appear in the infinite-volume equations. This allows us to calculate the $DD\pi$ and $DD^*$ energy levels and compare them to lattice $DD^*$ levels obtained from ref.~\cite{Padmanath:2022cvl}. Our approach reproduces the lattice spectrum closely, aside from states that can be interpreted as having primarily a $DD\pi$ composition, which were not determined in the lattice calculation. This agreement provides a test of the full formalism.

In refs.~\cite{Padmanath:2022cvl,Collins:2024sfi} the lattice energies are analyzed using the two-particle quantization for the $DD^*$ system, leading to results for the $J^P=1^+$ ($s$-wave) and $J^P=0^-$ ($p$-wave) phase shifts. As noted above, it is known that this approach fails near the left-hand cut, and we investigate this issue in the final part of this work. We do so by solving the two-particle quantization condition using the $J^P=1^+$ and $0^-$ phase shifts obtained from the solution to the integral equations. We observe that, as expected, the two- and three-particle quantization conditions give similar energy levels above the $DD^*$ threshold, while there are significant discrepancies between the levels lying near the left-hand cut. Consequently, several low-lying energy levels observed in the lattice simulations cannot be used as input into the two-body quantization condition but can be analyzed using the three-body one.

We organize this paper as follows. In \Cref{sec:inf-vol}, we describe the scattering kinematics and the details of the infinite-volume formalism: relativistic integral equations, partial-wave projection, and the models of two- and three-body $K$ matrices. We also discuss the singularities of the $DD\pi$ amplitude and its LSZ reduction to that for $DD^*$. In \Cref{sec:fin-vol}, we review the finite-volume formalism derived in ref.~\cite{Hansen:2024ffk}, as well as the two-body quantization condition for the $DD^*$ system, including partial-wave mixing. In \Cref{sec:results}, we present the results of the numerical implementation of the formalism. We show the $DD^*$ amplitude extracted from the solutions of the integral equations for $DD\pi$ in the $J^P=1^+$ and $0^-$ channels. We also present the finite-volume energies obtained from the quantization conditions and compare them to the original data set of ref.~\cite{Padmanath:2022cvl}. Based on these numerical results, we discuss the applicability of the three- and two-particle formalisms near the left-hand cut. We summarize this work in \Cref{sec:conclusion}. 

We include four appendices. In \Cref{app:pw}, we discuss formal aspects of the partial-wave projection of the OPE amplitude, the three-body $K$ matrix, and the two-particle amplitude $\cM_2$. In \Cref{app:pw_expr,app:pw_K3}, we summarize, respectively, the projected matrix elements of the OPE and the three-body $K$ matrix in the channels used in this work. In \Cref{app:chpt}, we compare our $D\pi$ amplitude model to the tree-level result in the heavy-light meson ChPT.

%%%%%%%%%%%%%%%%%%%%%%%%%%%%%%%%%%%%%%%%%
% SECTION
%%%%%%%%%%%%%%%%%%%%%%%%%%%%%%%%%%%%%%%%%
\section{Infinite-volume formalism for the $T^+_{\rm cc}$}
\label{sec:inf-vol}

To describe the $T_{cc}^+$, we consider an elastic three-body scattering process, ${D D \pi \to D D \pi}$, in the isospin $I=0$ channel. In line with existing lattice QCD results, the isospin limit is assumed throughout. We consider heavier-than-physical pions, such that the $D^*$ is a $D \pi$ bound state rather than a resonance.

The formalism that we use has been developed in the
Relativistic Field Theory (RFT) approach of refs.~\cite{\HSQCa,\HSQCb}. The integral equations relevant to the $DD\pi$ system were derived in ref.~\cite{\tetraquark}; here, we give a more detailed presentation, focusing on the aspects relevant to practical implementation. Methods of solution of the RFT integral equations have been developed in refs.~\cite{\IntegralEquations}. In the following, we present the extensions necessary for nondegenerate systems with multiple subchannels, as well as allow for the inclusion of three-particle K matrices.

%%%%%%%%%%%%%%%%%%%%%%%%%
%%%%%%%%%%%%%%%%%%%%%%%%%
\subsection{$DD\pi$ scattering amplitude and kinematics}
\label{sec:DDpi-kinematics}

The connected $3 \to 3$ scattering amplitude, $\cM_3$, depends on eight independent variables. To organize them conveniently, we divide the three particles into a \emph{pair} and a \emph{spectator}, analyzing the process as a quasi-two-body coupled-channel reaction. It is described by the so-called unsymmetrized amplitude\footnote{For simplicity, we drop the superscript $(u,u)$ used in ref.~\cite{Hansen:2024ffk}, as we work exclusively with the pair-spectator amplitudes.}, $\cM^{(ij)}_3$. Indices $i,j$ label, respectively, the final and initial pair-spectator channels. Both indices are drawn from $\{1,2\}$: the choice ``1'' indicates that the $D$ is a spectator and $D\pi$ forms the isospin $I=1/2$ pair, while ``2'' means that the $\pi$ spectates and the pair is $DD$, which has isospin $I=1$. Similarly, we denote the mass of the $D$ and $\pi$ mesons as $m_1 = m_D$ and $m_2 = m_\pi$, respectively. Once $\cM^{(ij)}_3$ is known, it can be symmetrized with respect to all possible pair-spectator choices to recover the genuine amplitude $\cM_3$. This procedure is described in Sec.~3.3 of ref.~\cite{Hansen:2024ffk}.

We consider the reaction in the three-body center-of-mass (c.m.)~frame, where the total four-momentum of three particles is $P = (P^0,\P) = (E, \bm{0})$. We label the momentum of the incoming and outgoing spectator as $\k_j$ and $\p_i$, respectively. In the following, to lighten the notation, we will drop the $i,j$ subscripts from the momentum variables as long as it does not lead to ambiguity. It should be kept in mind, however, that the masses of the spectator and pair particles depend upon the choice of indices. All momenta are on-shell, and their magnitudes are denoted $p = |\p|$ and $k = |\k|$. A spectator momentum $\q$ fixes the invariant mass of the corresponding pair to be
    %%%%%
    \begin{align}
    \label{eq:sigma_variable}
    \sigma_q^{(i)} \equiv \sigma_{q_i}  = (E - \omega_{q_i})^2 - {q_i}^2 \, .
    \end{align}
    %%%%%
Here, $\omega_q^{(i)} = \omega_{q_i} = \sqrt{m_i^2 + q_i^2}$, denotes the particle's energy. Again, when no ambiguity is present, we drop indices $i,j$.

In addition to the total energy and two-body invariant masses, the amplitude depends on five angular variables describing the orientation of the external momenta. Following ref.~\cite{Jackura:2023qtp}, we choose one of them to be the angle $\Theta$ between the incoming and outgoing spectator momenta in the three-body c.m.~frame, i.e.~$\hat{p} \cdot \hat{k} = \cos \Theta$. We specify the next two variables in the c.m.~frame of the initial pair, a frame defined by the condition $\P_{\k}^\star = \P^\star - \k^\star = 0$, where we use $\star$ to denote momenta boosted to that frame. The variables are the polar and azimuthal angle of the momentum of one of the particles in the pair,%
%%%%
\footnote{The chosen particle is denoted the ``primary'' member of the pair and is taken to be the $D$ in the $D\pi$ pair, following Refs.~\cite{\BStwoplusone, \implement, \tetraquark}.
For the $DD$ pair, the choice is arbitrary.}
%%%%
$\Omega_k^\star = (\vartheta^\star_k, \varphi_k^\star)$. These are defined in a coordinate system such that the $z$ axis in the pair c.m.~frame is aligned along the pair's momentum in the overall c.m.~frame, $\hat{z}_k = -\hat{k}$. In all these variables, the subscript, here $k$, indicates the momentum of the corresponding spectator. The dependence on $\Omega_k^\star$ is decomposed into spherical harmonics, with the angular momentum of the pair denoted $s$, and its $z$ component denoted $\lambda$. The latter is the helicity of the pair, given the choice of $z$ axis described above.%
%%%%
\footnote{Note that all previous papers using the RFT approach, e.g., refs.~\citep{\HSQCa,Hansen:2024ffk}, use the notation $\ell$ instead of $s$. Also, the azimuthal component $m$ in these works is defined relative to an unspecified choice of $z$ axis, whereas here we use the helicity $\lambda$.}
%%%%
The last two variables are analogous angles for the final-state pair, defined in the $\P_p^\star = \P^\star - \p^\star=0$ frame. For simplicity, we use the same $\star$ symbol to denote variables in the final and initial pair reference frames, even though they are different. It will always be clear from the context to which frame we refer. The angular variables are $\Omega_p^\star = ( \vartheta^\star_p,\varphi_p^\star)$, and the corresponding angular momentum and helicity are denoted $s'$ and $\lambda'$. The reader can find further details on the angular variables in Refs.~\cite{Jackura:2018xnx, Jackura:2023qtp} and in App.~\ref{app:pw}.

To summarize, given the pair-spectator division, the amplitude depends on a triplet of energy-like variables, either $(p,k, E)$, or, equivalently, $(\sigma_p, \sigma_k, E)$, and on five scattering angles, for which we choose $\Theta$, $(\vartheta^\star_p, \varphi_p^\star)$, and $(\vartheta^\star_k, \varphi_k^\star)$. Since, in the overall c.m.~frame, angular momentum and parity are good quantum numbers, we wish to project the amplitude onto definite $J^P$. The projected amplitude is obtained in three steps~\cite{Jackura:2018xnx, Jackura:2023qtp, Chung:1971ri}. The details are reviewed in \Cref{app:pw}; here, we provide a summary. First, we project onto the basis of definite spin and helicity of external pairs,
    %%%%%
    \begin{align}
    \cM^{(ij)}_{3, s'\lambda';s\lambda}(p,k; E,\Theta) = \frac{1}{4\pi} \int d\Omega_p^\star \int d\Omega_k^\star \, Y^*_{s'\lambda'}(\Omega_p^\star) \, \cM_3^{(ij)}(p, \Omega_p^\star, k, \Omega_k^\star; E,\Theta) \, 
    Y_{s\lambda}(\Omega_k^\star) \, .
    \label{eq:projectM}
    \end{align}
    %%%%%
Here $ Y_{s'\lambda'}$ are the standard spherical harmonics, and the normalization follows that of ref.~\cite{\HSQCa}.%
%%%%
\footnote{The choice of which spherical harmonic is complex-conjugated is opposite from that introduced in ref.~\cite{\HSQCa} and used in all subsequent works in the RFT approach. While this choice is ultimately a convention, we prefer that given here as it is consistent with the standard choice in Quantum Mechanics, namely $\braket{\hat n|\ell m} = Y_{\ell m}(\hat n)$.}
%%%%
The overall phase of the amplitude on the left-hand side depends on the choice of $x$ and $y$ axes used for the helicity projections. While the choice is arbitrary, a consistent convention must be used throughout, and one such is described in~\Cref{app:pw}. The second step projects the amplitude onto definite $J$ and is achieved by,
    %%%%%
    \begin{align}
    \cM^{(ij) J}_{3, s'\lambda';s\lambda}(p,k;E) = \frac{1}{2} \int d\!\cos\Theta \, \cM^{(ij)}_{3,s'\lambda';s\lambda}(p, k; E, \Theta) \, 
    d_{\lambda \lambda'}^J(\Theta) \, .
    \label{eq:Jprojection}
    \end{align}
    %%%%%
The final step is to transform from the spin-helicity basis to the LS basis, in which $\ell'$ and $\ell$ refer, respectively, to the relative orbital angular momenta between the final and initial spectator and the corresponding pair. This is done using the recoupling coefficients, 
    %%%%%
    \begin{align}
    \cC^{J}_{\ell s \lambda} = \sqrt{\frac{2\ell+1}{2J+1}} \, \langle J, -\!\lambda | \ell, 0; s, -\!\lambda \rangle \, , 
    \label{eq:recoupling}
    \end{align}
    %%%%%
leading to the result,
    %%%%%
    \begin{align}
    \cM^{(ij) J}_{3, \ell' s';\ell s}(p,k;E) = \sum_{\lambda'=-s'}^{s'} 
    \sum_{\lambda = -s}^{s} 
    C^J_{\ell' s' \lambda' } \, \cM^{(ij)J}_{3,s'\lambda' ; s \lambda}(p,k;E) \, 
    C^J_{\ell s \lambda } \, .
    \label{eq:M3JLS}
    \end{align}
    %%%%%
The negative signs appearing in \Cref{eq:recoupling} are explained in~\Cref{app:pw}. In fact, for the channels we consider here, the properties of Clebsch-Gordon coefficients are such that the result holds with these negative signs removed.

Given that the amplitude depends on many variables, we implement a compact notation in which it is treated as a generalized matrix in a multi-dimensional space,
    %%%%%
    \begin{align}
    \cM^{(i j) J}_{3, \ell' s';\ell s}(p,k; E) \equiv \cM^{(i j)}_{3, \ell' s';\ell s}(p,k)  = \left[\,\bm{\cM}_3 \,\right]_{i p \ell' s'; j k \ell s} \, ,
    \end{align}
    %%%%%
Moreover, since $E$ and $J$ are both conserved, we keep the dependence on these quantities implicit. We define matrix multiplication via the summation over discrete indices and integration over continuous momenta,
    %%%%%
    \begin{align}
    \label{eq:generalized-matrix}
    \left[\,\bm{\cA} \, \bm{\cB} \,\right]_{i p \ell' s'; j k \ell s} = \sum_{n=1}^2 \sum_{\ell''_n = 0}^{\ell_{\rm max}^{(n)}} \sum_{ s''_n = 0}^{s_{\rm max}^{(n)}} \int\limits_0^{q_{\rm max}^{(n)}} 
    \frac{dq_n \, q_n^2}{(2\pi)^2 \omega_{q_n}} 
    \cA^{(i n)}_{\ell' s';\ell''_n s''_n}(p,q_n) \, 
    \cB^{(n j)}_{\ell''_n s''_n;\ell s}(q_n,k) \, .
    \end{align}
    %%%%%
Here, we keep the intermediate on-shell pair-spectator channel index $n$ explicit in all related variables: the magnitude of the spectator's momentum, $q_n$, intermediate pair's spin $s''_n$, and pair-spectator's orbital angular momentum, $\ell_n''$. The momentum cutoff $q_{\text{max}}^{(n)}$ depends on the intermediate channel $n$ and is determined by the positions of the left-hand singularities in the two-body scattering amplitudes. We specify it below.

In principle, infinitely many $(\ell, s)$ combinations contribute to the sums in \Cref{eq:generalized-matrix}. In practice, they must be truncated to reduce the equations for $\cM_3$ to a numerically solvable problem. Such a truncation is justified by the fact that amplitudes with higher values of angular momentum are suppressed near the threshold. For the $J^P=1^+$ tetraquark channel considered here, we neglect spins higher than $s_{\rm max}^{(1)} =1$ in the $D\pi$ subchannels. Due to the presence of $D^*$, we assume dominance of the $D\pi$ $p$-wave interactions and include the lower $s$-wave scattering for consistency. The $DD$ channel is considered weakly interacting, and we restrict the spin space even further by including only the $s$-wave interaction in our description, $s_{\rm max}^{(2)} = 0$. Finally, we set $\ell_{\rm max}^{(n)}=2$. These restrictions lead to the following $(\ell, s)$ combinations in the $J^P=1^+$ channel, 
    %%%%%
    \begin{equation}
    \label{eq:angular_momenta_J1}
    (\ell_1, s_1) = (1,0), \, (0,1), \, (2,1), \quad  (\ell_2, s_2) = (1,0) \, .
    \end{equation}
    %%%%%
Note that this choice allows for the partial-wave mixing between the $LS$ quantum numbers $(\ell_1,s_1) = (0,1)$ and $(\ell_1,s_1) = (2,1)$.
To compare our results to those of ref.~\cite{Padmanath:2022cvl}, we also study scattering in the $J^P=0^-$ partial wave, for which the combinations considered are
    %%%%%%
    \begin{equation}
    (\ell_1, s_1) = (0,0), \, (1,1), \quad  (\ell_2, s_2) = (0,0) \, .
    \end{equation}
    %%%%%% 

%%%%%%%%%%%%%%%%%%%%%%%%%
%%%%%%%%%%%%%%%%%%%%%%%%%
\subsection{Three-body integral equations}

The pair-spectator amplitude $\bm{\cM}_3$ is given by a sum of two terms~\cite{Hansen:2015zga, Blanton:2021mih, Hansen:2024ffk},
    %%%%%
    \begin{align}
    \label{eq:two-terms-of-amplitude}
    \bm{\cM}_3 = \bm{\cD} + \bm{\cM}_{\text{df},3} \, .
    \end{align}
    %%%%%
The first is called the \emph{ladder} amplitude and is determined solely by the pair two-body amplitude, $\bm{\cM}_2$, and by the kernel describing one-particle exchange between the initial and final pair, $\bm{G}$. The second term, the ``divergence-free'' amplitude, requires the presence of at least one short-range three-body interaction, given by the three-particle $K$ matrix, $\cK_3$, and involves
final and initial two-body rescatterings.%
%%%%
\footnote{In previous RFT works, the three-particle $K$ matrix has been referred to as $\cK_{{\rm df},3}$~\cite{Hansen:2024ffk}. Here, for the sake of brevity, we drop the subscript ``${\rm df}$''.}
%%%%
In the following, we recall the expressions for $\bm{\cD}$ and $\bm{\cM}_{\text{df},3}$ using our compact notation.

%%%%%%%%%%%%%%%%%%%%%%%%%
%%%%%%%%%%%%%%%%%%%%%%%%%
\subsubsection{The ladder equation}
\label{sec:ladder}

The ladder amplitude is given by the on-shell partial-wave-projected integral equation, written here in generalized matrix form,
    %%%%%
    \begin{align}
    \bm{\cD}= - \bm{\cM}_2 \, \bm{G} \, \bm{\cM}_2 - \bm{\cM}_2 \, \bm{G} \, \bm{\cD} \, .
    \end{align}
    %%%%%
To simplify its solution, one typically removes the external two-body amplitudes, 
    %%%%%
    \begin{align}
    \label{eq:D-to-d}
    \bm{\cD} = \bm{\cM}_2 \, \bm{d} \, \bm{\cM}_2 \, ,
    \end{align}
    %%%%%
defining the ``amputated'' partial-wave projected ladder amplitude, $\bm{d}$, free of the two-body singularities in the $\sigma_p$ and $\sigma_k$ variables. It satisfies
    %%%%%
    \begin{align}
    \label{eq:d_formal}
    \bm{d}= - \bm{G} - \bm{G} \, \bm{\cM}_2 \, \bm{d} \, ,
    \end{align}
    %%%%%
whose formal solution is,
    %%%%%
    \begin{align}
    \label{eq:solution_formal}
    \bm{d} = - \left[ \bm{1} + \bm{G} \, \bm{\cM}_2 \right]^{-1} \bm{G} \, .
    \end{align}
    %%%%%

%%%%%%%%%%%%%%%%%%%%%%%%%
%%%%%%%%%%%%%%%%%%%%%%%%%
\paragraph{Two-body amplitudes:}

The ladder equation depends on the partial-wave-projected two-body amplitude $\bm{\cM}_2$, fully describing interactions in the  $D\pi$ and $DD$ subchannels. In our generalized notation, due to the energy and angular momentum conservation, it is a diagonal matrix,
    %%%%%
    \begin{align}
    \left[\,\bm{\cM}_2 \,\right]_{i p \ell' s'; j k \ell s} = \eta_i \, \delta_{i p \ell' s'; j k \ell s} \, \cM_{2,s}^{(i)}(p) \, , 
    \label{eq:M2new}
    \end{align}
    %%%%%
where the generalized Dirac delta function is,
    %%%%%
    \begin{equation}
    \delta_{i p \ell' s'; j k \ell s} = \tilde{\delta}(p-k) \, \delta_{ij} \, \delta_{\ell'\ell} \, \delta_{s's}  \,,
    \label{eq:generalized_delta}
    \end{equation}
    %%%%%
with the momentum part,
    %%%%%
    \begin{equation}
    \tilde{\delta}(p-k) = (2\pi)^2 \, (\omega_{p}/p^2) \, \delta(p-k)\,.
    \label{eq:delta_tilde}
    \end{equation}
    %%%%%
The symmetry factors are $\eta_1 = 1$ and $\eta_2 = 1/2$, corresponding to the fact that the pair is composed of identical particles if the spectator is a pion. The derivation of this projected form of $\cM_2$ is described in \Cref{app:projectM2}. We represent the diagonal elements of $\bm{\cM}_2$ in $K$-matrix form,
    %%%%%
    \begin{align}
    \label{eq:M2}
    [\cM_{2,s}^{(n)}(p)]^{-1} = [\cK_{2,s}^{(n)}(\sigma_p) ]^{-1} - i \rho^{(n)}(\sigma_p) \, .
    \end{align}
    %%%%%
Here the two-body phase space is,
    %%%%%
    \begin{align}
    \label{eq:phase-space}
    \rho^{(n)}(\sigma_p) &=  \frac{\eta_n q_p^\star }{8 \pi \sqrt{\sigma_p}} \, .
    \end{align}
    %%%%%
where,
    %%%%%
    \begin{align}
    \label{eq:qpstar}
    q_{p_n}^\star &= \frac{\lambda^{1/2}(\sigma_{p}^{(n)}, m^2_{n_1}, m^2_{n_2})}{2 \sqrt{\sigma_{p}^{(n)}}} \, ,
    \end{align}    
    %%%%%
and $\lambda(x,y,z)$ is the K\"allen triangle function. Equation~\eqref{eq:qpstar} gives the magnitude of the relative momentum (in the pair's rest frame) between particles in the pair corresponding to the spectator of momentum $p_n$ (here, we explicitly indicate the spectator's channel for clarity.) Note that it is different from the intermediate spectator integration momentum of \Cref{eq:generalized-matrix}.  The indices $n_1, n_2$ label the two particles in the pair corresponding to spectator $n$, which, in the conventions of ref.~\cite{\tetraquark}, are $\{n_1,n_2\}=\{D,\pi\}$ for $n = 1$, and $\{n_1,n_2\}=\{D,D\}$ for $n=2$. We orient the cuts of $\cM_2$ to run to complex infinity, i.e., we write,
    %%%%%
    \begin{align}
    \label{eq:two-body-cuts}
    \lambda^{1/2}(\sigma_{p}^{(n)}, m^2_{n_1}, m^2_{n_2}) = i \sqrt{\sigma_{p}^{(n)} - (m_{n_1} - m_{n_2})^2} \, \sqrt{(m_{n_1} + m_{n_2})^2 - \sigma_{p}^{(n)}}  \, ,
    \end{align}
    %%%%%
assuming the canonical definition of the complex square root. The branch point at $\sigma_{\rm thr}^{(n)} = (m_{n_1} + m_{n_2})^2$ is required by elastic two-body unitarity and corresponds to the two-body thresholds, $\sigma_{\rm thr}^{(1)} = (m_D+m_\pi)^2$ for $D\pi$ and $\sigma_{\rm thr}^{(2)} = 4 m_D^2$ for $DD$.

Dynamical information about two-body scattering is contained in the two-body $K$ matrix, $\cK_2$. We model these interactions using the effective-range expansion (ERE)~\cite{Blatt:1949zz, Bethe:1949yr}, truncating to two terms,
    %%%%%
    \begin{align}
    \label{eq:k-matrix}
    [\cK_{2,s}^{(n)}(\sigma_p)]^{-1} &=  \frac{\eta_n}{8 \pi \sqrt{\sigma_p} \, (q_p^\star)^{2s}} \, \left( - \frac{1}{a_s^{(n)}} + \frac{1}{2} \, r_s^{(n)} \, (q_p^\star)^2 \right) \, .
    \end{align}
    %%%%%
In the following, we choose values of the parameters $a_{s}^{(n)}$ (two-body scattering lengths), and $r_{s}^{(n)}$ (two-body effective ranges), such that 
there are no poles on the physical sheet in $s$-wave $D\pi$ or $DD$ scattering. For $p$-wave $D\pi$ scattering, however, we choose values such that a $D^*$ bound-state pole develops in  $\cM_{2,1}^{(1)}$. In general, a pole occurs when the relative momentum takes the purely imaginary value $q_p^\star = q_0 = i |q_0|$ satisfying,
    %%%%%
    \begin{align}
    \label{eq:pole-condition}
    - \frac{1}{a_s^{(n)}} + \frac{1}{2} \, r_s^{(n)} \, q_0^2 - i q_0^{2s+1} = 0 \,.
    \end{align}
    %%%%%
Furthermore, the left-hand side must be a decreasing function of $q_p^{\star 2}$ at $q_p=q_0$ for the pole to correspond to a physical state; see chapter 12 in ref.~\cite{Taylor:1972pty}. Explicitly, we tune the ERE parameters such that, 
    \begin{align}
    \label{eq:D-star-pole-spec}
    q_0 = \frac{\lambda^{1/2}(m_{D^*}^2, m_D^2, m_\pi^2)}{2 m_{D^*}} \,,
    \end{align}
    %%%%%
or,
    \begin{align}
    \label{eq:D-star-pole-sigma}
    m_{D^*} = \sqrt{ m_\pi^2 + q_0^2 } + \sqrt{ m_D^2 + q_0^2} \,.
    \end{align}    
    %%%%%
We discuss this tuning in more detail in \Cref{sec:results}.

To complete the discussion of the two-body amplitudes, we briefly comment on their other singularities. As can be seen from \Cref{eq:two-body-cuts}, our parametrization of $\cM_{2,s}^{(n)}$ has an (unphysical) pseudo-threshold left-hand branch point at $\sigma_{\rm psth}^{(n)}$, where $\sigma_{\rm psth}^{(1)} = (m_D - m_\pi)^2$ and $\sigma_{\rm psth}^{(2)} = 0$. Moreover, the partial-wave projected amplitudes should include two-pion exchange left-hand cuts at 
$\sigma_{2\pi,{\rm lhc}}^{(1)} = (m_D^2 - m_\pi^2)$ and $\sigma_{2\pi,{\rm lhc}}^{(2)} = 4(m_D^2 - m_\pi^2)$ that are absent from our model. All these singularities, however, lie outside of the allowed region for our choice of cutoff momentum $q_{\text{max}}^{(n)}$, as will be discussed below. Another issue concerns the fact that, as is well known, the ERE parametrization can produce unphysical poles for a wide range of scattering parameters, some of which can occur at complex energies on the first Riemann sheet of the two-body amplitude~\cite{Adhikari:1982bzs, Ebert:2021epn}. 
When choosing values of the ERE parameters, we ensure that such poles are distant from the $D\pi$ and $DD$ thresholds, so that they do not significantly affect the physics of the two-body subchannels.

%%%%%%%%%%%%%%%%%%%%%%%%%
%%%%%%%%%%%%%%%%%%%%%%%%%
\paragraph{The OPE amplitude:}

The second important quantity in the ladder equation is the kinematic one-particle exchange amplitude. In our generalized notation, it is,~\cite{\BStwoplusone,\tetraquark}
    %%%%%
    \begin{align}
    [\, \bm{G} \,]_{ip\ell's';jk\ell s} = c_{ij} \, 
    (-1)^{s' d_{ij}}G^{(ij)J}_{\ell's';\ell s}(p,k) (-1)^{s d_{ji}} \, ,
    \label{eq:Gmat}
    \end{align}
    %%%%%
where the double-index symmetry factor, $c_{ij}$, takes the values $c_{11} = 1, c_{12} = c_{21} = -\sqrt{2}$, and $c_{22} = 0$,
while the additional phases are controlled by
$d_{ij}$, which vanishes except for $d_{12}=1$.
Note that the $d$ on the right-hand side of $G^{(ij)}$ has indices reversed compared to that on the left-hand side.
The amplitude vanishes for the $(ij) = (22)$ channel since it is impossible to exchange a particle between two $DD$ pairs. In the spin-helicity basis, the one-particle exchange amplitude has the form,%
%%%%
\footnote{We remind the reader that we are using a convention for the complex conjugation of spherical harmonics that is opposite from that in the previous RFT literature. We also note that $G$ is usually presented using spherical harmonics corresponding to a space-fixed $z$ axis. The transformation to the helicity basis used here (defined relative to the pair's momentum) is unitary and can be consistently applied throughout the integral equations.}
    %%%%%
    \begin{align}
    \label{eq:ope-spin-helicity}
    G_{s' \lambda'; s \lambda}^{(i j)}(p,\Omega_p^\star;k,\Omega_k^\star;E,\Theta) =  \, Y_{s' \lambda'}^*(\hat{\bm{k}}_{p}^\star) \,
    \left( \frac{k_{p}^\star}{q_{p}^\star} \right)^{s'} 
    %%%
    \frac{ 4\pi \, H_{i j}(p,k)
     }{b_{p k}^2 - m_{i j}^2 + i \epsilon}
    %%%
    \left( \frac{p_{k}^\star}{q_{k}^\star} \right)^{s} \, Y_{s \lambda}(\hat{\bm{p}}_{k}^\star) \, ,
    \end{align}
    %%%%%
where $b_{p k}^2 = (E - \omega_{p} - \omega_{k})^2 - (\p + \k)^2$ and $m_{ij}$ is a mass of exchanged particle, $m_{11} = m_\pi$, $m_{12} = m_{21} = m_D$. The vector $\k_{p}^\star$ ($\p_{k}^\star$) denotes the momentum of the initial (final) spectator in the final (initial) pair's rest frame. 

The function $H_{ij}(p,k)$ is a smooth cutoff function that enters in the derivation of the three-body formalism in the RFT approach~\cite{Hansen:2014eka, Hansen:2015zga, Blanton:2021mih}, and takes the separated form $H_{i j}(p,k) = H_i(p) H_j(k)$.
It always equals unity at the position of the pole in \Cref{eq:ope-spin-helicity}, and smoothly drops to zero as $p,k$ increase. 
There is a family of allowed choices of $H(p)$ and the specific choice we use is\footnote{%
The cutoff function in \Cref{eq:ope-spin-helicity}, while equaling unity at the pole, serves as an effective form factor for the $D^* D\pi$ vertices that drops below unity when the pair invariant mass equals $m_{D^*}^2$. 
},
    %%%%%
    \begin{align}
    \label{eq:cutoff-fi}
    H_i(p) = J(f_i(\sigma_p)) ~~~\text{and}~~~f_i(\sigma_p) = (1+\epsilon_H) \, \frac{\sigma_p - \sigma_{\rm min}^{(i)} }{ \sigma_{\rm thr}^{(i)} - \sigma_{\rm min}^{(i)} } \, ,
    \end{align}
    %%%%%
with $\sigma_{\rm min}^{(1)} = \sigma_{2\pi,\rm lhc}^{(1)}$ and $\sigma_{\rm min}^{(2)} = \sigma_{2\pi, {\rm lhc}}^{(2)}$. 
The $J$ function is,
    %%%%%%
    \begin{align}
    \label{eq:cutoff-J}
    J(x) & =
    \begin{cases}
    0 \,, & x \le 0 \, , \\
    %%%
    \exp \left( - \frac{1}{x} \exp \left [-\frac{1}{1-x} \right] \right ) \,, 
    & 0<x \le 1 \, , \\
    %%%
    1 \,, & 1<x \, .
    \end{cases}
    \end{align}
    %%%%%
This form implies that the integration in \Cref{eq:generalized-matrix} is performed up to the energy-dependent momentum cutoff,
    %%%%%
    \begin{align}
    \label{eq:cutoff_moms}
    q_{\text{max}}^{(n)} = \frac{\lambda^{1/2}(E^2, \sigma_{\rm min}^{(n)}, m_D^2)}{2 E} \, .
    \end{align}
    %%%%%
As noted earlier, our choices of $\sigma_{\rm min}^{(i)}$ ensure that the $\pi D$ and $DD$ amplitudes are evaluated only above the left-hand cuts arising from a two-pion exchange.

The parameter $\epsilon_H$ entering in $f_i(\sigma_p)$ is a positive constant introduced to ensure that the cutoff function reaches unity below the pair threshold, thus avoiding a potential source of power-law finite-volume effects~\cite{Blanton:2020gmf}. In practice, for our choice of $J$, such effects are highly suppressed even if $\epsilon_H=0$, and we choose this value.

To convert $G^{(ij)}$ in \Cref{eq:ope-spin-helicity} to the $JLS$ basis appearing in \Cref{eq:Gmat}, we must perform the integrations and basis transformations applied to $\cM_3$ in \Cref{eq:projectM,eq:M3JLS} above. This non-trivial task was accomplished in ref.~\cite{Jackura:2023qtp} at a high level of generality. In~\Cref{app:pw}, we describe the relevant manipulations in the specific case of interest, and in~\Cref{app:pw_expr} we list all the relevant matrix elements for $J^P=1^+$ and $0^-$.

%%%%%%%%%%%%%%%%%%%%%%%%%
%%%%%%%%%%%%%%%%%%%%%%%%%
\subsubsection{Three-body forces}
\label{sec:threebodyforces}

In the previous sections, we considered the ladder amplitude alone. Here, we describe the second term of~\Cref{eq:two-terms-of-amplitude}, the matrix $\bm{\cM}_{\text{df},3}$, which parametrizes the effect of short-range interactions. It is given by~\cite{\HSQCb,\tetraquark},
    %%%%%
    \begin{align}
    \label{eq:l-t-r}
    \bm{\cM}_{\text{df},3} = \bm{\cL} \, \bm{\cT} \, \bm{\cR} \, ,
    \end{align}
    %%%%%
where the ``endcap'' matrices $\bm{\cL}$, $\bm{\cR}$ describe all rescattering processes within and between external pairs that do not involve short-range three-particle interactions. After projection to the $JLS$ basis, they are given by
    %%%%%
    \begin{align}
    [\, \bm{\cL} \,]_{ip\ell's';jk\ell s} &= \left(\frac{1}{3} - \tilde{\rho}^{(i)}_s(k) \, \cM_{2,s}^{(i)}(k) \right) \delta_{ip\ell's';jk\ell s} 
    - \cD^{(ij)}_{\ell's';\ell s}(p,k) \, \tilde{\rho}^{(j)}_s(k) \, , \\
    %%%
    [\, \bm{\cR} \,]_{ip\ell's';jk\ell s} &= \left(\frac{1}{3} - \tilde{\rho}^{(i)}_s(k) \, \cM_{2,s}^{(i)}(k) \right) 
    \delta_{ip\ell's';jk\ell s} 
    - \tilde{\rho}^{(i)}_{s'}(p) \, \cD^{(ij)}_{\ell's';\ell s}(p,k)  \, ,
    \end{align}
    %%%%%
and contain the modified two-body phase space,
    %%%%%
    \begin{align}
    \label{eq:generalized-rho}
    \tilde{\rho}^{(i)}_s(\sigma_p) = H_i(p)  \left( - i \rho^{(i)}(\sigma_p) 
    + \frac{\eta_i}{8 \pi \sqrt{\sigma_p}} \, \frac{c_{\text{PV},s}^{(i)}}{(q_p^\star)^{2s}} \right) \, .
    \end{align}
    %%%%%
Comparing this to the phase space factor in \Cref{eq:M2,eq:phase-space}, we note the presence of the cutoff function as well as the addition of the so-called $I_{\text{PV}}$ term (proportional to the coefficients $c_{\rm PV,s}^{(i)}$) introduced in ref.~\cite{Romero-Lopez:2019qrt}. The latter is needed in the derivation of the finite-volume quantization condition (to be discussed in \Cref{sec:fin-vol}) in the presence of bound-state poles and resonances in two-particle subchannels. Since we must use the same definition of $\cK_3$ in the finite-volume formalism and the integral equations, we need to include the $I_{\text{PV}}$ term here. In our case, non-zero $c_{\rm PV,s}^{(i)}$ are needed due to the presence of $D^*$ and $D^*_0$ poles in the $D\pi$ $p$- and $s$-wave amplitudes, respectively.

The $\bm{\cT}$ matrix is given by the integral equation,
    %%%%%
    \begin{align}
    \label{eq:t-matrix}
    \bm{\cT} = \bm{\cK}_3 - \bm{\cK}_3 \, \bm{\tilde{\rho}} \, \bm{\cL} \, \bm{\cT} \, ,
    \end{align}
    %%%%%
where we promote the modified phase space into a matrix,
    %%%%%
    \begin{align}
     [\, \bm{\tilde{\rho}} \,]_{ip\ell's';jk\ell s} &= \delta_{ip\ell's';jk\ell s} \, \tilde{\rho}^{(i)}_s(k) \, .
     \label{eq:rhonew}
    \end{align}
    %%%%%
    
The quantity $\bm{\cK}_3$ is the partial-wave projection of the symmetrized three-body $K$ matrix, 
    %%%%%
    \begin{align}
    [\, \bm{\cK}_3 \,]_{ip\ell's';jk\ell s} &= \cK_{3,\ell's';\ell s}^{(ij)J}(p,k;E) \equiv \cK_{3,\ell's';\ell s}^{(ij)}(p,k) \, ,
    \label{eq:K3firsttime}
    \end{align}
    %%%%%
which is a regularization-scheme-dependent object parameterizing short-range three-body interactions. It is closely related to the three-body element of the genuine on-shell multi-particle $K$ matrix~\cite{Jackura:2022gib} and allows for a convenient description of three-body forces in the RFT formalism. For the physical three-body kinematics, it is a real function of momenta that satisfies the same symmetries as the full $3\to 3$ amplitude, including Lorentz invariance. In particular, unlike the pair-spectator amplitudes $\cM_3^{(ij)}$ are different objects for different choices of $i,j$, the $\cK_3^{(ij)}$ all correspond to the same underlying amplitude, although expressed in a different coordinate system. This is the meaning of a ``symmetrized'' $K$ matrix.

In the following, we assume a form of $\bm{\cK}_3$ in which dependence on the initial and final channel variables separates between ``left'' and ``right'' functions,
    %%%%%
    \begin{align}
    \label{eq:k3-separable}
    \cK_{3,\ell's';\ell s}^{(ij)}(p,k) = \sum_{a=1}^{a_{\rm max}} \cK_{L,\ell's'}^{a,(i)}(p) \, \cK_{R,\ell s}^{a,(j)}(k) \, .
    \end{align}
    %%%%% 
Although not the most general parametrization of $\cK_3$, for large enough $a_{\rm max}$, and appropriately chosen functions $\cK_L^a$, $\cK_R^a$, it is capable of describing many practical models of interest, including those with energy- and momentum-dependent contact interactions and explicit resonance poles. Models satisfying this property have been suggested and successfully used for the three-body couplings in recent studies, see for instance~\cite{Mikhasenko:2019vhk, Sadasivan:2020syi, Sadasivan:2021emk, Garofalo:2022pux, Zhang:2024dth, Feng:2024wyg}. In our generalized matrix notation, the separable ansatz becomes a sum of the outer products of vectors,
    %%%%%
    \begin{align}
    \label{eq:sep-gen}
    \bm{\cK}_3 = \sum_a \bm{\vec{\cK}}_L^a  \otimes \bm{\vec{\cK}}_R^a \, .
    \end{align}
    %%%%%
Using this form, the solution of the integral equation for $\cT$ is reduced to the problem of computing a set of double integrals. This can be seen by employing the form~\eqref{eq:sep-gen} in \Cref{eq:t-matrix},
    %%%%%
    \begin{align}
    \label{eq:t-matrix-2}
    \bm{\cT} = \bm{\cK}_3 -
    \sum_a \bm{\vec{\cK}}_{L}^{a}
    \otimes \bm{\vec{\alpha}}^{a} \, .
    \end{align}
    %%%%%
where,
    %%%%%
    \begin{align}
    \bm{\vec{\alpha}}^{a} = \bm{\vec{\cK}}_{R}^a \, \bm{\tilde{\rho}} \, \bm{\cL} \, \bm{\cT} \, .
    \end{align}
    %%%%%
By multiplying \Cref{eq:t-matrix-2} by $\bm{\vec{\cK}}_{R}^b \, \bm{\tilde{\rho}} \, \bm{\cL}$ from the left, one arrives at an algebraic equation,
    %%%%%
    \begin{align}
    \label{eq:t-matrix-equation-2}
    \bm{\vec{\alpha}}^{b} = \bm{\vec{\beta}}^b  -
    \sum_a \cI_{ba} \, \bm{\vec{\alpha}}^{a} \, ,
    \end{align}
    %%%%%
where we introduced,
    %%%%%
    \begin{align}
    \label{eq:matrix-I}
    \cI_{ba} &= \bm{\vec{\cK}}_{R}^b \, \bm{\tilde{\rho}} \, \bm{\cL} \, \bm{\vec{\cK}}_{L}^{a} \, , \\
    %%%%
    \label{eq:vector-beta}
    \bm{\vec{\beta}}^a &= \bm{\vec{\cK}}_{R}^b \, \bm{\tilde{\rho}} \, \bm{\cL} \, \bm{\cK}_3 = \sum_b \cI_{ab} \, \bm{\vec{\cK}}_R^b ,
    \end{align}
    %%%%%
and $b \in\{1,\dots, a_{\rm max} \}$. Note that in the definition of $\cI$ the ``left'' function $\bm{\vec{\cK}}_L$ is on the right side of the expression while the ``right'' vector $\bm{\vec{\cK}}_L$ is on the left side. In the generalized matrix notation, $\cI$ is a scalar, i.e., it carries no dependence on external channel variables, which allowed us to drop the $\otimes$ sign in \Cref{eq:t-matrix-equation-2}. In consequence, the solution to \Cref{eq:t-matrix-equation-2} is obtained simply by inverting the $a_{\rm max} \times a_{\rm max}$ matrix,
    %%%%%
    \begin{align}
    \bm{\vec{\alpha}}^a = \sum_b \, (\bm{1} + \cI)^{-1}_{ab} \, \bm{\vec{\beta}}^b \, .
    \end{align}
    %%%%%
Finally, using the above formula in \Cref{eq:t-matrix-2} yields the solution for the $\bm{\cT}$ amplitude,
    %%%%%
    \begin{align}
    \bm{\cT} = \sum_{a,b} \bm{\vec{\cK}}_L^a  \otimes 
    (\bm{1} + \cI)^{-1}_{ab} \, \bm{\vec{\cK}}_R^b \, .
    \end{align}
    %%%%%
For completeness, the $\bm{\cM}_{3,{\rm df}}$ amplitude is given by the expression with the endcap matrices restored,
    %%%%%
    \begin{align}
    \label{eq:mdf3-sol}
    \bm{\cM}_{3,{\rm df}} = \sum_{a,b} \bm{\vec{L}}^a \otimes 
    (\bm{1} + \cI)^{-1}_{ab} \, \bm{\vec{R}}^b \, ,
    \end{align}
    %%%%%
where we introduced the ``left'' and ``right'' generalized endcap vectors,
    %%%%%
    \begin{align}
    \bm{\vec{L}}^a = \bm{\cL} \, \bm{\vec{\cK}}_L^a \, , \quad 
    %%%
    \bm{\vec{R}}^a = \bm{\vec{\cK}}_R^a \, \bm{\cR} \, .
    \end{align}
    %%%%%
To conclude, the task of solving~\Cref{eq:t-matrix} is reduced to an algebraic problem of inverting a small matrix entering~\Cref{eq:mdf3-sol}. In cases when the two- and three-body parameters can be fixed independently, the solution of the ladder integral equation, $\bm \cD$, is obtained just once, tabulated, and reused in~\Cref{eq:mdf3-sol} for any model of $\bm \cK_3$ satisfying the property~\eqref{eq:sep-gen}.

%%%%%%%%%%%%%%%%%%%%%%%%%
%%%%%%%%%%%%%%%%%%%%%%%%%
\paragraph{Threshold expansion:}

More concretely, in this work, we implement a model of the three-body $K$ matrix based on the threshold expansion of ref.~\cite{Blanton:2021eyf}. Working to given order in powers of an expansion about the three-particle threshold, it is the most general low-energy form of any smooth $\cK_3$ constrained by Lorentz invariance, $PT$ invariance, and the symmetry under the interchange of the $D$ mesons separately in the initial and final three-body state. Through linear order, the result is
    %%%%%
    \begin{align}
    \cK_3(\{\bm p\},\{\bm k\})
    = \cK_3^{\text{iso},0} 
    + \cK_3^{\text{iso},1} \Delta 
    + \cK_3^{B} \Delta_2^S
    + \cK_3^{E} \tilde{t}_{22} \, .
    \label{eq:K3thr}
    \end{align}
%%%%%
where $\{\bm p\}$ ($\{\bm k\}$) indicates the set of final (initial) state momenta,
$\cK^{\rm iso, 0}_3$, $\cK^{\rm iso, 1}_3$, $\cK_3^B$, $\cK_3^E$ are real coefficients, and
    %%%%%
    \begin{align}
    \begin{split}
    & \Delta = \frac{E^2 - M^2}{M^2} \, , 
    ~~~ \Delta_2^S = \Delta_2 + \Delta_2' \, ,
    ~~~ \tilde{t}_{22} = \frac{(k_\pi - p_\pi)^2}{M^2} \, , \\
    %%%
    & \Delta_2 = \frac{\sigma_{k, DD} - 4m_D^2}{M^2} \, , 
    ~~~ \Delta_2' = \frac{\sigma_{p, DD} - 4m_D^2}{M^2} \, ,
    \end{split}
    \label{eq:kinematics}
    \end{align}
    %%%%%
Here, $k_\pi, p_\pi$ are energy-momentum four-vectors of initial and final pion, respectively, while $\sigma_{k,DD}$, and $\sigma_{p,DD}$ are the two-body invariant masses squared of the initial and final $DD$ subsystems. The $DD\pi$ threshold energy, $M = 2 m_D + m_\pi$, is introduced to make all the above kinematic variables dimensionless. 

We note that, although, in the RFT approach, $\mathcal K_3$ is guaranteed to be a smooth function within the kinematic regime of the validity of the formalism, the low-order polynomial form of Eq.~(2.49) will become increasingly inaccurate as one approaches the boundaries of this regime. In particular, a breakdown is expected as one approaches the left-hand singularity associated with the two-particle exchange in $D^*D$ system. In the following, we stay away from these boundary regions.

To implement this form for $\cK_3$ in the integral equations described above, we have to partial-wave project it onto a definite $J$ in the $LS$ basis and decompose it into the separable form. We describe these steps in \Cref{app:pw} and provide an explicit expression for $J^P=1^+$ and $0^-$ in \Cref{app:pw_K3}. We find that all terms of~\Cref{eq:K3thr} contribute to $J^P=0^-$, while only the $\cK_3^E$ term contributes to the $J^P=1^+$ amplitude. Additional contributions from higher-order terms will be present but are not considered here. The only additional subtlety is the need to include symmetry factors in $\cK_3^{(ij)}$. These are given by multiplying the elements obtained in \Cref{app:pw_K3} by $e_{ij}$, where $e_{11}=1$, $e_{12}=e_{21}=-1/\sqrt2$ and $e_{22}=1/2$. Equivalently, the results for the factorized quantities $\cK_L^{a,(i)}$ and $\cK_R^{a,(i)}$ should both be multiplied by $e_i$, where $e_1=1$ and $e_2=-1/\sqrt2$.

%%%%%%%%%%%%%%%%%%%%%%%%%
%%%%%%%%%%%%%%%%%%%%%%%%%
\subsubsection{Numerical solution}
\label{sec:numerical_solution}

The solutions to integral equations for $\bm{\cM}_3$, \Cref{eq:solution_formal,eq:mdf3-sol}, can be obtained numerically by discretizing the momentum variables and rewriting the equations in a purely algebraic form. The discrete solutions approach the continuum one for increasingly dense meshes, and the asymptotic value is extracted using the techniques described in Refs.~\cite{Jackura:2020bsk, Dawid:2023jrj}. 

To improve the stability and convergence of the solution, it is useful to implement complex deformed contours for numerical integration. Note that, in principle, one is free to independently deform integration intervals in pair-spectator channels $n=1$ and $n=2$ in \Cref{eq:generalized-matrix}. The shape of a contour is chosen such that it avoids all singularities of the integration kernel, $\bm{G} \, \bm{\cM}_2$. These occur in $\bm{G}$ in the form of OPE cuts and in $\bm{\cM}_2$ as unitarity and unphysical cuts and the $D^*$ pole. Moreover, deformed contours must avoid so-called domains of non-analyticity generated by the motion of these momentum- and energy-dependent singularities in the kernel~\cite{Dawid:2023jrj}.

\begin{figure}[t!]
    \centering
    \includegraphics[width=0.9\textwidth]{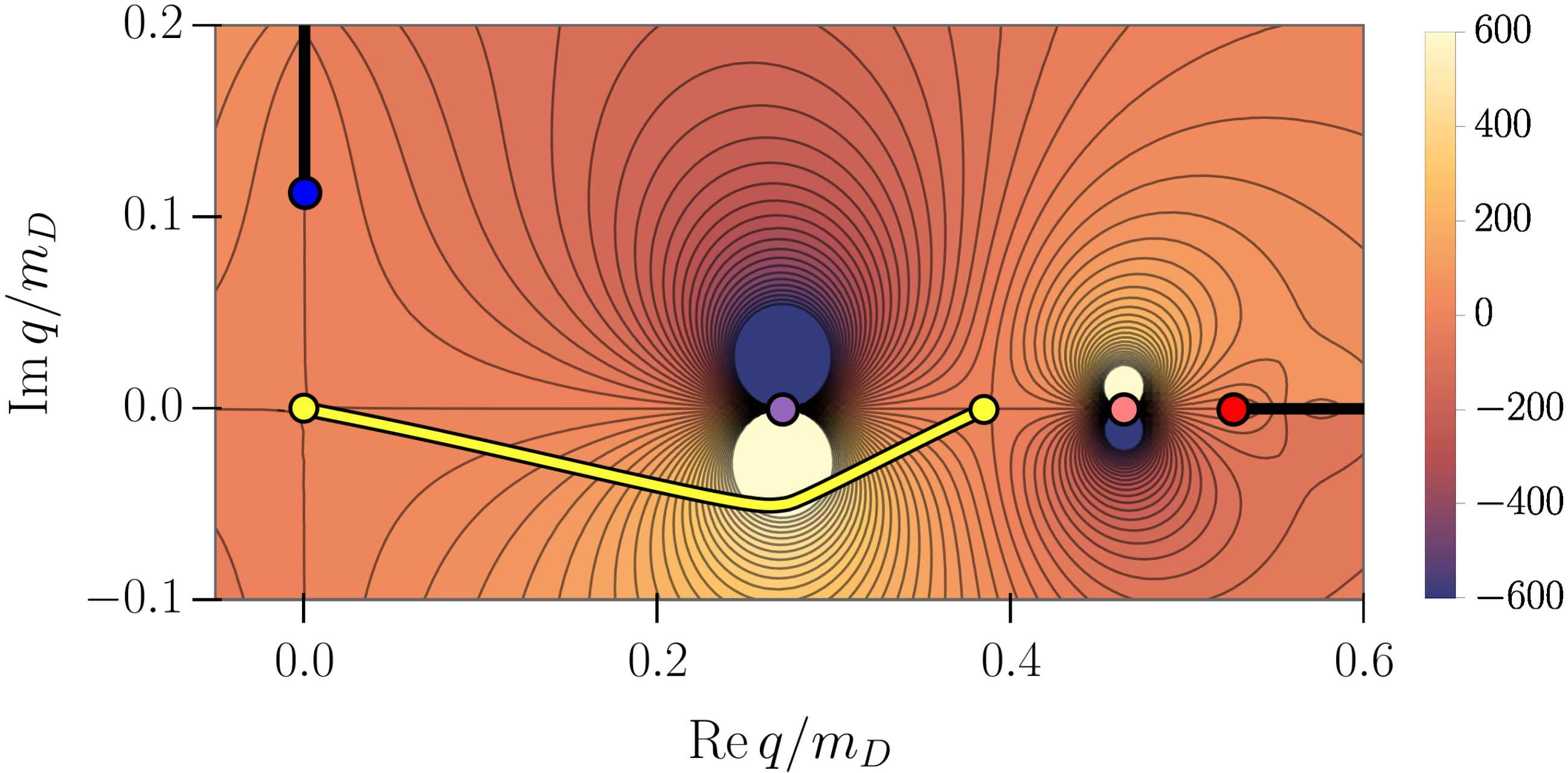}
    \caption{Imaginary part of $\cM_{2,1}^{(1)}(q)$ in the complex $q$ momentum plane for $E/m_D = 2.133$, $\kappa=0.1453$, $a_1^{(1)} m_D^3 = -7.9^3$, and $r_1^{(1)}=0$. The amplitude contributes singularities to the integration kernel $\bm{G} \bm{\cM}_2$ of the ladder equation in the intermediate channel $n=1$. Branch cuts are shown as black lines and singularities as colored points. The threshold branch point is shown in blue, the pseudo-threshold in red, the $D^*$ bound-state pole in purple, and the unphysical pole in pink. The deformed integration contour is shown in yellow. Note that the upper endpoint of the integration contour is placed below the unphysical singularities, as discussed in the text.}
    \label{fig:contour-deformation}
\end{figure}

In the context of $T_{cc}^+$, the energies of interest are real and near the $DD^*$ threshold, $E_{DD^*} = m_D + m_{D^*}$ since this is where the virtual-state pole is expected to appear. However, for energies that satisfy the following condition~\cite{Dawid:2023jrj},
    %%%%%
    \begin{align}
    E > \frac{1}{2 m_D} \left(m_D^2 - m_\pi^2 + m_{D*}^2 + \sqrt{4 m_D^2 + \lambda(m_D^2,m_\pi^2,m_{D^*}^2)} \right) \, ,
    \end{align}
    %%%%%
only the $D^*$ pole causes a relevant obstruction in the numerical solution. The above constraint is caused by the development of the circular cut in the integration kernel, which requires more sophisticated methods as discussed in ref.~\cite{Dawid:2023jrj}. We thus constrain our solutions to higher energies and deform the integration contour in the intermediate $D\pi$ $p$-wave channel, $n=1$, $s_n=1$, such that the $D^*$ pole is safely avoided, as shown in \Cref{fig:contour-deformation}. For the $s$-wave channels, both with $n=1$ and $n=2$, we use the straight, purely real contour. 

%%%%%%%%%%%%%%%%%%%%%%%%%
% FIGURE
%%%%%%%%%%%%%%%%%%%%%%%%%
%%%%%%%%%%%%%%%%%%%%%%%%%

Contour deformation allows one to probe the integration kernel away from the $D^*$ pole position, making the integrated functions smooth and slowly varying. In consequence, a good numerical convergence can be achieved by using Gauss-Legendre quadratures with meshes of size $N=50$ to $100$.

%%%%%%%%%%%%%%%%%%%%%%%%%
%%%%%%%%%%%%%%%%%%%%%%%%%
\subsection{Connecting $DD\pi$ and $DD^*$ scattering amplitudes}

In the vicinity of the $D^*$ pole, the $D\pi$ $p$-wave scattering amplitude, $\cM_{2,1}^{(1)}$ is,
    %%%%%
    \begin{equation}
    \cM_{2,1}^{(1)}(q) \simeq \frac{\zeta^2}{\sigma_q - m_{D^*}^2} \, ,
    \label{eq:LSZ-1}
    \end{equation}
    %%%%%
where $m_{D^*}$ is the $D^*$ mass and $\zeta^2$ is the residue. As discussed above, for a $p$-wave bound state, the residue is positive, and thus $\zeta$ is real and can be chosen positive.
This pole appears as a singularity of the $DD\pi$ amplitude through its dependence on the two-body scattering in external states. The LSZ reduction formula for bound states~\cite{Lehmann:1954rq, Zimmermann:1958hg, Fried:1958} implies that continuation of the $D(D\pi) \to D(D\pi)$ pair-spectator scattering amplitude to external invariant masses $\sigma_p = \sigma_k = m_{D^*}^2$ allows one to extract the particle–bound-state amplitude from the solution of the three-body integral equations. We apply this reasoning to the $i=j=1$, $s'=s=1$ matrix elements~\cite{Jackura:2020bsk, Dawid:2023kxu},
    %%%%%
    \begin{align}
    \label{eq:LSZ}
    \lim_{\sigma_p, \sigma_k \to m_{D^*}^2 } 
    \left(\frac{\sigma_p - m_{D^*}^2}\zeta \right) \, 
    \cM_{3,\ell'1, \ell 1}^{(11)}(p,k;E) \, 
    \left(\frac{\sigma_k - m_{D^*}^2}\zeta \right) = 
    \,
    - [\cM_{DD^*}(E)]^J_{\ell' 1,\ell 1,} \,  
     \, ,
    \end{align}
    %%%%%
to obtain the partial-wave projected $DD^*$ amplitude, $\cM_{DD^*}(E)$.

Since $D^*$ has a non-zero spin, we employ a generalized $K$ matrix parametrization of the $DD^*$ amplitude,
    %%%%%
    \begin{equation}
    \left[\cM_2^{-1} \right]^{J}_{\ell's', \ell s} = \left[ \cK_2^{-1} \right]^{J}_{\ell's', \ell s} - i \rho \, \delta_{\ell's', \ell s} \, ,
    \label{eq:M2DDs}
    \end{equation}
    %%%%%
where $\rho = q_b/(8\pi E)$ is the two-body phase space and $q_b = \lambda^{1/2}(E^2, m_D^2, m_{D^*}^{2})/2E$. Here, the two-body $K$ matrix, $\cK_2$, is diagonal in $J$, but the orbital angular momenta and spins are not conserved individually. However, since parity is conserved, even and odd values of $\ell$ do not mix in $\cK_2$. 

If the value of $J^P$ allows for only a single $LS$ combination, one may use one phase shift to describe the low-energy scattering,
    %%%%%
    \begin{equation}
    \label{eq:phase_shift_one}
    (8 \pi E) \,  q_b^{\ell} \left[ \mathcal{K}_2^{-1} \right]^{J}_{\ell s, \ell s}  q_b^{\ell} = q_b^{2\ell+1} \cot (\delta_J) \, .
    \end{equation}
    %%%%%
Here we have added the threshold factors $q_b^\ell$ such that the quantities are finite as $q_b\to 0$.
By contrast, if several $LS$ channels mix, more complicated forms are required, such as the Blatt-Biedenharn parametrization~\cite{Blatt:1952zz}. This is the case of $J^P=1^+$ $DD^*$ scattering, where transitions between $(\ell, s)=(0,1)$ and $(\ell,s)=(2,1)$ are possible. In this parametrization, the $[\cK_2]_{\ell's'=1,\ell s=1}^{J^P=1^+} \equiv [\cK_{DD^*}]_{\ell', \ell}$ matrix is described by three real energy-dependent angles: two phase shifts, $\delta_{\alpha}, \delta_\beta$, and a mixing angle $\epsilon$,
    %%%%%
    \begin{align}
    \label{eq:Blatt-BiedenharnA}
    (8 \pi E)  \left[\cK_{DD^*}^{-1} \right] &= 
    %%%
    \begin{pmatrix}
    \cos (\epsilon) & - \sin (\epsilon) \\ 
    \sin (\epsilon) & \cos (\epsilon)
    \end{pmatrix}
    %%%
    \left( \begin{array}{cc}
    q_b \cot (\delta_\alpha) & 0 \\
    0 & q_b \cot (\delta_\beta)
    \end{array} \right) 
    %%%
    \begin{pmatrix}
    \cos (\epsilon) &  \sin (\epsilon) \\
    - \sin (\epsilon) & \cos (\epsilon)
    \end{pmatrix} \, .
    \end{align}
    %%%%%
In practice, it is useful to rewrite this in terms of quantities that remain finite as $q_b\to 0$. Introducing a threshold-factor matrix $Q$ such that $[Q]_{\ell' \ell} = \delta_{\ell' \ell}\, q_b^\ell$, the expression \Cref{eq:Blatt-BiedenharnA} can be written in terms of finite quantities as
    %%%%%
    \begin{multline}
    \label{eq:Blatt-Biedenharn}
    (8 \pi E) [Q]   \left[\cK_{DD^*}^{-1} \right] [Q] = \\ 
    %%%
    \begin{pmatrix}
    \cos (\epsilon) & -\frac{1}{q_b^2} \sin (\epsilon) \\ 
    q_b^2 \sin (\epsilon) & \cos (\epsilon)
    \end{pmatrix}
    %%%
    \left( \begin{array}{cc}
    q_b \cot (\delta_\alpha) & 0 \\
    0 & q^{5}_{b} \cot (\delta_\beta)
    \end{array} \right) 
    %%%
    \begin{pmatrix}
    \cos (\epsilon) & q_b^2 \sin (\epsilon) \\
    -\frac{1}{q_b^2} \sin (\epsilon) & \cos (\epsilon)
    \end{pmatrix} \, .
    \end{multline}
    %%%%%
Note that, below the left-hand branch point at energy $E_{2}^{\rm lhc}$ (defined below), the scattering phase shifts and the mixing angle in general acquire non-zero imaginary parts. 

The $DD^*$ amplitude has several important singularities on the real energy axis. The threshold right-hand branch cut occurs at the energy $E_{DD^*}$. If the $T_{cc}^+$ is a bound state, then there will be a pole below the threshold on the physical Riemann sheet. It can be computed through the formula,
   %%%%%
    \begin{equation}
    \label{eq:amp-1st-sheet}
    \left[(\cM_{DD^*})^{-1} \right]_{\ell', \ell} = \left[ \cK_{DD^*}^{-1} \right]_{\ell \ell } - i \rho \, \delta_{\ell' \ell} \, .
    \end{equation}
    %%%%%
We note that the orthogonal transformation that diagonalizes $\cK_{DD^*}^{-1}$ (see \Cref{eq:Blatt-BiedenharnA}) also diagonalizes the phase-space term, and thus $\cM_{DD^*}$. It follows that the bound-state pole condition takes a form analogous to~\Cref{eq:pole-condition} separately for both $\cot\delta_\alpha$ and $\cot\delta_\beta$. If the $T_{cc}^+$ is a virtual state, the pole appears on the real axis of the second Riemann sheet, and can be computed via,
    %%%%%
    \begin{equation}
    \label{eq:amp-2nd-sheet}
    \left[(\cM_{DD^*}^{\rm II})^{-1} \right]_{\ell', \ell} = \left[ \cK_{DD^*}^{-1} \right]_{\ell \ell } + i \rho \, \delta_{\ell' \ell} \, .
    \end{equation}
    %%%%%
It is also possible that the pole moves into the complex plane on the second sheet, as discussed below.

Finally, the two-body $K$ matrix has a left-hand branch point below the threshold due to the virtual $\pi$ exchange between the interacting mesons. In our LSZ-reduction-based approach this singularity is inherited from the $J^P=1^+$ OPE amplitude $G_{\ell'1,\ell 1}^{(11)}(p,k)$ continued to the momentum $p=k=q_b$. The cut is placed between branch points at
    %%%%%
    \begin{align}
    E^{\text{lhc}}_1 &= \frac{m_D^2 - m_{D^*}^2}{m_\pi} \, , ~~~ E^{\text{lhc}}_2 = \sqrt{2(m_D^2 + m_{D^*}^2) - m_\pi^2} \, .
    \end{align}
    %%%%%
In general, scattering phase shifts and the mixing angle are complex for energies between these two values.

%%%%%%%%%%%%%%%%%%%%%%%%%
%%%%%%%%%%%%%%%%%%%%%%%%%
\section{Finite-volume formalism}
\label{sec:fin-vol}

We aim to demonstrate how the formalism from ref.~\cite{\tetraquark} maps lattice QCD results for $DD^*$ and $DD\pi$ energy levels to conclusions about the $T_{cc}^+$. Here, we focus on the first step: using quantization conditions to constrain the $K$ matrices from the lattice spectrum, which in a typical application would be then used as input for the integral equations discussed earlier.

Specifically, we summarize the two quantization conditions that we use in this work: first the standard L\"uscher QC2 for $D^*D$ systems, and then the QC3 for $DD\pi$. We emphasize that our main focus in this paper is to demonstrate the applicability of the QC3 approach, while we use the QC2 to show where and how badly it breaks down near the left-hand cut. We also note, as mentioned already in the introduction, that there are versions or extensions of the QC2 that are able to solve the left-hand cut problem~\cite{Raposo:2023oru,Bubna:2024izx,Meng:2021uhz}. While it would be interesting to compare the results of the QC3 to those from these alternate QC2s, this is beyond the scope of the present work.

We also note that the QC2 and QC3 are valid up to terms that are exponentially suppressed in the size of the spatial box (here taken to be cubic of length $L$). Dropped terms are generically suppressed by $O(e^{- m_\pi L})$. In the $DD^*$ case, the suppression factor is determined by the binding momentum of the $D^*$.

%%%%%%%%%%%%%%%%%%%%%%%%%
%%%%%%%%%%%%%%%%%%%%%%%%%
\subsection{QC2 for $DD^*$ scattering}
\label{sec:QC2}

We consider heavier than physical pion masses such that the $D^*$ meson is stable, i.e.~a $D\pi$ bound state with binding momentum $q_0$, defined via ${m_{D^*} = \sqrt{m_D^2-|q_0|^2} + \sqrt{m_\pi^2-|q_0|^2}}$. Thus, the mass of the $D^*$ in finite volume receives corrections that are proportional to $\exp(- |q_0| L)$~\cite{Luscher:1985dn}. For near-physical pion masses, $|q_0| \lesssim m_\pi$, and so these corrections are the dominant terms that are dropped in the $D D^*$ finite-volume formalism. Thus, up to these exponentially-suppressed effects, the $DD^*$ spectrum can be described by the two-body finite-volume formalism of ref.~\cite{Luscher:1986pf} extended to include particles with spin~\cite{Briceno:2014oea}. The range of applicability of this method is given by,
    %%%%%
    \begin{equation}
    E^{\rm lhc}_2 < E^* < {\rm min}[ 2 m_D + m_\pi \, , 2 m_{D^*}] \, ,
    \end{equation}
    %%%%%
where $E^*=\sqrt{E^2-\bm P^2}$ is the c.m.~frame energy, with $\bm P$ the total momentum in the finite-volume frame. The upper bound is set by the first inelastic threshold, either $DD\pi$ or $D^*D^*$, depending on the pion mass.\footnote{If the first inelastic threshold is $D^*D^*$, a two-body multi-channel approach is possible~\cite{Briceno:2014oea}.} The lower bound is determined by the location of the one-pion exchange left-hand cut. For energies $E \lesssim E^{\rm lhc}_2$, the non-analytic behavior of the two-meson Bethe-Salpeter kernel invalidates the derivation of the formalism, as explained in detail in ref.~\cite{Raposo:2023oru}. 

In the presence of spin, the QC2 can be written as,
    %%%%%
    \begin{equation}
    \det_{J m_J \ell} \left[ \widehat{\mathcal{K}}_2^{-1}(E, \boldsymbol{P})+ \widehat{U} \cdot \widehat{F}_2 (E, \boldsymbol{P}, L) \cdot \widehat{U}^{\dagger}\right]=0 \, .
    \label{eq:QC2}
    \end{equation}
    %%%%%
The hats indicate that the quantities are matrices in the
space corresponding to the combination of orbital angular momentum $\ell$ with the spin, $s=1$, of the $D^*$, to give total angular momentum $J$, and azimuthal component $m_J$. This space can be spanned either by the $\{J m_J \ell\}$ basis  (in which $\cK_2$ is most naturally expressed) or the $\{\ell m m_s\}$ basis (most natural for $F_2$). The unitary matrix $\widehat{U}$ transforms between these bases, and is given by
    %%%%%
    \begin{equation}
    [\, \widehat{U} \, ]_{J m_J \ell ; \ell m m_s} = \left\langle J, m_J \mid \ell, m; s=1, m_s\right\rangle \, .
    \end{equation}
    %%%%%
In practice, one must truncate the QC2 by keeping partial waves $J<J_{\rm max}$, such that all matrices have finite dimensionality. For the $T_{\rm cc}^+$ case, we set $J_{\rm max}=2$, which means keeping $J^P=0^-, 1^+ $ and $1^-$.

We now define the objects in the QC2. $\widehat{\mathcal{K}}_2$ is the $DD^*$ $K$ matrix, whose form has been discussed above; see in particular \Cref{eq:M2DDs} and subsequent discussion. The matrix $\hat{F}_2$ is a known kinematic function,  most naturally expressed in the $\{\ell m m_s\}$ basis, 
and takes the form,\footnote{%
%%%%
In this section, we revert to the standard RFT convention of which of the spherical harmonics to complex conjugate. We stress that the resulting energy levels do not depend on this choice.}
    %%%%%
    \begin{equation}
    \left[\widehat{F}_2 \right]_{\ell^{\prime} m^{\prime}  m_s^{\prime} ; \ell m m_s} = 
    \delta_{m_s^{\prime}  m_s}
    \left[\frac{1}{L^3} \sum_{\boldsymbol{a}}^{\rm UV}-\text{p.v.} \!\! \int\limits^{\rm UV}  \frac{d^3 a}{(2\pi)^3} \right] \frac{\mathcal{Y}_{\ell^{\prime} m^{\prime}}\left(\boldsymbol{a}^*\right) \mathcal{Y}_{\ell m}^*\left(\boldsymbol{a}^*\right)}{2 \omega^D_a\left(b^2 - m_{D^*}^2 \right)} \frac{1}{\left(q^*\right)^{\ell^{\prime}+\ell}} \,.
    \label{eq:F2def}
    \end{equation}
    %%%%%
Here $(\omega_a^D,\bm a)$ is the four-momentum of the $D$ meson, four-vector $b=P-a$ where $P^\mu = (E, \boldsymbol{P})$. The vector $\boldsymbol{a^*}$ is the spatial part of a four-vector resulting from boosting $a$ to the overall c.m. frame. Harmonic polynomials are defined as $\mathcal Y_{\ell m}(\boldsymbol{a}) = \sqrt{4\pi} a^\ell Y_{\ell m}(\hat a)$. The sum over $\bm a$ runs over the finite-volume set, $\bm a= (2\pi/L) \bm n$, with $\bm n$ a vector of integers. The integral over the pole is regulated using the principal value prescription, and a consistent ultraviolet regulator (denoted by ``UV'') is implicit for both sum and integral. Finally, $q^* = \lambda^{1/2}(E^{*2},m_D^2,m_{D^*}^2)/(2 E^*)$, where $E^*=\sqrt{E^2-\bm P^2}$ is the c.m. frame energy. $\widehat F_2$ can be evaluated following the method described in appendix B of ref.~\cite{Blanton:2019igq}.

%%%%%%%%%%%%%%%%%%%%%%%%%
%%%%%%%%%%%%%%%%%%%%%%%%%
\subsection{QC3 for $DD\pi$ scattering}
\label{sec:QC3}

In this section, we describe the finite-volume three-particle formalism of ref.~\cite{Hansen:2024ffk}, which connects the finite-volume energy levels of the $DD\pi$ system to the elastic scattering amplitude. We focus on the $T_{cc}^+$ channel, with overall isospin ${I=0}$. The formalism is agnostic as to whether the $D^*$ meson is stable or not---this is determined by the $I=1/2$ $D\pi$ scattering amplitude, which is one of the inputs. Assuming a stable $D^*$ meson, the range of validity of the formalism is,
    %%%%%
    \begin{equation}
    E^{\rm lhc}_{2\pi}=\sqrt{2m_{D^*}^2+2m_D^2-4 m_\pi^2} < E^* < \min \left[ 2m_D + 2m_\pi, 2m_{D^*} \right]\,.
    \end{equation}
    %%%%%
The lower limit is the position of the left-hand branch point associated with a virtual two-pion exchange between $D$ and $D^*$. It is typically relatively far from the physical region of interest. The upper limit is given by the lowest inelastic threshold, either $DD\pi\pi$ or $D^*D^*$, depending on the pion mass.

The QC3 takes the form,
    %%%%%
    \begin{equation}
    \underset{i, \boldsymbol{k}, s m}{\operatorname{det}}
    \left[\mone + \widehat{\mathcal{K}}_{3} \,   \widehat{F}_3\right]=0\,.
    \label{eq:QC3}
    \end{equation}
    %%%%%
As for the integral equations discussed in the previous section, the three mesons are divided into a pair and a spectator. The flavor of the spectator is labeled by $i$ (either $D$ or $\pi$), with three-momentum denoted by $\k$, drawn from the finite-volume set. Discrete indices $s, m$ label the orbital angular momentum of the pair in its c.m.~frame, defined as in~\Cref{sec:DDpi-kinematics} except that here we revert to using a space-fixed coordinate system to define the azimuthal component. This is just a convention, but it has become the standard choice in implementations of the finite-volume quantization condition~\cite{Blanton:2019igq}. The determinant in \Cref{eq:QC3} is evaluated on matrices living in the multidimensional space with all these indices. The cutoff function $H_i(k)$ (which appears in $\widehat F_3$, as seen below) ensures that the sum over $\bm k$ is truncated, while we truncate in $s$ in the same manner as for the integral equations---see the end of \Cref{sec:DDpi-kinematics}.
In the following, we describe the building blocks of the above equation in detail. 

First, the matrix $\widehat{F}_3$ is given in terms of three components, $\widehat F$, $\widehat G$ and $\widehat{\cK}_2$, as,
    %%%%%
    \begin{equation}
    \widehat{F}_3 \equiv \frac{\widehat{F}}{3}-\widehat{F} \frac{1}{1+\widehat{\mathcal{M}}_{2, L}
    \widehat{G}} \, \widehat{\mathcal{M}}_{2, L} \widehat{F}, \quad \quad \widehat{\mathcal{M}}_{2, L}
    \equiv \frac{1}{\widehat{\mathcal{K}}_{2,L}^{-1}+\widehat{F}}\,.
    \label{eq:F3def}
    \end{equation}
    %%%%%
The first component is,
    %%%%%
    \begin{equation}
    \widehat F = \text{diag}\left( \widetilde F^{D}, \widetilde F^{\pi} \right)\,,
    \end{equation}
    %%%%%
where flavor indices are displayed explicitly, and
    %%%%%
    \begin{multline}
    \left[\wt F^{(i)}\right]_{p' s' m';p s m} =
    \delta_{\bm p' \bm p} \, 
    \frac{H^{(i)}(p)}{2\omega^{(i)}_{p} L^3} 
    \left\{\rule{0cm}{0.9cm}\right.
    \left[ \frac1{L^3} \sum_{\bm a}^{\rm UV} - {\rm p.v.} \!\! \int\limits^{\rm UV} \frac{d^3 a}{(2\pi)^3} \right]
    \\
    \times \left[
    \frac{\cY_{s' m'}(\bm a_{p'}^{\star})}{\big(q_{p'}^{\star}\big)^{s'}}
    \frac1{4\omega_{a}^{(j)} \omega_{b}^{(k)}
    \big(E\!-\!\omega_{p}^{(i)}\!-\!\omega_{a}^{(j)}\!-\!\omega_{b}^{(k)}\big)}
    \frac{\cY^*_{s m}(\bm a_p^{\star})}{\big(q_{p}^{\star}\big)^{s}}
    \right] 
    + \frac1{8\pi \sqrt{\sigma^{(i)}_p}} 
    \frac{c_{{\rm PV},s}^{(i)}}{(q_p^\star)^{2s}} 
    \left.\rule{0cm}{0.9cm}\right\}
    \, .
    \label{eq:Ft}
    \end{multline}
    %%%%%
Much of the notation is as for $F_2$ above---see \Cref{eq:F2def}---while the $c_{\rm PV}$ coefficients have been introduced in \Cref{eq:generalized-rho}. Here, however, there are three particles: the initial spectator with momentum $\bm p$ and flavor $i$, and the corresponding pair, consisting of a primary with momentum $\bm a$ and flavor $j$, and a secondary with momentum $\bm b=\bm P-\bm p - \bm a$ and flavor $k$. The momentum $\bm a_p^\star$ is the result of boosting $a=(\omega_a^{(j)},\bm a)$ to the c.m.~frame of the pair, while $q_p^\star$ is defined in \Cref{eq:qpstar}. We have kept flavor indices explicit, as there is an ambiguity as to which member of the pair is primary if $i=D$. Our choice in this case is $j=D$, $k=\pi$. If $i=\pi$, then we have $j=k=D$.

The flavor structure of $\widehat G$ is given as,
    %%%%%
    \begin{equation}
    \widehat{G}  =
    \begin{pmatrix}
    \wt G^{DD} & -\sqrt2 P^{(s)}\, \wt G^{D\pi} \\
    -\sqrt2\, \wt G^{\pi D} P^{(s)} & 0
    \end{pmatrix}\,,
    \end{equation}
    %%%%%
where the entries are defined by,
    %%%%%
    \begin{equation}
    \left[\wt G^{(ij)}\right]_{p s' m';r s m} =
    \frac1{2\omega^{(i)}_{p} L^3}
    \frac{\cY_{s' m'}(\bm r_p^\star)}{\big(q_{p}^{\star}\big)^{s'}}
    \frac{H_i(p) H_j(r)}{b_{pr}^2-m_{ij}^2}
    \frac{\cY^*_{s m}(\bm p_r^{\star})}{\big(q_r^{\star}\big)^{s}}
    \frac1{2\omega^{(j)}_{r} L^3} \, ,
    \label{eq:Gt}
    \end{equation}
    %%%%%
and,
    %%%%%
    \begin{equation}
    [P^{(s)}]_{p s' m';r s m} = \delta_{\bm p \bm r} \, \delta_{s's} \delta_{m' m} \, (-1)^s \,.
    \label{eq:Pelldef}
    \end{equation}
    %%%%%
We note that \Cref{eq:Gt} is closely related to \Cref{eq:ope-spin-helicity}, with two main differences: an overall $2\omega_p L^3$ factor, and the fact that the spherical harmonics in \Cref{eq:Gmat} are defined with respect to the azimuthal component, and not in the spin-helicity basis. 

The final component of $\widehat F_3$ is $\widehat{\mathcal K}_{2,L}$, given by
    %%%%%
    \begin{equation}
    \widehat{\cK}_{2,L} =
    \text{diag}\left( {\cK}_{2,L}^{D\pi, I=1/2}, \tfrac{1}{2} \, {\cK}_{2,L}^{DD, I=1} \right),
    \label{eq:KhatI0}
    \end{equation}
    %%%%%
where each entry corresponds to a modified $K$ matrix for the $\{jk\}$ system with isospin $I$,
    %%%%%
    \begin{align}
    \left[{\cK}_{2,L}^{jk, I\,}\right]_{ps'm'; r s m} &=
    \delta_{\bm p \bm r} 2\omega^{(i)}_{p} L^3 \delta_{s' s} \delta_{m' m} \cK^{jk, I}_{2,s}(\bm p) \,,
    \label{eq:K2L}
    \\
    %%%
    \left[\cK^{jk,I\,}_{2,s}(\bm p)\right]^{-1}
    &=
    \frac{ \eta_i}{8 \pi \sqrt{\sigma^{(i)}_p}}
    \left\{ q_{p}^{\star} \cot [\delta_s^{(jk), I}(q_{p}^{\star})] + |q_p^{\star}| [1-H^{(i)}(p)] - H^{(i)}(p) \frac{c_{{\rm PV},s}^{(i)}}{(q_p^\star)^{2s}} \right\} \,,
    \label{eq:K2s}
    \end{align}
    %%%%%%
where $\delta_s^{(jk),I}$ is the corresponding phase shift for angular momentum $s$. The constants $c_{\rm PV,s}^{(i)}$ are to be determined so that $\cK^{jk,I\,}_{2,s}$ does not have a pole in the kinematically allowed range $\bm p$~\cite{Romero-Lopez:2019qrt}.

Finally, we describe how the three-particle $K$ matrix, $\cK_3$, is converted to the matrix form that enters the QC3. One begins with the chosen model form in terms of the momenta, e.g., the threshold expansion given in \Cref{eq:K3thr}. Here, we label the final momenta $\{\bm p\}\equiv \{\bm p_1, \bm p_2, \bm p_3\}$, where the flavors are $1=2=D$ and $3=\pi$, and similarly for the initial momenta. One then applies operators that project onto the pairs' angular momenta $s', s$. The flavor structure of the result takes the form of an outer product~\cite{\tetraquark},
    %%%%%
    \begin{align}
    \widehat{\cK}_\text{3} &=
    {\boldsymbol{\cY}}^{[I=0]}\circ \cK_{3}(\{\bm p\}; \{\bm k\})\circ {\boldsymbol{\cY}^{[I=0]}}^\dagger\,,
    \label{eq:KdfI0} \\
    {\boldsymbol{\cY}}^{[I=0]} & = \begin{pmatrix}
    -\YR123 \\
    \sqrt{\tfrac12} \YR312
    \end{pmatrix}\,,
    %%%
    \quad
    {\boldsymbol{\cY}^{[I=0]}}^\dagger = \begin{pmatrix}
    -\YL123, & \sqrt{\frac12} \YL312
    \end{pmatrix}\,,
    \label{eq:YI0}
    \end{align}
    %%%%%
where the operator $\boldsymbol{\mathcal Y}^{[kab]}_{\boldsymbol \sigma}$ is defined through is action on functions $g(\{p\})$ of three on-shell momenta,  
    %%%%%
    \begin{align}
    \left[{\boldsymbol {\mathcal Y}}_{{{\boldsymbol \sigma}}}^{[kab]} \circ g \right]_{k s m}
    &=
    \frac{1}{4\pi} \int d\Omega_{a^*} Y_{s m}(\hat a^*)
    g(\{ p_i \})\bigg|_{p_{\sigma(1)}\to k,\ p_{\sigma(2)}\to a, \ p_{\sigma(3)}\to b} \,,
    \label{eq:YRdef}
    \end{align}
    %%%%%
where $\boldsymbol{\sigma}$ is a permutation of $\{123\}$. The explicit decomposition into the $k s m$ basis for the quantities entering the threshold expansion, i.e. those in \Cref{eq:kinematics}, has been presented in ref.~\cite{Blanton:2021eyf} for the $\pi^+\pi^+ K^+$ and $K^+K^+\pi^+$ systems and implemented numerically in ref.~\cite{Draper:2023boj}. This work carries over to the $I=0$ $DD\pi$ system, aside from trivial changes due to differences in meson masses and the nontrivial appearance of negative signs in~\Cref{eq:YI0} (which are absent in the $\pi^+\pi^+ K^+$ and $K^+K^+\pi^+$ systems).

%%%%%%%%%%%%%%%%%%%%%%%%%%%%%%%%%
% SECTION
%%%%%%%%%%%%%%%%%%%%%%%%%%%%%%%%%
\section{Numerical results}
\label{sec:results}

In practice, one wishes to apply the framework presented above to the QCD finite-volume $DD\pi$ energies to first constrain parametrizations of $\cK_2$ and $\cK_3$, then solve the integral equations, and finally determine the pole position of $T_{cc}^+$. Given that no $DD\pi$ lattice data exists yet, we illustrate the application of the three-body formalism by making reasonable assumptions about the interactions of the $DD\pi$, $DD$, and $D\pi$ systems. We study their behavior for different fixed choices of the two- and three-particle parameters and estimate a value for the three-body $K$ matrix.

Numerical solutions of the partial-wave projected three-body integral equations, described in \Cref{eq:d_formal,eq:t-matrix}, are obtained by fixing the two-body $D\pi$ and $DD$ interactions from the available lattice results and adjusting the unknown three-body $K$ matrix to match our LSZ-reduced amplitude to the result of ref.~\cite{Padmanath:2022cvl}. We then use our determination of $\cK_3$ in the three-body finite-volume quantization condition to derive corresponding $DD\pi$ and $DD^*$ energy levels and compare them to the analogous two-body results. Finally, we discuss the significance of the left-hand cuts in both the infinite- and finite-volume numerical solutions.

We start with~\Cref{sec:res-latt-two-body}, where we provide the values of particle masses, thresholds, and the ERE parameters used to describe interactions in the two-body subchannels. We use a combination of lattice results and a comparison with the heavy-light meson ChPT to interpolate scattering lengths and effective ranges to the pion mass used in ref.~\cite{Padmanath:2022cvl}. Next, in~\Cref{sec:res-k3-0}, we focus on the solution to the ladder equation, where the states interact only in pairs and by one-particle exchanges. We first discuss an approximate solution, where the partial-wave mixing is ignored by artificially truncating the $DD^*$ angular momentum space. Then, we present the partial-wave mixing result. In both cases, we find that the ladder amplitude alone cannot describe the lattice data. Thus, in~\Cref{sec:res-full-res}, we include a non-zero $\cK_3$, trying to tune the partial-wave projected integral equations to describe the $DD^*$ scattering. We find that by adjusting a single parameter, $\cK^E_3$, we can match the $DD^*$ phase shift to the lattice results reasonably well. Again, two models are considered: one neglecting and one including the partial-wave mixing. In both cases, we observe the $T_{cc}^+$ as a pair of complex poles and study their trajectory as a function of the three-body coupling. The analytic properties of the amplitude obtained in our approach are similar to and confirm those reported in Refs.~\cite{Du:2023hlu, Collins:2024sfi, Meng:2023bmz, Abolnikov:2024key}. In \Cref{sec:res-j-0}, we produce the $DD^*$ amplitude in the $J^P = 0^-$ wave, employing the set of parameters fixed in the $J^P=1^+$ case. We find no additional bound or virtual states with this quantum number and observe a qualitative agreement between our result and the infinite-volume parametrization extracted from the lattice calculation. Finally, in \Cref{sec:res-fin-vol}, we determine the form of the three-particle spectrum using the determined scattering parameters as input in QC2 and QC3. We observe a breakdown of the QC2 in the vicinity of the left-hand cut in agreement with the numerical results of ref.~\cite{Dawid:2023jrj}.

%%%%%%%%%%%%%%%%%%%%%%%%%
%%%%%%%%%%%%%%%%%%%%%%%%%
\subsection{Physical setup and two-body interactions}
\label{sec:res-latt-two-body}

We compute the $DD^*$ amplitude for one of the two sets of masses reported in ref.~\cite{Padmanath:2022cvl},
    %%%%%
    \begin{align}
    \label{eq:masses}
    m_\pi \approx 280~\text{MeV} \, ,~~~ m_D \approx 1927~\text{MeV} \, ,~~~ m_{D^*} \approx 2049~\text{MeV} \, .
    \end{align}
    %%%%%
Throughout this section, we use $m_D$ as a unit of energy. Above values correspond to the ratios $\kappa = m_\pi/m_D = 0.1453$, and $m_{D^*}/m_D = 1.0663$. The three-body threshold is placed at $E_{DD\pi}/m_D = 2.1453$, the $DD^*$ threshold at $E_{DD^*}/m_D = 2.0663$ and the nearest OPE branch point in the $DD^*$ amplitude at $E_{2}^{\rm lhc}/m_D = 2.0592$. The authors of ref.~\cite{Padmanath:2022cvl} determined the $S$-wave $DD^*$ phase shift from the L\"uscher method and implemented the ERE parametrization to identify the $T_{cc}^+$ as a virtual state around the energy ${E_{T_{cc}}/m_D = 2.0582}$, slightly below the left-hand cut branch point. Subsequent studies reanalyzed the same data set using chiral-EFT-based dynamical equations and incorporating the presence of the OPE in the amplitude~\cite{Du:2023hlu, Collins:2024sfi, Meng:2023bmz, Abolnikov:2024key}. Instead of a single $T_{cc}^+$ state, they found two complex poles on the second Riemann sheet of the $DD^*$ amplitude, at energies around $E_{T_{cc}}/m_D = 2.0597 \pm 0.0043i$.

To apply the three-body integral equations to the $DD\pi$ scattering, one must specify interactions in the $D\pi$ and $DD$ subchannels. For consistency, the two-body amplitudes should be given for the particle masses of~\Cref{eq:masses}. While existing lattice studies do not match these values, the available data is sufficient to place nontrivial constraints on the required two-body amplitudes, as we now explain.

Following the description of~\Cref{sec:ladder}, we assume that the $I=1$ $DD$ channel is weakly repulsive and well-approximated by neglecting all but the lowest partial wave, $s=0$. To our knowledge, no lattice computations have been performed for this channel, but, like the $I=1$ $KK$ system, it is not expected to exhibit bound states or resonances. Our model of the reaction depends on a single parameter, the two-body scattering length $a_{0}^{(2)}$, and our solutions show a negligible dependence on its value. 
This is due in part to the assumed weakness of the interactions in this channel. More important, however, is that, for our choice of cutoff function, the $DD$ subsystem decouples when its invariant mass falls below the two-pion exchange branch cut. It happens when,
    %%%%%
    \begin{align}
    E/m_D < 2 \sqrt{1 - \kappa^2 } + \kappa = 2.1241 \, , 
    \end{align}
    %%%%%
which lies significantly above the $DD^*$ threshold and slightly below the $D^*D^*$ and $DD\pi$ thresholds. Therefore, from the perspective of $DD^*$ scattering and $T_{cc}^+$ physics, the $DD$ interactions contribute only at energies near the inelastic thresholds. In practice, we set $m_D \, a_0^{(2)} = -10^{-2}$.

%%%%%%%%%%%%%%%%%%%%%%%%%%%%%%
\begin{figure}[t]
    \centering
    \includegraphics[width=0.8\textwidth]{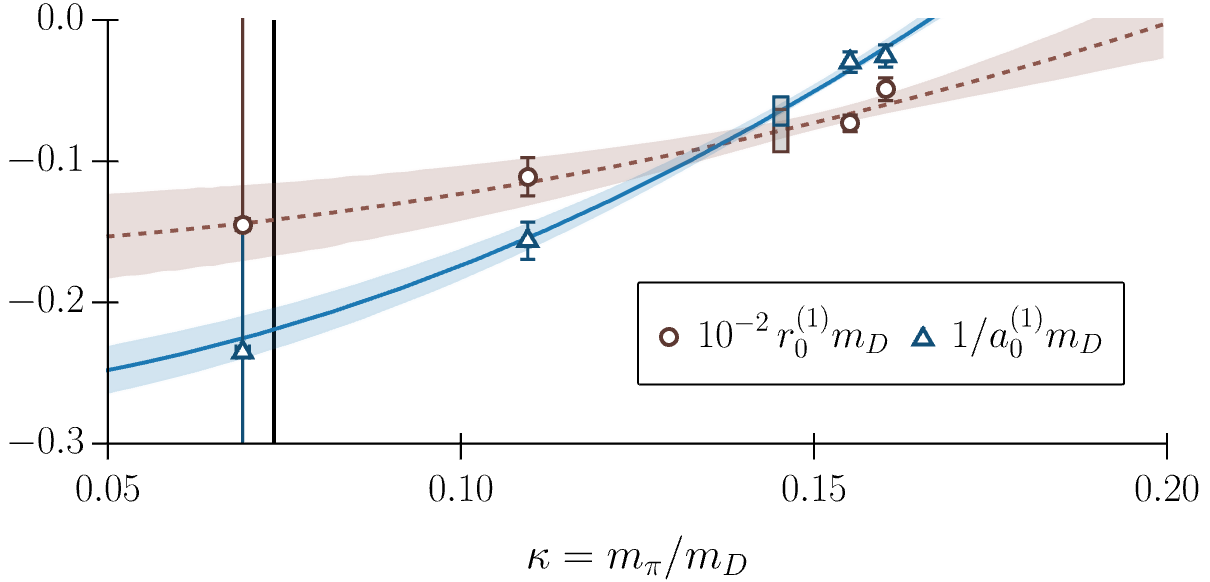}
    \caption{Determination of the $D\pi$ $s$-wave scattering parameters. Triangles and circles represent entries of Table II in ref.~\cite{Yan:2024yuq}, while solid curves represent chiral extrapolation of the form $x = c_0 + c_1 \kappa^2$, where the dimensionless parameter $x = 1/a_0^{(1)} m_D$ (solid, blue) or $r_0^{(1)} m_D$ (dashed, brown). Dark blue and brown boxes at $\kappa = 0.1453$ depict the intervals of parameters considered in this work. The black vertical line at $\kappa = 0.0735$ corresponds to the physical ratio, $m_\pi^{\rm phys}/m_D^{\rm phys}$.
    }
    \label{fig:two-body-params}
\end{figure}
%%%%%%%%%%%%%%%%%%%%%%%%%%%%%%

The $I=1/2$ $D\pi$ channel is described with four parameters: the $s$-wave scattering length $a_0^{(1)}$ and effective range $r_0^{(1)}$, together with the $p$-wave scattering length $a_1^{(1)}$ and effective range $r_1^{(1)}$. Several finite-volume studies of the $D\pi$ reaction are available~\citep{Mohler:2012na, Becirevic:2012pf, Moir:2016srx, Gayer:2021xzv, Yan:2024yuq} with the most recent work, ref.~\cite{Yan:2024yuq}, exploring the $s$- and $p$-wave scattering in ensembles corresponding to four different pion masses. In principle, no precise connection can be assumed between the $DD^*$ system considered in ref.~\cite{Padmanath:2022cvl} and the $D\pi$ lattice studies mentioned above due to different physical parameters, finite-volume and discretization contributions, and other systematic effects, associated, e.g., with the set of the employed interpolation operators or the approximations made in applying the L\"uscher method. For this reason, we use the above studies to determine a plausible interval of the two-body parameters rather than trying to fix them exactly. We then study the dependence of our solutions on these values within the constrained ranges. 

As in the $DD$ case, we assume weak interactions in the $D\pi$ $s$-wave scattering and consider two models for the amplitude. In the ``spartan'' approach, we set $r_0^{(1)}=0$ and constrain $|a_0^{(1)}| < 1/m_\pi$ such that no bound or virtual states appear in the vicinity of the $D\pi$ threshold. For these values, the physical and spurious pole of the LO ERE move to the complex plane on the first (for positive $a_0^{(1)}$) or the second Riemann sheet (for negative $a_0^{(1)}$), away from the threshold, and do not introduce any important physical effects. The $D\pi$ interactions are expected to be attractive, so we explore only positive values of $a_0^{(1)}$.

In the second approach, we consider an amplitude including a state analogous to the physical $D_0^*(2300)$ meson. This system has been studied extensively on the lattice in refs.~\cite{Moir:2016srx, Gayer:2021xzv, Yan:2024yuq}. To specify this model of the amplitude, we use physical parameters from Tables~II and~VIII of ref.~\cite{Yan:2024yuq}. Note that we follow the opposite sign convention for the LO term of the ERE than this work. We use the provided values to interpolate to our mass of choice, $\kappa = 0.1453$, by applying the chiral formula $x = c_0 + c_1 \kappa^2$, where the dimensionless quantity $x$ is either $1/a_0^{(1)} m_D$ or $r_0^{(1)} m_D$. The results are shown in \Cref{fig:two-body-params}. The estimated scattering length and effective range are,
    %%%%%%
    \begin{align}
    \label{eq:Dpi-params-s}
    a_0^{(1)} m_D = -15.5 \pm 2.5 \, , ~~~ r_0^{(1)} m_D = -7.84 \pm 1.50  \, ,
    \end{align}
    %%%%%%
where the chosen ranges are conservative.  Assuming the above values of the parameters, the $D_0^*$ appears as a virtual state below the $D\pi$ threshold, in agreement with Fig.~3 of ref.~\cite{Yan:2024yuq}.

Regardless of the choice of model, we find that the $J^P=1^+$ three-body amplitude is largely insensitive to the assumptions about the $s$-wave interactions. In the subsequent sections, we use the central values from~\Cref{eq:Dpi-params-s}. The situation changes for the $J^P=0^-$ $DD^*$ scattering, for which we find a noticeable dependence on the $s$-wave parameters. We discuss this further in~\Cref{sec:res-j-0}.

We now turn to the $p$-wave $D\pi$ interactions. As discussed above, we use a two-term ERE expansion,~\Cref{eq:k-matrix}. Under the assumption that the $D^*$ pole position of $\cM_{2,1}^{(1)}$ is fixed at a mass given in~\Cref{eq:masses}, the parameters $a_1^{(1)}$ and $r_1^{(1)}$ are related to each other by~\Cref{eq:pole-condition}, which depends on the (purely imaginary) binding momentum $q_0$, \Cref{eq:D-star-pole-spec}. For our chosen masses, this momentum is $q_0 = i \, 0.87 m_\pi$.
We choose $r_{1}^{(1)}$ to be the free parameter and fix it by equating the value of the residue of $\cM_{2,1}^{(1)}$ at the pole to the residue computed in terms of the $D^*D\pi$ coupling, $g_{D^* D \pi}$, defined in the continuum literature~\cite{Belyaev:1994zk}. As explained in \Cref{app:chpt}, we find,
    %%%%%%
    \begin{align}
    \label{eq:Dpi-r}
    r_1^{(1)} = -3 |q_0| - \frac{128 \pi}{g_{DD^*\pi}^2} \frac{m_{D^*}^5}{ m_{D^*}^4 - (m_\pi^2 - m_D^2)^2 } \,.
    \end{align}
    %%%%%%
In the lattice QCD literature, it is more common to use the coupling $g$ appearing in heavy-quark effective theory. The relation between couplings is  $g_{DD^*\pi}^2 = 4 g^2 m_D m_{D^*}/f_\pi^2$, where the pion decay constant is taken to be $f_\pi = \sqrt{2}F_\pi = 130\;$MeV. Using experimental decay widths, one finds $g_{\rm exp} \approx 0.57$~\cite{BaBar:2013zgp}, while lattice simulations find the range $g_{\rm lat} \in [0.50,0.61]$, depending on the pion mass and lattice spacing~\cite{Becirevic:2012pf}.

We consider values of $g$ in this range, which translates to $r_{1}^{(1)} /m_D \in [-7.7, -5.3]$ and $a_1^{(1)} M_D^3 \in [24.7, 16.8]$ with the central values $r_{1}^{(1)}/m_D = -6.5$, $a_1^{(1)} m_D^3 = 19.98296$. The solutions to the integral equations depend noticeably on the assumed values of the $p$-wave parameters. Therefore, in the following, to provide an idea of the parameter dependence of our $DD^*$ amplitude, we show our main results for three choices of the coupling, $g = 0.4996 \approx 0.50$, $g = 0.5464 \approx \, 0.55$, and $g = 0.6094 \approx \, 0.61$. We note that for this set of parameters, the $p$-wave amplitude has additional poles; a virtual state with $q_0/m_D \approx - 0.12 \, i$, as well as a distant physical-sheet pole with $q_0/m_D \approx 3 \, i$. These singularities do not affect the amplitude significantly in the integration region (e.g., above the threshold) due to the dominance of the $D^*$ pole.

Finally as explained in ref.~\cite{Romero-Lopez:2019qrt}, to compute the finite-volume spectrum, we need to fix constants introduced in~\Cref{eq:generalized-rho}. We set $c_{\text{PV},1}^{(1)}/m^3_D  = -0.25$ in the $p$-wave $D\pi$ channel to account for the real-axis $D^*$ pole and $c_{\text{PV},0}^{(1)}/m_D  = -0.2$ in the $s$-wave $D\pi$ channel due to the lattice equivalent of the $D^*_0$ pole.

%%%%%%%%%%%%%%%%%%%%%%%%%
%%%%%%%%%%%%%%%%%%%%%%%%%
\subsection{The $J^P=1^+$ ladder amplitude}
\label{sec:res-k3-0}

%%%%%%%%%%%%%%%%%%%%%%%%%
\begin{figure}[t]
    \centering
    \includegraphics[trim={0 0 0 3pt},clip, width=0.98\textwidth]{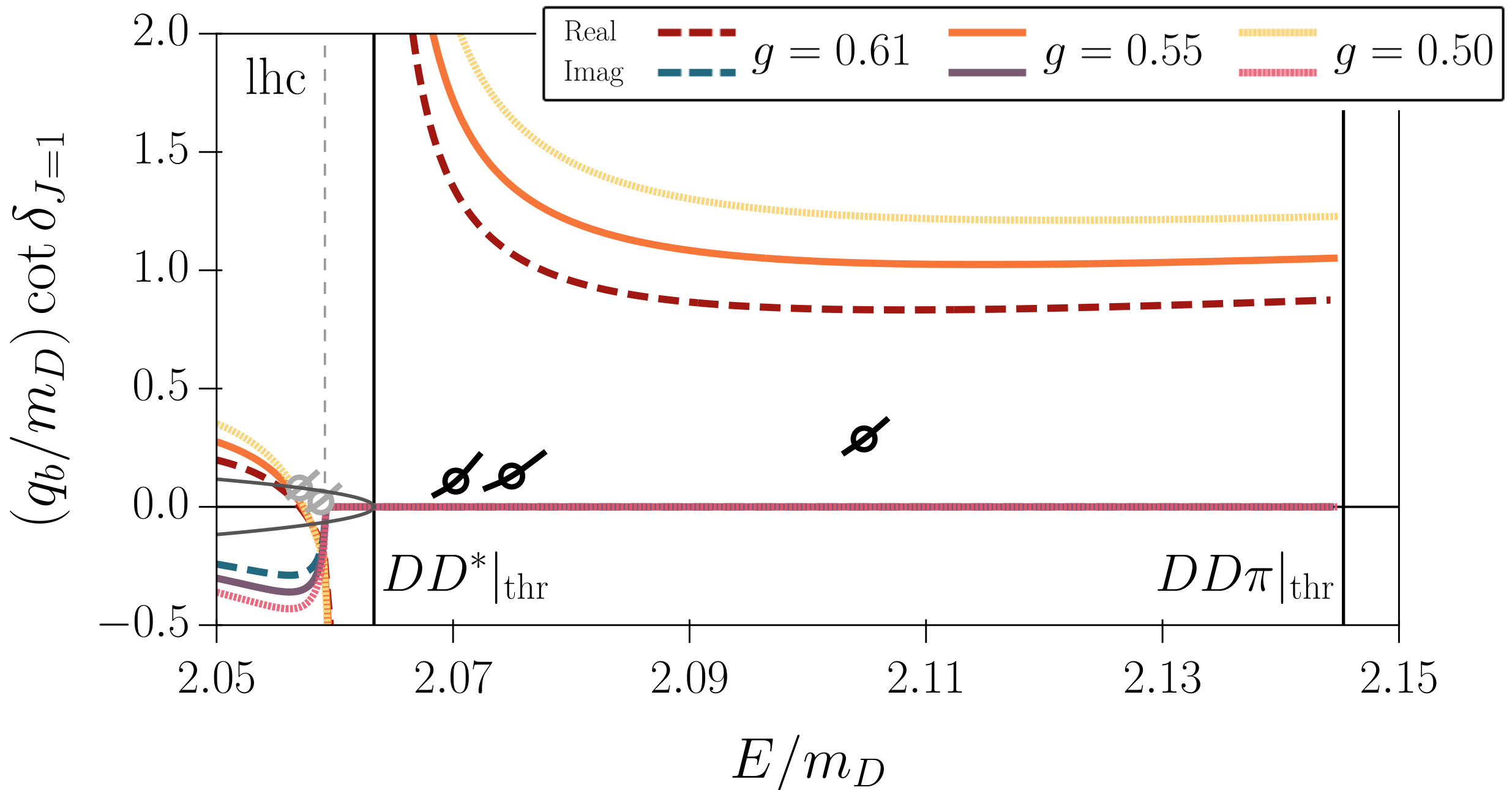}
    \caption{The $J^P=1^+$ $DD^*$ scattering phase shift, as defined in~\Cref{eq:phase_shift_one}, 
    obtained by $s$-wave truncation of the ladder amplitude. We plot $q_b \cot \delta$ as a function of energy in the vicinity of $DD^*$ and $DD\pi$ thresholds (black, vertical lines). The solution is given for three different values of the $D^* \to D\pi$ coupling $g$,
    as shown in the legend.
    The phase shift acquires a non-zero imaginary part below the left-hand cut branch point (dashed vertical line). Black and gray points are from the $s$-wave phase shift extracted using the QC2 in ref.~\cite{Padmanath:2022cvl}. Points below $E_{2}^{\rm lhc}$ are highlighted by a lighter, gray color, as they are not expected to be reliable.  
    Gray, solid lines below the $DD^*$ threshold are $\pm |q_b/m_D|$. 
    The absence of a crossing of these lines above the left-hand cut shows the absence of bound or virtual states in this channel.
    }
    \label{fig:ladder-approx}
\end{figure}
%%%%%%%%%%%%%%%%%%%%%%%%%

%%%%%%%%%%%%%%%%%%%%%%%%%
\begin{figure}[t]
    \centering
    \includegraphics[trim={0 0 0 3pt},clip, width=0.98\textwidth]{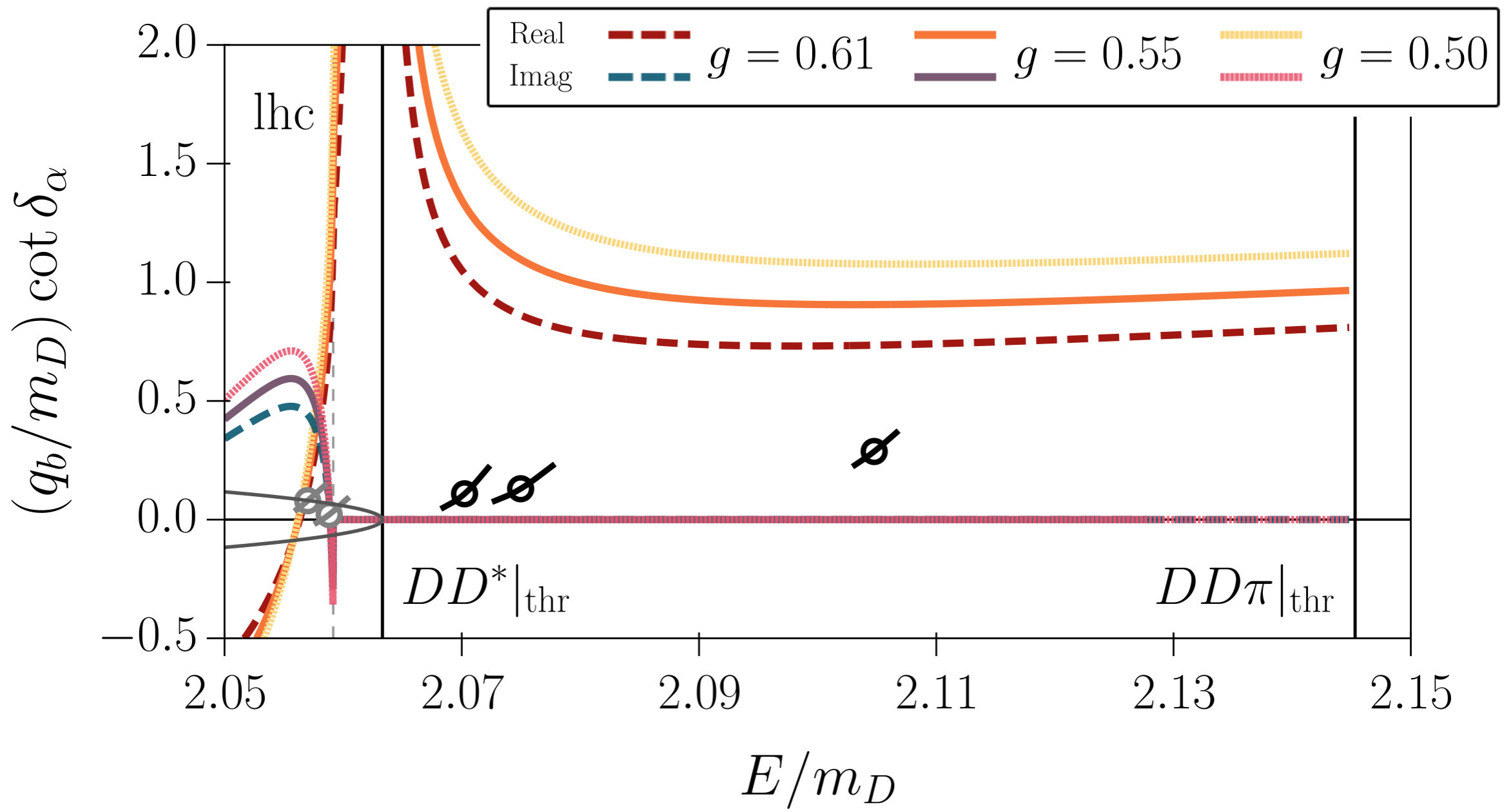}
    \caption{Results for $q_b\cot\delta_\alpha$, as defined in the Blatt-Biedenharn parametrization of ~\Cref{eq:Blatt-Biedenharn}. It is obtained from the LSZ-reduced ladder amplitude including the $s$- and $d$-wave mixing.
    Notation and conventions are as in~\Cref{fig:ladder-approx}. Note that the enhancement in the real parts below the threshold is not a pole, instead peaking at a finite value.
    The imaginary parts display a sharp dip just below the left-hand cut.
    Gray, solid lines are $\pm |q_b/m_D|$. 
    The absence of a crossing of these lines above the left-hand cut shows the absence of bound or virtual states in this channel.
    }
    \label{fig:ladder-full}
\end{figure}
%%%%%%%%%%%%%%%%%%%%%%%%%

In the first step, we solve the ladder equation,~\Cref{eq:d_formal}, using methods of~\Cref{sec:numerical_solution} and assuming two-body parameters from the previous subsection. The short-range interactions are ignored by setting $\bm \cK_3 = 0$ to investigate the amplitude driven purely by the pair-wise interactions and one-particle exchanges.%
%%%%
\footnote{%
We remind the reader that this is a scheme-dependent statement that changes its meaning whenever a different choice of the cutoff function is made.}
%%%
We perform the LSZ reduction of the $DD\pi$ amplitude and extract the $DD^*$ $K$ matrix.

We first present the approximate solution obtained by restricting the angular momentum space to the ${}^3S_1 \to {}^3 S_1$ transitions. This is obtained by artificially removing (setting to zero) entries of the integral equation corresponding to $(\ell,s) = (2,1)$ or $(\ell',s') = (2,1)$. Although this approximation explicitly violates the unitarity of the amplitude, it is justified, provided this system exhibits small partial-wave mixing. It has been assumed in many preceding studies~\cite{Padmanath:2022cvl, Lyu:2023xro, Chen:2022vpo, Collins:2024sfi} and was additionally investigated in ref.~\cite{Whyte:2024ihh} where the authors discussed in detail the effect of the higher partial waves. Our results are shown in~\Cref{fig:ladder-approx}, displayed as $q_b \cot \delta_{J=1}(E)$, as defined in~\Cref{eq:phase_shift_one}. They are compared to the lattice data points of ref.~\cite{Padmanath:2022cvl}. The visible disagreement indicates that a non-zero $\cK_3$ must be included to describe the data. We note that the amplitude exhibits a zero below the threshold, which leads to a pole in the inverse of the $DD^*$ $K$ matrix. This agrees with the studies of this setting based on non-relativistic chiral EFT~\cite{Sasa:2024}. 
The appearance of this zero has also been argued on the grounds of general $N/D$ parametrization of amplitudes constructed from the one-particle exchange interactions~\cite{Du:2024snq}, independently of the value of the short-range couplings. Moreover, the amplitude has no virtual or bound state pole near the threshold that could be identified with $T_{cc}^+$. This can be inferred from the figure, since the real part of $q_b \cot \delta_{J=1}$ does not cross the parabola $\pm |q_b/m_D|$ (gray, solid line) above the $E_2^{\rm lhc}$. 

We display in \Cref{fig:ladder-approx} and subsequent figures both the $DD^*$ and $DD\pi$ thresholds, shown as black vertical lines, as well as the left-hand cut (dashed vertical line). In principle, our formalism is valid above the $DD\pi$ threshold, as well as below the left-hand cut. However, for the $D^*$ mass considered here, another threshold, namely that of the $D^*D^*$, is placed below $E_{DD\pi}$, at $E_{D^*D^*}/m_D = 2.1266$. We cannot investigate the effect of the opening of this new channel, as this would require a four-particle formalism in our approach. Our results effectively assume that it is closed, and are insensitive to its presence. Thus, to avoid clutter, we do not show this threshold on the figures. We do note, however, that it has been recently studied (at a heavier pion mass) in ref.~\cite{Whyte:2024ihh} using a multichannel QC2 approach.

The solution of the full ladder equation, including  ${}^3S_1$--${}^3D_1$ mixing, is shown in~\Cref{fig:ladder-full} in the form of the $DD^*$ $K$ matrix $\alpha$ eigenvalue, as defined in the Blatt-Biedenharn parametrization,~\Cref{eq:Blatt-Biedenharn}.
It is similar to the truncated ladder solution above the $DD^*$ threshold, which hints at the insignificance of the partial-wave mixing in this system. We include the lattice $\cot \delta_{J=1}$ points in the figure for reference, even though they are not, strictly speaking, comparable with the outcome of the partial-wave mixing calculation. Nevertheless, since $\cot \delta_\alpha = \cot \delta_{J=1} + \cO(\epsilon^2)$~\cite{Briceno:2013bda}, and we find small values for the mixing angle $\epsilon$ (not shown), a qualitative comparison can be made, and we see again that our results do not lie close to the lattice points.

The amplitude exhibits a zero below the threshold in the $(\ell,s)=(0,1)$ partial wave, which leads to a peak (but no pole) in $\cot \delta_\alpha$ around the same energy. Other matrix elements of the amplitude vanish at the threshold in agreement with the expected $\propto q_b^{2(\ell'+\ell)}$ behavior. Generally speaking, the behavior of the alpha eigenvalue significantly differs from the $\cot \delta_{J=0}$ of the truncated solution below the elastic threshold. Not only has the pole disappeared, but the imaginary part below the lhc changed its sign from negative to positive. This can be explained by the rapid increase of the mixing angle below the threshold, which does not contradict the na\"ive expectation that $\cot \delta_{\alpha} \approx \cot \delta_{J=1}$ for energies above $E_{DD^*}$.
We postpone showing results for the mixing angle and $\cot\delta_\beta$ until we consider the full solution with nonzero $\cK_3$.

%%%%%%%%%%%%%%%%%%%%%%%%%
\begin{figure}[t]
    \centering
    \includegraphics[trim={0 0 0 3pt},clip, width=0.98\textwidth]{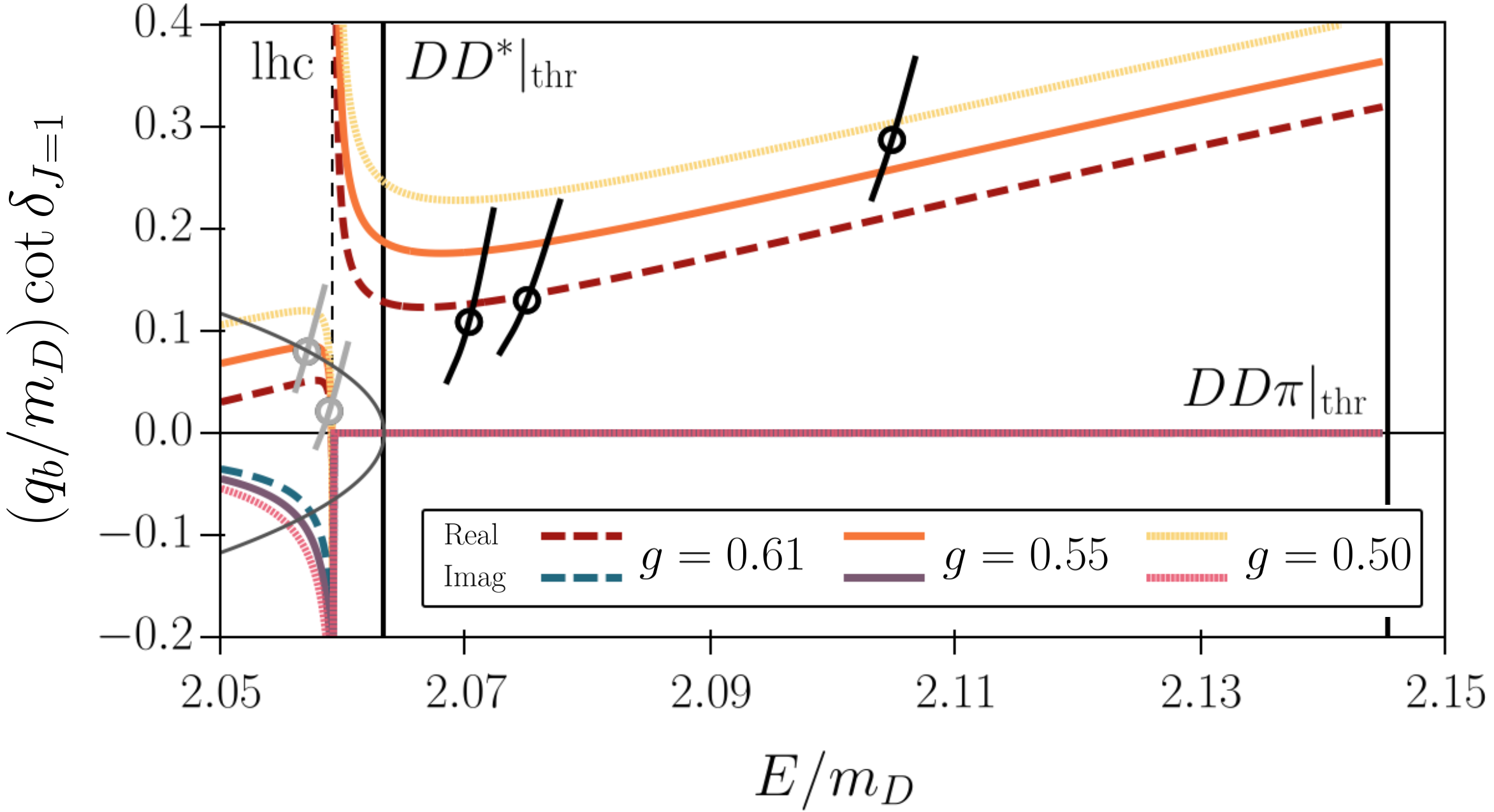}
    \caption{%
    Results for $q_b\cot\delta_{J=1}$ for the $s$-wave truncated solution with a nonzero value for the three-particle $K$ matrix, $m_D^2 \cK_3^E = 1.9 \cdot 10^5$.
    This value is chosen so that the $g=0.55$ curve passes through the lattice data points lying above the left-hand cut.
    Gray, solid lines are $\pm |q_b/m_D|$. 
    Notation as in \Cref{fig:ladder-approx}.
    }
    \label{fig:m3-trunc}
\end{figure}
%%%%%%%%%%%%%%%%%%%%%%%%%

%%%%%%%%%%%%%%%%%%%%%%%%%
%%%%%%%%%%%%%%%%%%%%%%%%%
\subsection{The $J^P=1^+$ amplitude with non-zero $\cK_3$}
\label{sec:res-full-res}

After this preliminary discussion, we turn to the complete solution of the combined three-body~\Cref{eq:d_formal,eq:t-matrix}, obtained for non-zero $\cK_3$. 
We emphasize again that, although the threshold expansion of $\cK_3$ contains several terms (see \Cref{eq:K3thr}), at the order we work only the $\cK_3^E$ term contributes to the $J^P=1^+$ channel.
As for the ladder amplitude discussed previously,  we study the result both in the $s$-wave truncated and the partial-wave mixing cases.

The results for $\cot \delta_{J=1}$ from the truncated solution are shown in~\Cref{fig:m3-trunc},
at a fixed strength of the three-body parameter, $m_D^2 \, \cK_3^E = 1.9 \cdot 10^5$, and for our standard three values of $g$. This is the value of the three-body $K$ matrix for which the $g=0.55$ curve matches reasonably well to the lattice data of ref.~\cite{Padmanath:2022cvl}.%
%%%%%
\footnote{%
We stress again that we do not perform a fit to the available points, but choose to match our solution by eye. As mentioned at the beginning of ~\Cref{sec:numerical_solution}, our goal is not to perform a rigorous analysis of the lattice data but rather to benchmark the RFT formalism and estimate reasonable values for $\cK_3^E$.
} 
%%%%%
It is possible to match the $g=0.50$ result more closely to the data by decreasing its value to $m_D^2 \, \cK_3^E \approx 1.4 \cdot 10^5$ and to match the $g=0.61$ result by increasing it to $m_D^2 \, \cK_3^E \approx 2.4 \cdot 10^5$. 

A characteristic feature of the $q_b \cot\delta$ curves for $s$-wave truncated solutions is the presence of a pole between the left-hand cut and the $DD^*$ threshold,
corresponding to a zero of the amplitude.
This agrees qualitatively with the results obtained using $s$-wave truncated Lippmann-Schwinger equations in refs.~\cite{Du:2023hlu, Meng:2023bmz, Collins:2024sfi, Abolnikov:2024key}. We find that this zero moves to lower energies as $\cK_3^E$ increases, approaching, but not crossing, the left-hand cut. Clearly, this pole invalidates the use of the ERE to describe the amplitude near the left-hand cut, as emphasized in~\cite{Du:2023hlu}.
The half-parabola $|q_b/m_D|$ is not crossed by our solutions, so we find no virtual bound states, in agreement with most examples in the literature~\cite{Meng:2023bmz, Abolnikov:2024key, Du:2023hlu, Collins:2024sfi}. 
By continuing the amplitude to the unphysical Riemann sheet, using \Cref{eq:amp-2nd-sheet}, we find two subthreshold conjugate complex poles.
These have been identified in the above-mentioned studies as corresponding to the $T_{cc}^+$. The positions of these poles are given in \Cref{tab:poles} in the first row. 

%%%%%%%%%%%%%%%%%%%%%%%%
\begin{table}[t]
    \centering
    \begin{tabular}{c|c|c|c}
     Model & $g=0.50$ & $g=0.55$ & $g=0.61$ \\ \hline \hline
     truncated & $2.05755 \pm 0.00457 \, i$ & $2.05396 \pm 0.00589 \, i$ & $2.04941 \pm 0.00662 \, i$ \\
     p.w.~mixing & $2.05797 \pm 0.00463 \, i$ & $2.05469 \pm 0.00578 \,i$ & $2.05035 \pm 0.00626 \, i$
    \end{tabular}
    \caption{
    Positions of putative $T_{cc}^+$ poles on the second sheet for the three-body coupling $m_D^2 \, \cK^E_3 = 1.9 \cdot 10^{5}$ and our standard choices of $g$. The ``truncated'' model includes only $s$ waves in the $D D^*$ system, while the ``p.w. mixing'' model includes both $s$ and $d$ waves.
    }
    \label{tab:poles}
\end{table}
%%%%%%%%%%%%%%%%%%%%%%%%%

%%%%%%%%%%%%%%%%%%%%%%%%%
\begin{figure}[t]
    \centering
    \includegraphics[trim={0 0 0 3pt},clip, width=0.98\textwidth]{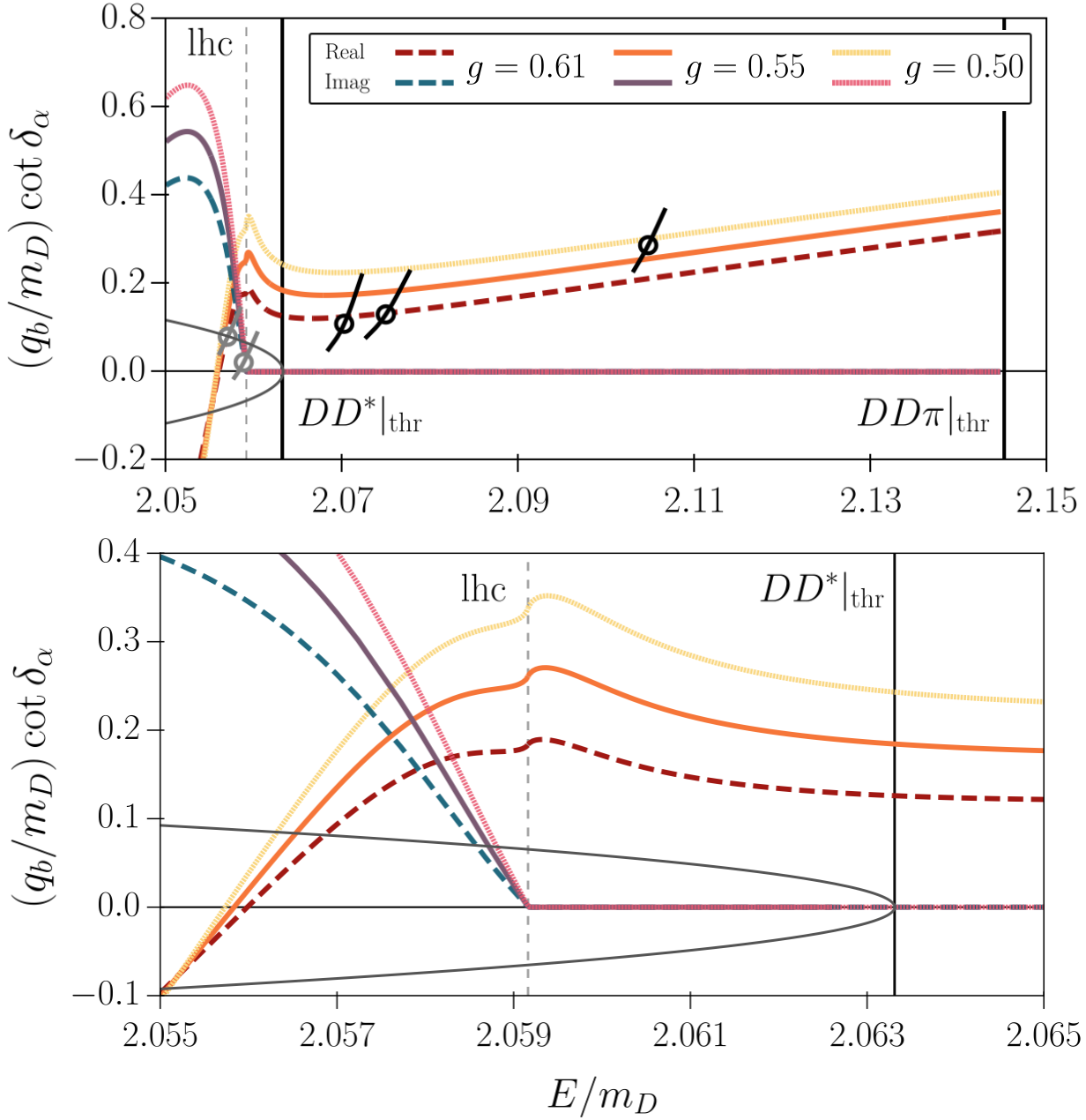}
    \caption{%
    Results for $q_b \cot\delta_\alpha$, obtained from the solution with  a nonzero three-body $K$ matrix, $m_D^2 \, \cK_3^E = 1.9 \cdot 10^5$, the same value as used in \Cref{fig:m3-trunc}.
    Notation as in \Cref{fig:ladder-full}. 
    The lower panel zooms in near the left-hand cut. 
    Gray, solid lines are $\pm |q_b/m_D|$. 
    }
    \label{fig:m3-full}
\end{figure}
%%%%%%%%%%%%%%%%%%%%%%%%%

%%%%%%%%%%%%%%%%%%%%%%%%%
\begin{figure}[t]
    \centering
    \includegraphics[trim={0 0 0 0},clip, width=0.98\textwidth]{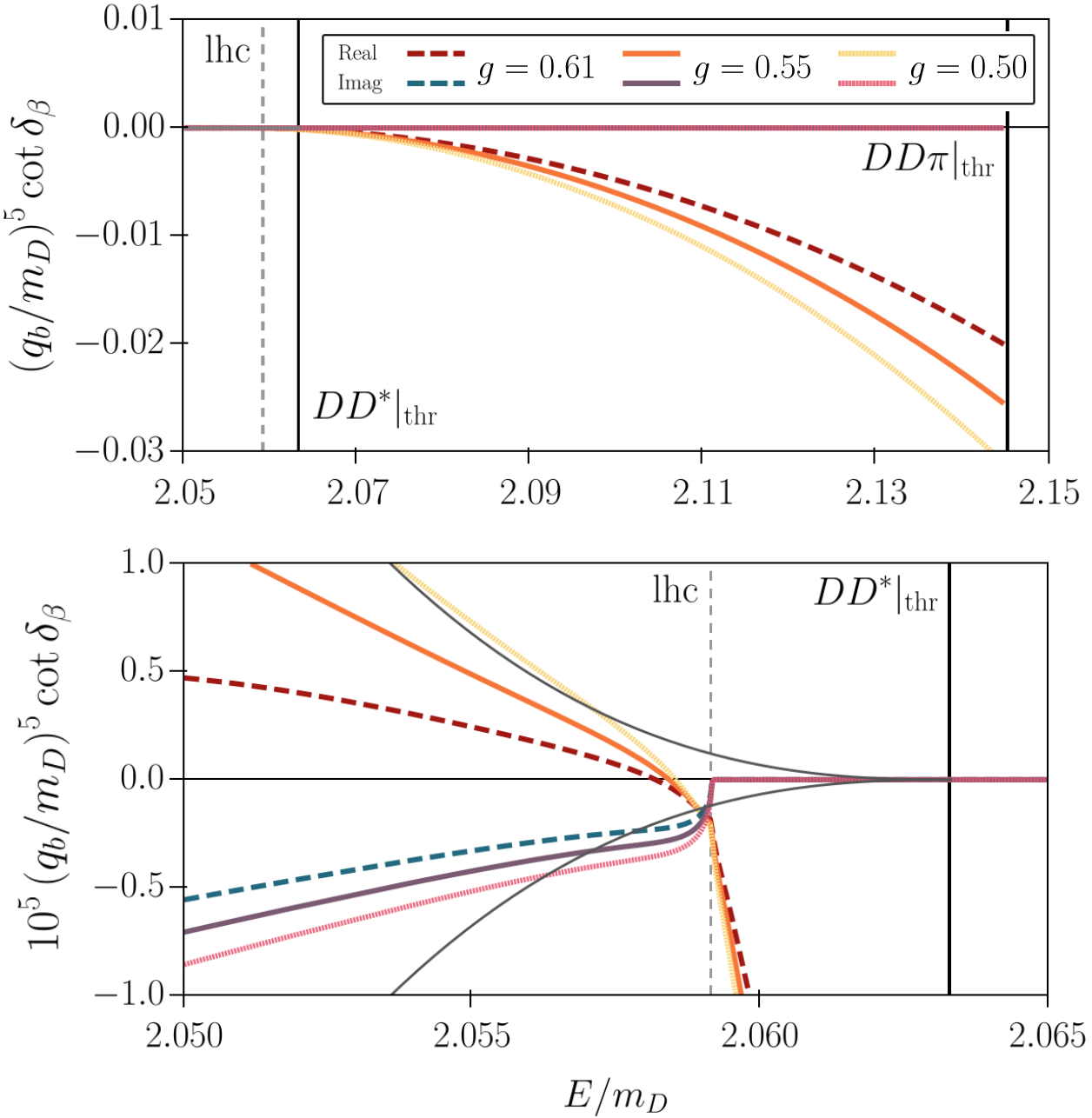}
    \caption{%
    Results for $q_b^5 \, \cot \delta_\beta$ from the same solution used to display $q_b \, \cot \delta_\alpha$ in
    \Cref{fig:m3-full}, using the same format. The lower panel zooms in near the left-hand cut. Gray, solid lines are $\pm |q_b/m_D|^5$. 
    }
    \label{fig:m3-beta}
\end{figure}
%%%%%%%%%%%%%%%%%%%%%%%%%

%%%%%%%%%%%%%%%%%%%%%%%%%
\begin{figure}[t]
    \centering
    \includegraphics[trim={0 0 0 0},clip, width=0.98\textwidth]{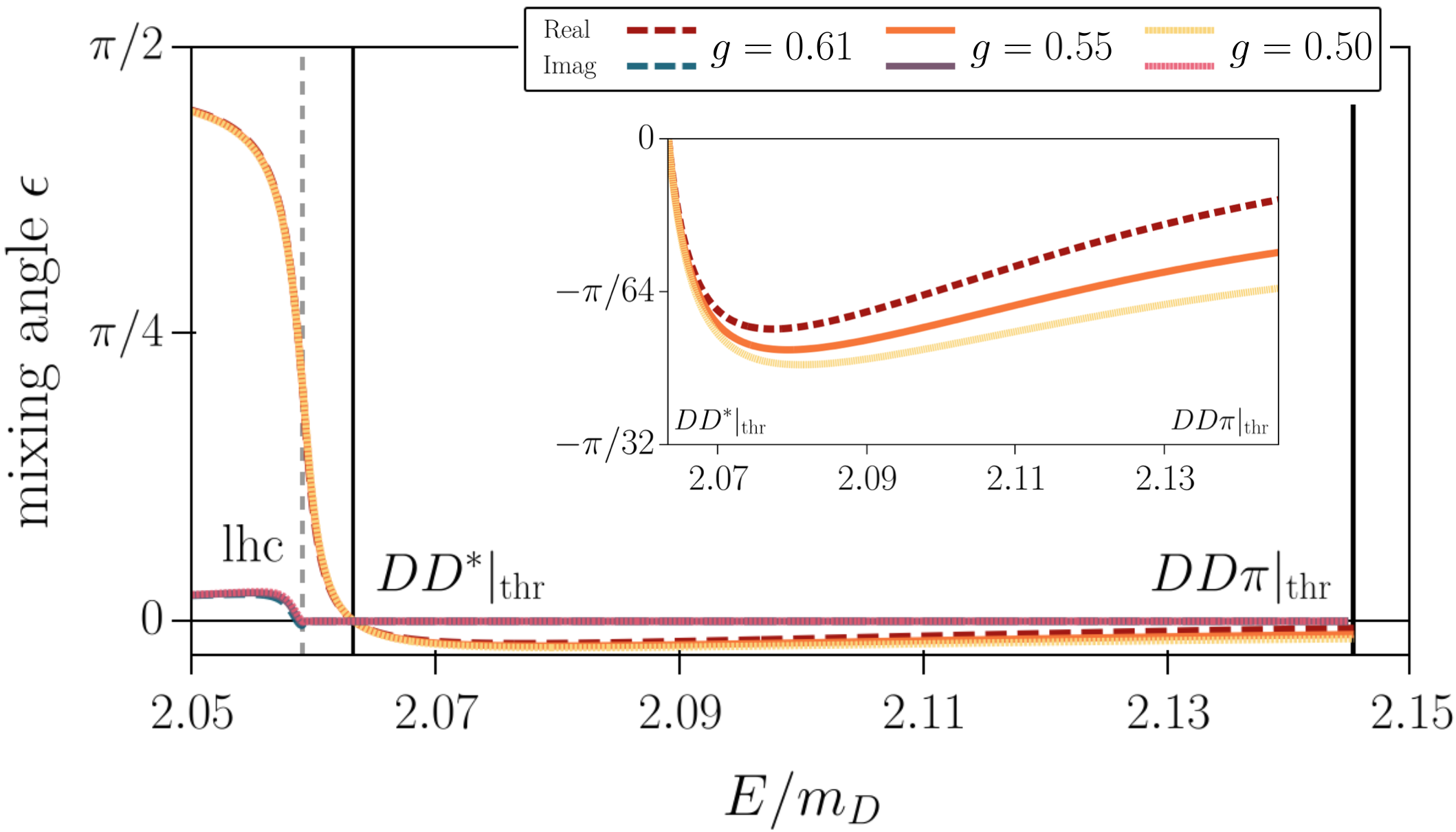}
    \caption{%
    Results for the mixing angle $\epsilon$ from the same solution as that used for~\Cref{fig:m3-full,fig:m3-beta}, and using the same format. The inset shows the result for $\epsilon$ in the physical region, where it is purely real.}
    \label{fig:m3-epsilon}
\end{figure}
%%%%%%%%%%%%%%%%%%%%%%%%%

Our result for the full, partial-wave mixing amplitude, obtained for the same non-zero value of the three-body $K$ matrix, $m_D^2 \, \cK_3^E = 1.9 \cdot 10^5$, is presented in three plots,~\Cref{fig:m3-full,fig:m3-beta,fig:m3-epsilon}. They show the alpha and beta eigenvalues of the Blatt-Biedenharn $\cK_{DD^*}$ matrix, and the mixing angle, respectively---see \Cref{eq:Blatt-Biedenharn}.

We first note that, above the $DD^*$ threshold, $\cot \delta_\alpha$ (\Cref{fig:m3-full}) is almost identical to the truncated model's $\cot \delta_{J=1}$ (\Cref{fig:m3-trunc}). This can be understood by the fact that the mixing angle $\epsilon \approx -\pi/64$ for $E > E_{DD^*}$, nearly decoupling the ${}^3S_1$ and ${}^3D_1$ sectors of the $K$ matrix. Furthermore, we find that the $\beta$ eigenvalue, shown in~\Cref{fig:m3-beta}, is an order of magnitude smaller than the $\alpha$ eigenvalue.

Below the $DD^*$ threshold, however, the situation is quite different. The mixing angle rises rapidly and the behavior of $\cot\delta_\alpha$ differs markedly from that of $\cot\delta_{J=1}$ in the truncated model. In particular, $\cot \delta_\alpha$ has no pole, behaving fairly smoothly, and its imaginary part below the left-hand cut has the opposite sign to that in the truncated model. These results show the importance of including the $d$-wave channel in this energy regime.

Neither $\cot\delta_\alpha$ nor $\cot\delta_\beta$ satisfies the corresponding bound or virtual state conditions, as can be seen from the lower panels of \Cref{fig:m3-full,fig:m3-beta}. Instead, as in the truncated model, we find two subthreshold complex poles. Their positions are listed in~\Cref{tab:poles} in the bottom row. We observe relatively small shifts between the truncated and partial-wave mixing cases. We also note that these values are of the same order as those presented in Fig.~7 of ref.~\cite{Abolnikov:2024key}.

\begin{figure}[t]
    \centering
    \includegraphics[trim={0 0 0 0},clip, width=0.98\textwidth]{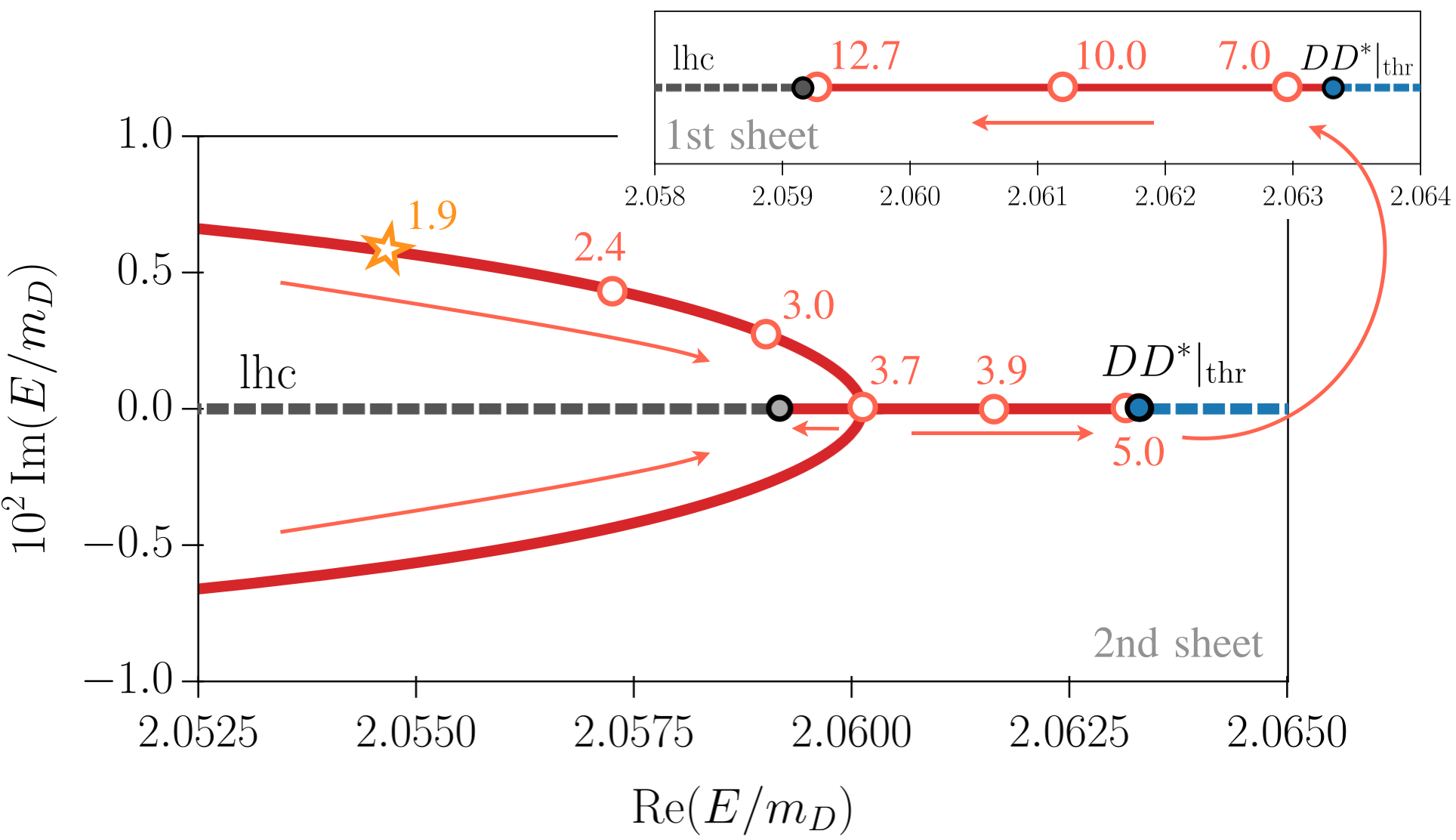}
    \caption{Trajectories of the putative $T_{cc}^+$ poles (red, solid lines) in the partial-wave mixing model, as a function of $\cK_3^E$, for the central value of the $D\pi$ $p$-wave coupling, $g=0.55$. Pink dots represent the position of the poles at the given values of $m_D^2\cK_3^E/10^5$. The orange star indicates the position of the pole for $m_D^2\cK_3^E = 1.9 \cdot 10^5$, the value chosen so that our $DD^*$ amplitude matches the data of ref.~\cite{Padmanath:2022cvl}. 
    The arrows indicate the motion of the poles as $\cK_3^E$ is increased. Dashed lines indicate the placement of the right- and left-hand cuts, which emerge respectively from the $DD^*$ threshold (blue dot) and left-hand branch point (grey dot). For further discussion, see the text.}
    \label{fig:pole}
\end{figure}

We have extended the above investigations to a large range of values of $\cK_3^E$, to explore the possible manifestations of the $T_{cc}^+$ pole. Our results are summarized in~\Cref{fig:pole}, where we show the pole trajectories for $g=0.55$. Starting from negative $\cK_3^E$, increasing its value leads to an evolution of the model's spectrum from two subthreshold complex poles to a pair of virtual states. These separate on the real axis, with one moving to the $DD^*$ threshold, and subsequently moving onto the first sheet and becoming, for large enough $\cK_3^E$, a bound state. As $\cK_3^E$ increases further, its binding energy increases, and the pole eventually moves below the left-hand cut, where we no longer trace its evolution. The companion pole remains a virtual state. In the $s$-wave truncated model, its position asymptotes with increasing $\cK_3^E$ to the left-hand branch point and the pole does not cross this singularity. By contrast, in the partial-wave mixing case, we find that this pole keeps moving to the left and crosses $E_{2}^{\rm lhc}$ at $m_D^2 \, \cK_3^E \approx 4 \cdot 10^5$. This is another example of an important difference between the two models below the $DD^*$ threshold.

Overall, the determined trajectory
resembles the results reported in the literature (see figures~11 and~12 in ref.~\cite{Collins:2024sfi} and figure~3 in ref.~\citep{Abolnikov:2024key}) although there different model parameters (charm quark and pion mass) were varied.

%%%%%%%%%%%%%%%%%%%%%%%%%
%%%%%%%%%%%%%%%%%%%%%%%%%

%%%%%%%%%%%%%%%%%%%%%%%%%
\begin{figure}[t!]
    \centering
    \includegraphics[trim={0 0 0 0},clip, width=0.97\textwidth]{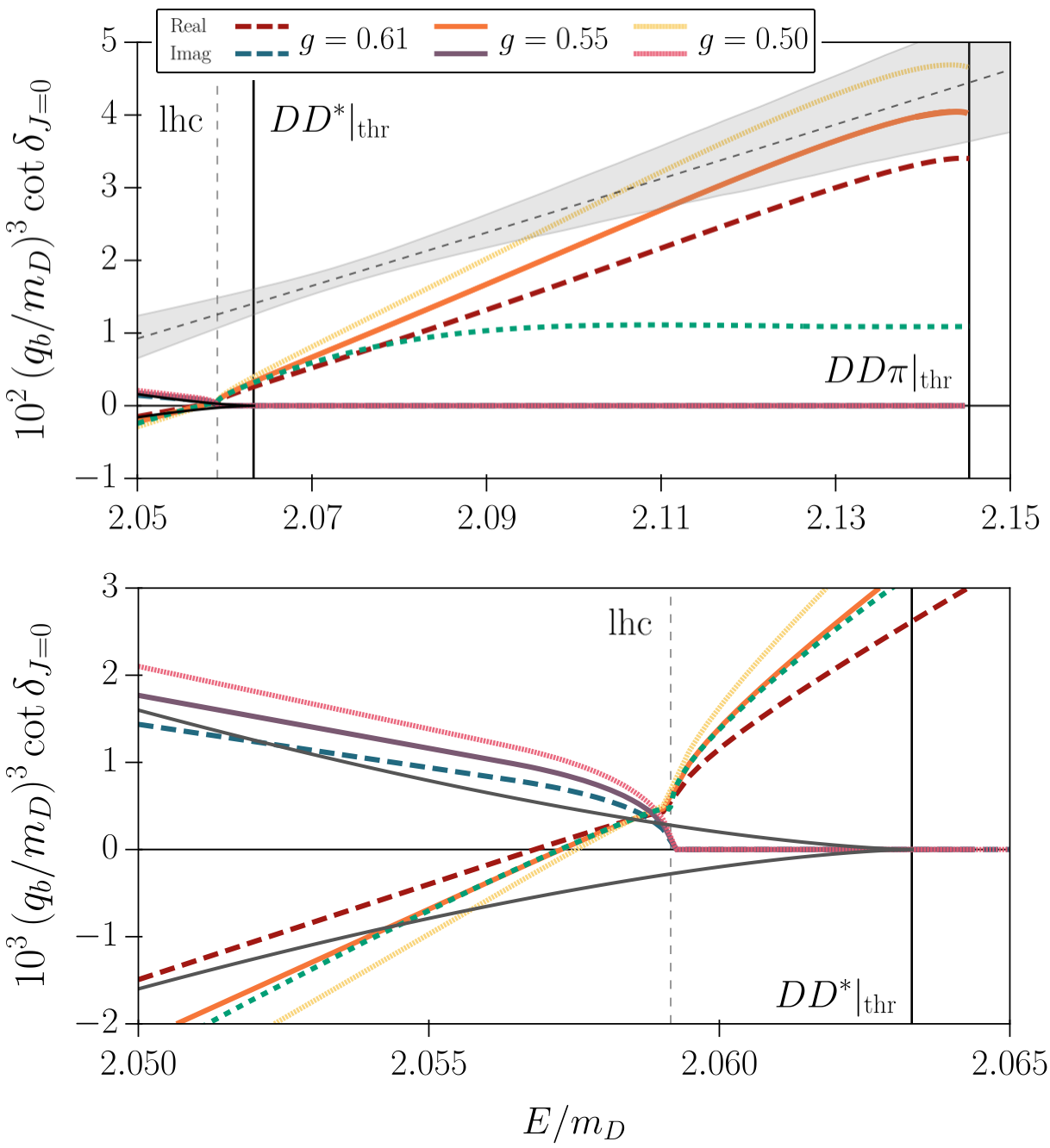}
    \caption{%
    Results for $q_b^3 \, \cot \delta_{J=0}$, i.e. the $DD^*$ interaction in the $\ell=1$, $J^P=0^-$ channel. As denoted in the legend, these are obtained for our standard values of the coupling $g$, with $m_D^2 \, \cK_3^E = 1.9 \cdot 10^5$, $\cK_3^{{\rm iso},0}=\cK_3^{
    {\rm iso},1}=\cK_3^B=0$, and with our standard choice of the $D\pi$ $s$-wave scattering parameters (\Cref{eq:Dpi-params-s}). In addition, we present as a green dashed line, the result obtained using the spartan model for the $D\pi$ $s$-wave interaction, as described in the text. The gray band centered around the dashed gray line is the result of a two-parameter ERE fit to the lattice spectrum using the QC2, as reported in ref.~\cite{Padmanath:2022cvl}. Dark gray lines correspond to $\pm |q_b/m_D|^3$, crossings of which above the left-hand cut correspond to bound or virtual states. The lower plot zooms in on the subthreshold region, showing the absence of any such crossings.}
    \label{fig:m3-j0}
\end{figure}
%%%%%%%%%%%%%%%%%%%%%%%%%

%%%%%%%%%%%%%%%%%%%%%%%%%
%%%%%%%%%%%%%%%%%%%%%%%%%
\subsection{The $J^P = 0^-$ amplitude}
\label{sec:res-j-0}

The lattice calculations of ref.~\cite{Padmanath:2022cvl} also report results for the phase shift in the $\ell=1$, $J^P=0^-$ $DD^*$ channel. Indeed, in the finite-volume analysis, in moving frames, the energies of states in certain irreducible representations (irreps) depend on the interactions in both the $J^P=1^+$ and $0^-$ channels. Thus, at some level of precision, one must include the $0^-$ channel in the calculation.

Thus motivated, we have solved the integral equations also for this channel. This requires performing the LSZ reduction on the ${}^3 P_0 \to {}^3 P_0 $ element of the $DD\pi$ amplitude, $\cM_{3;11;11}^{(11)0}$. We employ the parametrization of~\Cref{eq:phase_shift_one} to present our result in the form of $q_b^3 \, \cot\delta_{J=0}$ in~\Cref{fig:m3-j0}, providing also a comparison with the result from ref.~\cite{Padmanath:2022cvl}. A new feature of this channel is that the $\cK_{3}^{\rm iso, 0}$, $\cK_{3}^{\rm iso, 1}$, and $\cK_3^B$ terms in the threshold expansion of the three-particle $K$ matrix, \Cref{eq:K3thr}, contribute in addition to $\cK_3^E$. However, we find that the resulting $J^P=0^-$ phase shift is largely unaffected by any of the three-particle couplings at the order in the threshold expansion we consider---the dominant contribution is from the ladder amplitude. In particular, the clear disagreement in slope with the result from ref.~\cite{Padmanath:2022cvl} is not resolved by adding the $\cK_{3}^{\rm iso, 0}$, $\cK_{3}^{\rm iso, 1}$, or $\cK_3^B$ contributions. We note, however, that our result is consistent with that reported in ref.~\cite{Meng:2023bmz}---in both approaches $q_b^3 \, \cot \delta_{J=0}$ has a considerably larger slope than that from ref.~\cite{Padmanath:2022cvl}. We speculate that this difference may be due to the fact that the lattice data are fit using an ERE parametrization that does not include the effects of the one-pion exchange. We do find agreement with ref.~\cite{Padmanath:2022cvl} on the absence of virtual or bound states in this channel.

Finally, we note that the $J^P=0^-$ solution, unlike that for $J^P=1^+$, displays considerable dependence on the $s$-wave $D\pi$ scattering parameters, particularly at higher energies. 
This is true even given our assumption,
discussed in~\Cref{sec:res-latt-two-body}, of weak, attractive interactions in this subchannel. We recall that our main solutions are based on the ERE parameters in \Cref{eq:Dpi-params-s}, which lead to a $D^*_0$ resonance. To illustrate the dependence on the choice of interaction, we have repeated the calculation (for $g=0.55$) using the spartan $D\pi$ $s$-wave model, with $m_D \, a_0^{(1)} = 3.0$, and $r_{0}^{(1)} = 0$. The result is shown in~\Cref{fig:m3-j0} by the green, dashed curve. Although for this case the $J^P=1^+$ phase shifts do not deviate visibly from the results presented above, the slope of $q_b^3 \, \cot \delta_{J=0}$ significantly decreases at higher energies. We stress, however, that the behavior near the $DD^*$ threshold and the left-hand cut remains largely unchanged.

%%%%%%%%%%%%%%%%%%%%%%%%%
%%%%%%%%%%%%%%%%%%%%%%%%%
\subsection{Comparison with the finite-volume energies}
\label{sec:res-fin-vol}

In the previous sections, we presented numerical results obtained by solving infinite-volume integral equations at a pion mass $m_\pi = 280$~MeV corresponding to the lattice ensemble of ref.~\cite{Padmanath:2022cvl}. We used values for the two-body scattering parameters that included the $D^*$ and $D^*_0$ in the $D\pi$ subchannel and were motivated by existing lattice results. We then tuned the three-body $K$ matrix to match the $D D^*$ $s$-wave phase shift results from ref.~\cite{Padmanath:2022cvl}.

In this section, we present the numerical results obtained from the finite-volume formalism described in \Cref{sec:fin-vol}, in particular the three-particle $DD\pi$ quantization condition QC3 and the two-particle $DD^*$ QC2. The motivations for these two calculations are somewhat different. The aims in applying the QC3 are several. First, we wish to demonstrate that the formalism developed in ref.~\cite{\tetraquark}, in which the $D^*$ is not treated as a separate external field, but rather appears as a bound state in the $p$-wave $D\pi$ system, works as expected. This approach avoids the need to use a $2\leftrightarrow3$ formalism along the lines of ref.~\cite{Briceno:2017tce}. The approach followed here has been previously used successfully in model systems of scalar particles with a shallow bound state~\cite{Romero-Lopez:2019qrt, Jackura:2020bsk, Dawid:2023jrj}, but here the new features are a binding momentum that is comparable to $m_\pi$ and a bound state having spin.

%%%%%%%%%%%%%%%%%%%%%%%%%
\begin{figure}[t!]
    \centering
    \includegraphics[width=0.98\textwidth]{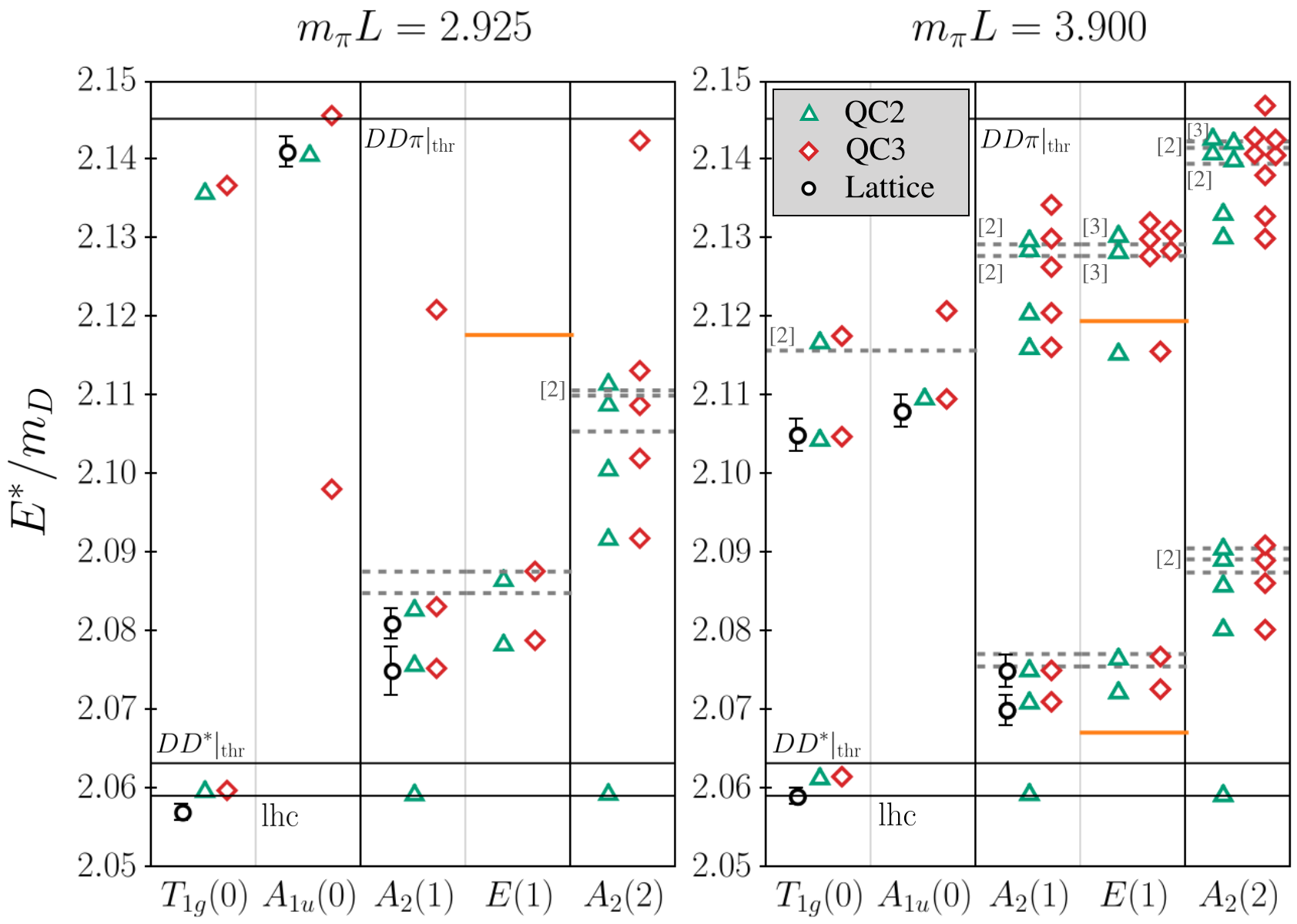}
    \caption{
    Finite-volume energy levels predicted by the QC3 (red diamonds) and QC2 (green triangles), compared to the levels found in a lattice computation~\cite{Padmanath:2022cvl} (black circles, with error bars.)
    Results are given for the c.m. frame energies.
    Levels are sorted horizontally according to their finite-volume irreps, using a notation described in the text. Details of the QC3 and QC2 methodology are described in the text. For the QC3 and QC2, only levels lying at or below the $DD\pi$ threshold are shown. Solid lines correspond to the positions of thresholds and the left-hand cut. Dashed lines correspond to the noninteracting $DD^*$ energy levels. For the $T_{1g}(0)$ irrep, there is a non-interacting level lying underneath the $DD^*$ threshold line for both values of $L$. The degeneracy of the noninteracting levels is denoted by a number next to the line, except if the level is nondegenerate. $DD$ levels appear in the $E(1)$ irrep, and their noninteracting energies are shown as orange lines. Note the extra energy levels produced by the QC2 near the left-hand cut.}
    \label{fig:fin-vol-spec}
\end{figure}
%%%%%%%%%%%%%%%%%%%%%%%%%

Second, we wish to show that the QC3 successfully predicts the energy levels that lie at or below the left-hand cut due to single-pion exchange (those that are greyed out in the figures above). This is the major reason for using the three-particle formalism to describe the $DD^*$ system~\cite{\tetraquark}. 

Third, we want to see how well the levels predicted by the QC3 match those obtained in ref.~\cite{Padmanath:2022cvl}. We stress that we are not attempting a fit of this data using the QC3, but aiming for an ``eyeball'' comparison. This does, however, provide a consistency check of the entire QC3 formalism. And, finally, we wish to investigate which energy levels are the most important for future lattice studies to focus on, in order to best constrain the underlying $K$ matrix parameters.

As concerns the $DD^*$ QC2, our main aim here is to investigate the manner in which it breaks down as one approaches the left-hand cut. A subsidiary aim is to see how well it matches the predictions of the QC3 in regions (i.e. above the $DD^*$ threshold) where both QC2 and QC3 are valid. Differences between the results can arise from the dropped exponentially suppressed terms.  For the quark masses we consider, the exponentially suppressed effects in the QC2 are dominated by the binding momentum of the $D^*$, $|q_0| \approx 0.87 m_\pi$. These effects, due to $D\pi$ loops, are automatically included in the QC3. However, the dropped finite-volume effects appear to be numerically small for the parameters used in the lattice simulations of ref.~\cite{Padmanath:2022cvl, Collins:2024sfi}. The measured $D^*$ masses on the two lattices used in these works are consistent within part per mille errors~\cite{Padmanath:private}. Furthermore, applying the QC2 to the $D\pi$ system with the $p$-wave ERE interaction used below, we find that the $D^*$ on the smaller lattice is heavier, but only by about a part per mille. Angular-momentum truncation in the two-body quantization condition can also lead to differences between QC2 and QC3, but these are small in practice due to threshold suppression.

We now turn to the results, shown in~\Cref{fig:fin-vol-spec}. We determine the finite-volume spectrum for two values of the box size, $m_\pi L = 2.925$ and $m_\pi L = 3.9$, equal to those used in ref.~\cite{Padmanath:2022cvl}.\footnote{%
%%%%
The exponentially-suppressed terms that are not included in the QC3 are generically of size $\exp(-m_\pi L)$. The rule of thumb used in lattice simulations is that one should take $m_\pi L \gtrsim 4$ to minimize these effects. The smaller volume used in ref.~\cite{Padmanath:2022cvl} clearly violates this inequality. However, as noted in \Cref{sec:QC2}, there is numerical and theoretical evidence that these dropped effects have an additional suppression in systems including $D$ mesons. 
%%%%
}
For the QC3 results, we set the $K$ matrices to the central values used in the final results from the integral equations, i.e. those displayed in \Cref{fig:m3-trunc,fig:m3-full,fig:m3-beta,fig:m3-epsilon,fig:m3-j0}. Specifically, we set $g=0.55$, $m_D^2 \, \cK_3^E = 1.9\cdot 10^5$, $\cK_3^{\rm iso,0}=\cK_3^{\rm iso,1}=\cK_3^B=0$, and $m_D \, a_0^{(2)} =-10^{-2}$, and use the $s$-wave $D\pi$ parameters of \Cref{eq:Dpi-params-s}.
For the QC2, we use as input the phase shifts for the $J^P = 0^-, 1^+$ $DD^*$ channels obtained from the solution to the integral equations using the same central values for the underlying parameters.

The results displayed in \Cref{fig:fin-vol-spec} are for the three irreps for which results are provided by ref.~\cite{Padmanath:2022cvl}, together with two additional irreps, $E(1)$ and $A_2(2)$, to which the $T_{cc}^+$ channel contributes. The notation for irreps is standard in the lattice literature; see, e.g., ref.~\cite{Morningstar:2013bda}. For example, $T_{1g}(0)$ indicates the $T_{1g}$ irrep of the little group associated with a dimensionless total squared momentum $d^2=0$, where $\bm d = L/(2\pi)\, \bm P$. Of the quantum numbers studied above in infinite volume, the $J^P=1^+$ irrep subduces into $T_{1g}(0)$, $A_{2}(1)$, $E(1)$, and $A_2(2)$, while $J^P=0^-$ subduces into $A_{1u}(0)$, $A_{2}(1)$, and $A_2(2)$. Higher partial waves contribute to all irreps, but we do not consider such waves in this work.

The QC3 predictions in the figure are shown as red diamonds and can be compared to the results of ref.~\cite{Padmanath:2022cvl}, shown as black circles with error bars. We note that, while with the QC3 we show levels up to the $DD\pi$ threshold, lattice levels are available in general only up to a lower energy. Thus the absence of a lattice level corresponding to one predicted by the QC3 does not necessarily indicate a disagreement. As discussed in \Cref{sec:QC3}, the QC3 is strictly valid only up to the $D^*D^*$ threshold, which lies at $E^*/m_D=2.127$, slightly below the $DD\pi$ threshold. We expect the impact on the higher levels shown in the figure to be small since they lie close to the threshold.

Consider first the $T_{1g}(0)$ irrep on the two lattices. To good approximation, this channel picks out the tetraquark quantum numbers. The lowest level corresponds, in the noninteracting limit, to a $DD^*$ pair with both at rest, whose energy is that of the $DD^*$ threshold (which is noted in the figure). The lattice and QC3 levels lie below this, as expected for an attractive interaction, with the former lying at (for the larger volume) or below (the smaller volume) the left-hand cut. For our chosen parameters, the QC3 levels do not reach this low. However, were we to use the parameters corresponding to $g=0.61$, the QC3 level on the smaller lattice would lie below the left-hand cut. This illustrates that the QC3 can predict levels at or below this cut.

The excited $T_{1g}(0)$ levels correspond in the noninteracting limit to a $D$ and $D^*$ with opposite momenta of unit dimensionless length. This ``free'' level is doubly degenerate, and the QC3 predicts that one of these levels is lowered by interactions, while the other is slightly pushed up. This can be seen on the larger volume, by comparing to the horizontal dashed line.\footnote{%
%%%%%%%
The free energies are given by $E^*(\p,\k) = \sqrt{ E(\p,\k)^2 - (\p+\k)^2 }$ where $E(\p,\k) = \omega_{p}^{D} + \omega_k^{D^*}$, and $\bm p$ and $\bm k$ are the finite-volume momenta of the $D$ and $D^*$, respectively.
%%%%%%%
}
Only the lower of the two levels is available from the lattice, and the agreement is good.
On the smaller volume, only the lower level falls below the $DD\pi$ threshold (with the free energy lying above this threshold) and the corresponding lattice level is not available.

We next consider the $A_{1u}(0)$ irrep, which is sensitive primarily to the $J^P=0^-$ channel, with no contributions from the tetraquark channel. On the larger volume, the QC3 finds two levels, the lower of which corresponds to the same free $D D^*$ level with moving $D$ and $D^*$ as appears in the $T_{1g}(0)$ irrep, although this level has degeneracy $1$ in the $A_{1u}(0)$ case. The upper level corresponds to the free $DD\pi$ state with all particles at rest, lowered substantially from the noninteracting energy by the subchannel interactions. We make this identification by gradually turning on the interactions and tracking the levels. On the smaller volume, the order of the two levels is inverted, with the lower level (around $E^*/m_D=2.1$) being the (much lowered) $DD\pi$ state. The absence of lattice states corresponding to these ``$DD\pi$'' levels on the two volumes is likely due to ref.~\cite{Padmanath:2022cvl} not including $DD\pi$ operators. 

Next, we consider the irreps in moving frames. Lattice levels are available only for the $A_2(1)$ case, which we discuss first. The lower pair of nearly degenerate free levels correspond to a moving $D$ with a $D^*$ at rest, or vice versa. Both levels are lowered by interactions, and the matching of QC3 results to those from the lattice is good on both volumes. On the larger volume, the first excited $A_2(1)$ levels fall within our energy range, with each free level being doubly degenerate. The QC3 finds these four levels, and also a fifth level that corresponds to a $DD\pi$ level with one of the $D$s moving, the free energy of which lies just above the displayed energy range at $E^*/m_D \approx 2.16$. On the smaller volume, the single level at $E^*/m_D\approx 2.12$ is also a lowered $DD\pi$ level (whose free energy is $E^*/M_D\approx 2.171$).

The next irrep is the $E(1)$, which we include as the tetraquark contributes to this channel, but $J^P=0^-$ does not. Thus it is more sensitive to tetraquark properties than the $A_2(1)$, which has contributions from both channels. The pattern of predicted levels is similar to that for the $A_2(1)$, with two main differences. First, we find that the lower pair of levels lie above the corresponding levels in the $A_2(1)$ irrep. Second, the first excited pair of levels, whose free energies match those in the $A_2(1)$ irrep, have a degeneracy of $3$ rather than $2$. The QC3 finds all these levels, some shifted down and others up relative to the free energies.

Parity is not conserved in moving frames, and this leads to mixing between $J^P=1^+$ and $J^P=1^-$ states in some irreps, such as the $E(1)$ irrep. This is an issue in the three-body formalism, since $DD$ and $DD\pi$ states with $J^P=1^-$ can mix, and two-to-three transitions are not included in the QC3~\cite{\tetraquark} (although doing so would be possible with an extended formalism). However, this mixing is likely to be small, and we only expect it to induce important finite-volume effects if the $DD$ and $DD\pi$ levels lie close to each other. For this reason, we show in the figure the position of the noninteracting $DD$ levels in the $E(1)$ irrep. On the smaller volume, the $DD$ level is well separated from the QC3 levels, while on the larger volume, mixing may be more of an issue.

The final irrep we consider is the $A_2(2)$, which we include to show how moving frames with larger momenta can provide additional constraints. The tetraquark contributes to this channel, as does $J^P=0^-$, but there is no mixing with $DD$ states. This irrep offers more states (four) in the lower band than $A_2(1)$. In the upper band (only present on the larger volume) the topmost level is likely a lowered $DD\pi$ state.

We have investigated the sensitivity of these results to the parameters in $\cK_3$. By far the greatest sensitivity is to $\cK_3^E$, which impacts all levels. If we set $\cK_3^E=0$ (rather than the standard value of $1.9\cdot 10^5$) then the QC3 prediction for the lightest $T_{1g}(0)$ state for the smaller volume moves above the $DD^*$ threshold, and the splitting between the two lowest $A_2(1)$ states is more than halved on both volumes.
This demonstrates that a nonzero $\cK_3$ is needed to have a reasonable description of the lattice data, just as was the case when solving the integral equations.

The other parameters in $\cK_3$, namely $\cK_3^{\rm iso,0}$, $\cK_3^{\rm iso,1}$, and $\cK_3^B$, do not contribute to the $T_{1g}(0)$ and $E(1)$ levels. For the other irreps, their contributions to the energies of the levels shown are an order of magnitude smaller than those of $\cK_3^E$, if we choose these parameters to have the same size as $\cK_3^E$.

Finally, we turn to the results from the QC2, which are shown as green triangles in \Cref{fig:fin-vol-spec}. Again we have displayed all levels up to the $DD\pi$ threshold, ignoring the coupling to the $D^*D^*$ channel. We observe that there is a good matching with the QC3 (and lattice) levels in the expected region, i.e., away from the left-hand cut, with the exception that the QC2 is missing states in the $A_{1u}(0)$ and $A_2(1)$ irreps that are predicted by the QC3. These missing states are, however, those that, as discussed above, correspond to $DD\pi$ states shifted down by interactions. Thus it is not surprising that they are missed by the QC2, which does not have a $DD\pi$ component. Additionally, the QC2 misses some excited $E(1)$ and $A_2(2)$ states, which is likely due to our truncation of the QC2 to $J<2$.
Overall, however, the use of the QC2 in refs.~\cite{Padmanath:2022cvl, Collins:2024sfi} for the levels above the $DD^*$ threshold appears justified.

The situation is different close to the left-hand cut. We know that the QC2 breaks down at, and likely slightly above, this cut. We find evidence of this in the appearance of solutions in the $A_2(1)$ and $A_2(2)$ irreps slightly above this cut, that are not present in the QC3 and thus clearly spurious. On the other hand, we note that the levels slightly above the cut that are predicted in the $T_{1g}(0)$ irrep do match well with those from the QC2.

%%%%%%%%%%%%%%%%%%%%%%%%%%%%%%%%%%%
% SECTION
%%%%%%%%%%%%%%%%%%%%%%%%%%%%%%%%%%%
\section{Summary and outlook}
\label{sec:conclusion}

In this work, we have described and implemented numerically the infinite-volume integral equations and the three-body finite-volume quantization condition for the $I=0$ $DD\pi$ system. These steps complete the formalism proposed in ref.~\citep{Hansen:2024ffk} and 
provide a ready-to-use methodology for the rigorous lattice determination of the properties of the $T_{cc}^+$. The approach applies to computations at both physical and heavier quark masses. In particular, unlike the traditional two-body formalism, it remains valid for energies near or on the left-hand cut associated with one-pion exchange in the $DD^*$ system.

The first part of our analysis involves solving the RFT partial-wave projected three-body integral equations, extending previous studies. Specifically, this is the first application of these equations to hadrons with unequal masses, which requires the consideration of two distinct pair-spectator channels. In addition, we include multiple two-body and three-body partial waves, using the results of ref.~\cite{Jackura:2023qtp} and~\Cref{app:pw}. We also consider for the first time terms beyond the isotropic approximation in the threshold expansion of the three-body $K$ matrix. This requires an extension of previous implementations of the integral equations. These methods are easily generalizable to other three-body systems of interest.

In the integral equations, we use a model of $D\pi$ and $DD$ interactions based on existing lattice data and effective theories, and estimate values for the unknown three-body parameter $\cK_3^E$ by matching the amplitude to existing $DD^*$ results~\cite{Padmanath:2022cvl}. By implementing LSZ reduction, we obtain the $DD^*$ elastic amplitude and find that the $T_{cc}^+$ manifests as a pair of subthreshold complex poles on the unphysical Riemann sheet, in agreement with the literature. 
We investigate the $DD^*$ phase shifts and the $T_{\rm cc}^+$ pole position for different interaction models, including and excluding partial-wave mixing. We find that partial-wave mixing between $s$ and $d$ waves in the $DD^*$ system is unimportant for physical kinematics, but its inclusion has a significant impact on the phase shifts below the $DD^*$ threshold.

In the second part, we implement the three-particle quantization condition (QC3) for the $DD\pi$ system, using it to describe the $DD^*$ energy levels down to and below the onset of the left-hand cut. This tests the strategy proposed in ref.~\cite{\tetraquark}, in which the $D^*$ is introduced in the QC3 as a pole in the $D\pi$ $p$-wave amplitude. Our predicted energies match those obtained from lattice QCD. We also explore the limitations of the two-particle $DD^*$ quantization condition. While it produces energy levels consistent with those from the QC3 above the elastic threshold, it breaks down near the left-hand cut, as expected. This confirms the advantage of using the three-particle approach in this system.

Looking forward, we suggest several extensions of the lattice calculations that would allow for a more robust study of the $T_{cc}^+$. First, our results show that $DD\pi$-like states are intermingled with those that are qualitatively $DD^*$-like. Since in finite volume these states mix, it is important to include $DD\pi$ interpolating operators in the lattice calculations. Second, extending the spectrum up to the $D^*D^*$ or $DD\pi$ thresholds would lead to additional constraints on the $K$ matrices. Similarly, additional finite-volume irreps could be studied, which may be good windows into the properties of $T_{cc}^+$. Examples are the $E(1)$ and $A_2(2)$ irreps in \Cref{fig:fin-vol-spec}. Finally, results for the $I=1$ $DD$ and $I=1/2$ $D\pi$ channels (the latter for both $s$ and $p$ waves) are needed in the three-body description of the $T_{cc}^+$.

Other infinite-volume approaches~\cite{Meng:2021uhz, Baru:2011rs, Du:2021zzh, Zhang:2024dth} have been developed for the analysis of the $T_{\rm cc}^+$ and related systems. While these methods differ from ours in their technical implementation,  
preliminary investigations indicate that the nonrelativistic limit of our equations leads to a formalism that is essentially equivalent to that of these approaches. A careful comparison is left for future work.
Moreover, we note that the integral equations described here provide a general parametrization for the three-meson amplitudes and may be also used in the phenomenological description of experimental data for the $T_{cc}^+ \to DD\pi$ decay, similar to refs.~\cite{Du:2021zzh, Zhang:2024dth}. We also leave this for future work.

In summary, this proof-of-concept study presents key ideas in the three-body treatment of $T_{cc}^+$ and opens the door for further studies in Lattice QCD. Although the parametrizations of the K matrices can be refined as more lattice QCD data becomes available, the overall workflow established in this work will remain unchanged. 
Given the recent advances in lattice QCD results for three-hadron scattering~\cite{Beane:2007es,Detmold:2008fn,Detmold:2008yn,Detmold:2011kw,Mai:2018djl,Horz:2019rrn,Blanton:2019vdk,Culver:2019vvu,Mai:2019fba,Fischer:2020jzp,Hansen:2020otl,Alexandru:2020xqf,Brett:2021wyd,Blanton:2021llb,NPLQCD:2020ozd,Mai:2021nul,Garofalo:2022pux,Yan:2024gwp}, 
we expect substantial progress for the $DD\pi$ system in the near term,
and anticipate that the strategy outlined here will play a central role.

\clearpage

%%%%%%%%%%%%%%%%%%%%%%%%%%%%%%%%%%%
%%%%%%%%%%%%%%%%%%%%%%%%%%%%%%%%%%%
\section*{Acknowledgments}

We wish to thank Saša Prelovšek for pointing us a discrepancy between our preliminary results and the literature, which helped to resolve a critical issue in our study. We also thank Ra\'ul Brice\~no and Andrew Jackura for insightful conversations about the partial-wave projection of the OPE amplitude. We thank Jeremy Green for discussions related to finite-volume irreps. Furthermore, we are grateful to Vadim Baru, Ra\'ul Brice\~no, Meng Lin Du, Max Hansen, Andrew Jackura, Saša Prelovšek, and Haobo Yan
for many other valuable discussions regarding this work. We also thank Saša Prelovšek and Madanagopalan Padmanath for sharing with us data from ref.~\cite{Padmanath:2022cvl}. SMD and SRS acknowledge the financial support through the U.S. Department of Energy Contract No. DE-SC0011637. FRL acknowledges partial support by the USDOE Contract No. DE-SC0011090 and DE-SC0021006, and the Mauricio and Carlota Botton Fellowship. This work contributes to the goals of the USDOE ExoHad Topical Collaboration, contract DE-SC0023598.

\appendix

%%%%%%%%%%%%%%%%%%%%%%%%%%%%%%%%%%%
% SECTION
%%%%%%%%%%%%%%%%%%%%%%%%%%%%%%%%%%%
\section{Partial-wave projection of OPE and $\cK_3$}
\label{app:pw}

As described in the main text, we project the integral equations onto definite total angular momentum in the three-body c.m.~frame prior to solving them. The rotation invariance of each quantity in the integral equations then propagates this projection separately onto each term. Thus we need to determine the projection of the OPE kernel $G$, the two-particle amplitude $\cM_2$, and the three-particle $K$ matrix. We discuss details of this projection in this appendix. We note that a very general formula for the projection of $G$ has been derived in ref.~\cite{Jackura:2023qtp},
including results that we need. 
Our discussion here offers an alternative presentation, which we hope will be useful to the readers. Our explicit results agree with those of ref.~\cite{Jackura:2023qtp}.
We make extensive use of the discussion of angular momentum projections 
given in ref.~\cite{Chung:1971ri}.

%%%%%%%%%%%%%%%%%%%%%%%%%%%%%%%%%%%
%%%%%%%%%%%%%%%%%%%%%%%%%%%%%%%%%%%
\subsection{Basic results}

Following the discussion in the main text, on-shell states of three spinless particles of given $E$, and having $\bm P=0$, can be described by the momentum of the spectator, $\bm k$, the pair's spin $s$, and the helicity $\lambda$ of the pair relative to its momentum, $-\bm k$. Here we consider the kinematic variables used in the main text for the initial state, although the following considerations apply equally well to the final state. The following discussion is independent of the flavor of the spectator, which enters only into the kinematics, so we keep the flavor index implicit. Since the three-particle state is analogous to a solid body, we can obtain a general state from that in which $\bm k$ is aligned along the $z$ axis by applying a rotation specified in the standard manner by three Euler angles. To be completely explicit, this state is given by
    %%%%%
    \begin{equation}
    \ket{\alpha,\beta,\gamma;-\lambda} \equiv \cU[R(\alpha,\beta,\gamma)]\{
    \cU[L_z(k)] \ket{0,0} \otimes \cU[L_z(-k)] \ket{s, -\!\lambda}
    \}\,,
    \label{eq:starting_state}
    \end{equation}
    %%%%%
where $\ket{0,0}$ and $\ket{s,-\!\lambda}$ are, respectively,
the states in which the spectator and the pair are at rest.
We use the standard notation for angular momentum eigenstates, $\ket{s,m}$, with $m$ the azimuthal component of the spin along the $z$ axis. For the pair, we have $m=-\lambda$, since the helicity is determined relative to $-\bm k = k (-\hat z)$. 
The unitary operators in \Cref{eq:starting_state} implement the boosts and rotations (defined in an active sense), with, for example, $L_z(k)$ indicating the boost that brings the spectator from rest to the momentum $k \hat z$.

Following ref.~\cite{Chung:1971ri}, the projection onto a state of definite total angular momentum $J$ and
azimuthal component $M$ is achieved by integrating with an appropriate Wigner D function,\footnote{%
We use a different normalization from ref.~\cite{Chung:1971ri}, chosen so that a state
that is independent of the angles projects with unchanged normalization onto $J=0$.
}
    %%%%%%
    \begin{equation}
    \ket{J M,-\!\lambda} = \frac{\sqrt{2J+1}}{8\pi^2}
    \int d\alpha  dc_\beta d\gamma\,  D_{M\mu}^{J*}(\alpha,\beta,\gamma) \ket{\alpha,\beta,\gamma; -\!\lambda}\,,
    \label{eq:projecJMhel}
    \end{equation}
    %%%%%
where $c_\beta\equiv \cos\beta$, and $\mu$ can, at this state, take any value satisfying $|\mu|\le J$. Using the result
    %%%%%
    \begin{equation}
    D^{J*}_{M\mu}(\alpha,\beta,\gamma) = D^{J*}_{M\mu}(\alpha,\beta,0) e^{i\gamma\mu}\,,
    \end{equation}
    %%%%%
together with,
    %%%%%
    \begin{equation}
    \ket{\alpha,\beta,\gamma; -\!\lambda} = \ket{\alpha,\beta,0; -\lambda} e^{i\gamma \lambda}\,,
    \end{equation}
    %%%%%
one can do the integral over $\gamma$; this sets $\mu=-\lambda$ and yields,
    %%%%%
    \begin{equation}
    \ket{J M,-\!\lambda} = \frac{\sqrt{2J+1}}{4\pi}
    \int d\alpha  dc_\beta \,  D_{M -\!\lambda}^{J*}(\alpha,\beta,0) \ket{\alpha,\beta,0; -\!\lambda} \,.
    \end{equation}
    %%%%%
We stress that these states transform correctly under rotations for any allowed choice of $\lambda$. Conversion to the LS basis is then achieved by the following unitary transformation~\cite{Chung:1971ri},
    %%%%%
    \begin{equation}
    \ket{JM;\ell s} = \sqrt{\frac{2\ell+1}{2J+1}}\sum_{\lambda} \ket{JM;-\!\lambda} 
    \braket{J, -\! \lambda | \ell, m_\ell=0 ; s, -\! \lambda}\,.
    \label{eq:toLS}
    \end{equation}
    %%%%%

To implement the above-described projection, and in particular the integral over the Euler angles $\alpha$ and $\beta$ for both initial and final states, we will need to determine the dependence of the quantities to be projected ($G$, $\cM_2$, and $\kdf$) on the angles denoted $\Omega_k^\star = (\vartheta_k^\star,\varphi_k^\star)$ and $\Omega_p^\star = (\vartheta_p^\star,\varphi_p^\star)$ in the main text (see \Cref{sec:DDpi-kinematics}). Focusing on the initial spectator, we recall that the angles $\Omega_k^\star$ are determined relative to a coordinate system in which the $z$ axis is aligned with the direction of the pair's momentum in the c.m.~frame. Thus we need to define this system. To do so, we start with an arbitrarily chosen right-handed triplet of axes (i.e.~a space-fixed coordinate system). We then apply the Euler rotation $R(\phi,\theta,0)$ that brings $\hat z$ to the direction of the pair, $\hat q \equiv -\hat k$. (We use $\hat q$ for brevity and generality.) This implies that the resulting $y$ axis (whose direction is denoted $\hat y_q$) lies in the original $xy$ plane. The explicit forms are given by\footnote{%
These expressions are ambiguous as $\hat q \to \pm \hat z$, but this ambiguity cancels in the final results.
}
    %%%%%
    \begin{equation}
    \hat z_q = \hat q\,, \ \
    \hat y_q = \frac{\hat z \times \hat q}{\sqrt{1- (\hat q\cdot \hat z)^2}}\,,
    \ \ {\rm and} \ \
    \hat x_q = \hat y_q \times \hat z_q= \frac{\hat q (\hat q\cdot \hat z) - \hat z}{\sqrt{1- (\hat q\cdot \hat z)^2}}\,.
    \label{eq:axes_q}
    \end{equation}
    %%%%%
We note that the ``$q$ basis'' triplet $\{\hat x_q,\hat y_q,\hat z_q\}$
is invariant under the boost along $\hat q$ that connects the pair's rest frame to the three-body c.m.~frame.

Further kinematic results that we will need are the components of the quantities $\bm k_p^\star$ ($\bm p_k^\star$) relative to the above-described coordinate systems for the final (initial) pair. 
These quantities, defined in the main text below \Cref{eq:ope-spin-helicity}, are, respectively, the initial (final) spectator momentum boosted to the final (initial) pair's c.m.~frame. 
Focusing on the initial spectator, 
and keeping in mind that the final pair momentum is $\hat{q} = -\hat{p}$, we find,
    %%%%%
    \begin{equation}
    \begin{split}
    (\bm k^\star_p)_{z_{-p}} &= -\gamma_p (\beta_p \omega_k + k c_\Theta)\,,
    \\
    \frac{(\bm k^\star_p)_{x_{-p}}}k &= 
    \frac{(\hat p c_\Theta - \hat k) \cdot \hat z}{\sqrt{1- (\hat p\cdot \hat z)^2}}\,,
    \\
    \frac{(\bm k^\star_p)_{y_{-p}}}k &= - \frac{\hat z \cdot \hat p \times \hat k}{\sqrt{1- (\hat p\cdot \hat z)^2}}\,.
    \end{split}
    \label{eq:kstarp_comps}
    \end{equation}
    %%%%%
    Here $c_{\Theta} = \cos \Theta$, with $\Theta$ the angle between $\p$ and $\k$.
Analogous equations hold for the components of $\bm p^\star_k$ relative to coordinate system with $\hat q=-\hat k$, and are obtained by $p\leftrightarrow k$ interchange. 

%%%%%%%%%%%%%%%%%%%%%%%%%%%%%%%%%%%
%%%%%%%%%%%%%%%%%%%%%%%%%%%%%%%%%%%
\subsection{Projection formula for OPE kernel $G$}

We first project onto $JM$ states with given pair helicities, and later convert to the $LS$ basis.
Using the results above, the first step yields,
    %%%%%
    \begin{equation}
    G^{(ij)JM}_{s' \lambda'; s \lambda}(p,k) = 
    \frac{2J+1}{(4\pi)^2} \int_{\Omega_p}\int_{\Omega_k}
    D^{J}_{M,-\lambda'}(\Omega_p)   D^{J*}_{M,-\lambda}(\Omega_k)
    G^{(ij)}_{s' \lambda';s \lambda}(\bm p, \bm k) \,,
    \label{eq:Gt_hel}
    \end{equation}
    %%%%%
where we are using the shorthand $\int_{\Omega_p} \equiv \int d\alpha_p d\cos\beta_p$,  while $\Omega_p$ stands for the rotation $R(\alpha_p,\beta_p,0)$ that brings $\hat z$ into the direction $\hat p$, and similarly with $p\to k$. 
The equality of the initial and final values of $J$ and $M$, which is built into this expression, is due to the overall rotation invariance of $ G^{(ij)}$.

Pulling out factors that are independent of the orientations of $\bm p$ and $\bm k$, 
we can rewrite \Cref{eq:Gt_hel} as,
    %%%%%
    \begin{align}
    G^{(ij)JM}_{s' \lambda'; s\lambda}(p,k) &= 
    \frac{H_{ij}(p,k)}{(q_p^{\star})^{s'} (q_{k}^{\star})^s} \,
    \mathcal G^{(ij) JM}_{s' \lambda' ; s \lambda}(p,k)\,,
    \label{eq:Gt_hel_decomp}
    \end{align}
    %%%%%
where,
    %%%%%
    \begin{align}
    \mathcal G^{(ij)JM}_{s'\lambda'; s \lambda}(p,k)
    &= 
    \frac{2J+1}{(4\pi)^2} \int_{\Omega_p}\int_{\Omega_k}
    D^{J}_{M,-\lambda'}(\Omega_p) \,
    D^{J*}_{M,-\lambda}(\Omega_k) \,
    \frac{\cY^*_{s' \lambda'}(\bm k^{\star}_p) \cY_{s \, \lambda}(\bm p^\star_k)}
    {b_{\bm{p k}}^2 - m_{ij}^2 + i\epsilon}
    \,.
    \end{align}
    %%%%%
Here $\cY_{\ell m}(\bm p) = \sqrt{4\pi} \, p\, Y_{\ell m}(\hat p) $ are harmonic polynomials normalized following the RFT conventions~\cite{\tetraquark}. It follows from rotation invariance that this expression is independent of $M$, as can be checked explicitly. Thus, we can replace the factor of $2J+1$ with a sum over $M$, which allows the two Wigner D matrices to be combined, using their unitarity. The result (with superscript $M$ now dropped) is,
    %%%%%
    \begin{align}
    \mathcal G^{(ij)J}_{s'\lambda'; s \lambda}(p,k)
    &= 
    \frac{1}{(4\pi)^2} \int_{\Omega_p}\int_{\Omega_k}
    D^{J*}_{-\lambda',-\lambda}(\Omega_p^{-1}\Omega_k)
    \frac{\cY^*_{s' \lambda'}(\bm k^{\star}_p) \cY_{s \, \lambda}(\bm p^\star_k)}
    {b_{\bm{p k}}^2 - m_{ij}^2 + i\epsilon}
    \,.
    \label{eq:calG_hel}
    \end{align}
    %%%%%
    
To perform the integrals over the orientations of $\hat p$ and $\hat k$ we proceed as follows. First, we treat the reaction plane as a solid body defined by the ordered vectors $\bm p$ and $\bm k$,
and integrate over its orientation with the standard Euler angles.%
%%%
\footnote{Denoted $\alpha$, $\beta$, and $\gamma$, which should not be confused with the earlier use of the same symbols for different angles in \Cref{eq:projecJMhel}.
}
%%%
In order to make it easier to use in the results given above for the orientation of the helicity axes, the $\alpha=\beta = \gamma=0$ configuration is chosen to be
that in which $\hat p$ is aligned along the $z$ axis, while $\hat k$ lies in the $xz$ plane with positive $x$ component. Second, we integrate the angle $\Theta$ between $\bm p$ and $\bm k$. These steps correspond to the variable transformation,
    %%%%%
    \begin{equation}
    \int_{\Omega_p}\int_{\Omega_k} = \int d c_\Theta\, d\alpha\, dc_\beta\, d\gamma \,.
    \end{equation}
    %%%%%
We thus need to express the rotations characterized by $\Omega_p$ and $\Omega_k$ in terms of the new angles. For $\Omega_p$ this is straightforward,
    %%%%%
    \begin{equation}
    R(\Omega_p) = R(\alpha,\beta,0) \,.
    \label{eq:ROmegap}
    \end{equation}
    %%%%%
To obtain $\Omega_k$, we can first rotate from $\hat p$ to $\hat k$ while they are in the $xz$ plane and then do the full solid-body rotation. This does not, however, lead to an orientation, i.e. a rotation described by two Euler angles; to obtain this, we must add an initial rotation about the lab $z$ axis,
    %%%%%
    \begin{equation}
    R(\Omega_k) = R(\alpha',\beta',0) =  R(\alpha,\beta,\gamma) R(0, \Theta, 0) R(0,0,-\delta)\,.
    \label{eq:ROmegak}
    \end{equation}
    %%%%%
Explicit expressions for $\alpha'$, $\beta'$ and $\delta$ in terms of $\alpha$, $\beta$, $\gamma$, and $\Theta$ can be obtained, but will not be needed in the following. What we do require is the result,
    %%%%%
    \begin{equation}
    R(\Omega_p^{-1} \Omega_k) = R(\gamma,\Theta,-\delta)\,.
    \end{equation}
    %%%%%
which follows from \Cref{eq:ROmegap,eq:ROmegak}.
Thus the Wigner D matrix needed in \Cref{eq:calG_hel} is given by,
    %%%%%
    \begin{equation}
    D^{J*}_{-\lambda', -\lambda} (\Omega_p^{-1}\Omega_k) = 
    e^{-i\gamma\lambda'} d^{J}_{-\lambda', -\lambda}(\Theta)
    e^{i \delta \lambda}\,,
    \label{eq:final_Wigner}
    \end{equation}
    %%%%%
where we have used the reality of the $d^J$ matrices. 

We also need to express the $x$ and $y$ components of $\bm k^\star_p$ in its helicity basis, given by \Cref{eq:kstarp_comps}, 
in terms of the new angles, and similarly for $\bm p^\star_k$. After some effort, we find,
    %%%%%
    \begin{equation}
    (\bm k^\star_p)_{x_{-p}} =  k s_\Theta c_\gamma\,,
    \qquad
    (\bm k^\star_p)_{y_{-p}} = - k s_\Theta s_\gamma\,,
    \label{eq:kstarp_final}
    \end{equation}
    %%%%%
where $s_\Theta = \sin\Theta$, etc., and,
    %%%%%
    \begin{equation}
    (\bm p^\star_k)_{x_{-k}} =
     - p s_\Theta c_\delta 
    \,,
    \qquad
    (\bm p^\star_k)_{y_{-k}} = 
    p s_\Theta s_\delta
    \,.
    \label{eq:pstark_finalb}
    \end{equation}
    %%%%%
In follows that the phases of the harmonic polynomials entering $\mathcal G^{(ij)}$, \Cref{eq:calG_hel}, are
    %%%%%
    \begin{align}
    \mathcal Y^*_{s'\lambda'}(\bm k^\star_p) 
    \propto  e^{ i \gamma \lambda'}\,,
    \qquad
    \mathcal Y_{s \lambda} (\bm p^\star_k) 
    \propto e^{ -i \delta \lambda}\,,
    \label{eq:harmonic_phases}
    \end{align} 
    %%%%%
with the proportionality constants being functions only of $\Theta$.
This shows that the unprojected $G^{(ij)}$ depends on angles other than $\Theta$, although not on $\alpha$. However, in the projected version \Cref{eq:calG_hel} the dependence on $\gamma$ and $\delta$
cancels between the phases in the harmonic polynomials and those in the Wigner D functions, as can be seen by comparing \Cref{eq:final_Wigner,eq:harmonic_phases}.

The conclusion is that the integrand in \Cref{eq:calG_hel} depends only on $\Theta$, allowing the integral over the other three angles to be done trivially. Using the result,
    %%%%%
    \begin{equation}
    d^J_{-\lambda', -\lambda}(\Theta) = d^J_{\lambda, \lambda'}(\Theta)\,,
    \end{equation}
    %%%%%
we obtain the final equation,
    %%%%%
    \begin{equation}
    \mathcal G^{(ij)J}_{s'\lambda'; s \lambda}(p,k) = \frac{1}{2}
    \int dc_\Theta \,
    d_{\lambda,\lambda'}^J(\Theta)
    \frac{\cY^*_{s' \lambda'}(\bm k^\star_p) \cY_{s \, \lambda}(\bm p^\star_k)}
    {b_{ij}^2 - m_k^2 + i\epsilon}\bigg|_{\alpha=\beta=\gamma=0} \,.
    \label{eq:calG_hel_f}
    \end{equation}
    %%%%%
Since the full integrand of \Cref{eq:calG_hel} is independent of $\alpha$, $\beta$, and $\gamma$, we can set these angles to any value. The choice $\alpha=\beta=\gamma=0$ is the simplest, since it sets $\delta=0$ and thus removes the phase dependence from
the Wigner D matrix in~\Cref{eq:final_Wigner}.

Finally, converting to the LS basis using \Cref{eq:toLS}, we obtain,
    %%%%%%
    \begin{multline}
    \mathcal G^{(ij)J}_{\ell' s'; \ell s} (p,k) = 
    \frac{\sqrt{(2\ell'+1)(2\ell+1)}}{2J+1} \sum_{\lambda' \lambda}
    \braket{J,-\!\lambda' | \ell', 0; s', -\!\lambda'} \braket{J,-\!\lambda | \ell, 0; s, -\!\lambda}
    \\
    \times \frac{1}{2} \int dc_\Theta \,
    d_{\lambda, \lambda'}^J(\Theta) \,
    \frac{\cY^*_{s' \lambda'}(\bm k^\star_p) 
    \cY_{s \, \lambda}(\bm p^\star_k)}
    {b_{\bm p \bm k}^2 - m_{ij}^2 + i\epsilon}\bigg|_{\alpha=\beta=\gamma=0}
    \,,
    \label{eq:calG_hel_LS}
    \end{multline}
    %%%%%
leading to the final form for the projection of $G$ in the LS basis,
    %%%%%
    \begin{align}
    G^{(ij)J}_{\ell' s' ; \ell s}(p,k) &= 
    \frac{H_{ij}(p,k)}{(q_p^{\star})^{s'} (q_{k}^{\star})^s} \,
    \mathcal G^{(ij)J}_{\ell' s' ; \ell s } (p,k)\,.
    \label{eq:GijJ}
    \end{align}
    %%%%%

The integrals that we require in~\Cref{eq:calG_hel_LS} can be brought to the form,
    %%%%%
    \begin{align}
    I^{(n)}(z_{ij})  
    &=  \int \frac{dc_\Theta}2
    \frac{c_\Theta^n}
    {(z_{ij}(p,k) - c_\Theta + i\epsilon)}\,,
    \end{align}
    %%%%%
where
\begin{equation}
z_{ij}(p,k) = \frac{(E-\omega_p-\omega_k)^2 - p^2 -k^2 -m_{ij}^2}{2pk}\,.
\label{eq:zijdef}
\end{equation}
The cases that we need are,
\begin{align}
I^{(0)}(z) &=  
\frac12 \log\left(\frac{z+1+i\epsilon}{z -1+i\epsilon}\right)\,,
\label{eq:I0}
\\
I^{(1)}(z) &= z I^{(0)}(z) - 1\,,
\label{eq:I1}
\\
I^{(2)}(z) &=  z^2 I^{(0)}(z) -  z\,,
\label{eq:I2}
\\
I^{(3)}(z) &= z^3 I^{(0)} (z)-  z^2 - \tfrac13\,.
\label{eq:I3}
\end{align}
These integrals can be expressed in terms of Legendre polynomials of the second kind as,
\begin{equation}
\begin{split}
I^{(0)}(z) &=  Q_0(z)\,,
\\
I^{(1)}(z) &= Q_1(z)\,,
\\
I^{(2)}(z) &= \tfrac23 Q_2(z) + \tfrac13 Q_0(z)\,,
\\
I^{(3)}(z) &= \tfrac25 Q_3(z) + \tfrac35 Q_1(z)\,,
\end{split}
\label{eq:ItoQ}
\end{equation}
with the proviso that the logarithmic cut runs between the singular points at $z=-1$ and $z=1$. This ensures that the OPE singularity leads to an imaginary part in the kinematic regime where the exchanged particle can go on shell. We note that all Legendre polynomials have their logarithmic cut at the same position in the $z$ variable, implying that inclusion of higher partial waves in the integral equations does not introduce additional singularities compared to the purely $s$-wave scenario (defined as $J=0$, $(\ell's';\ell s) = (00;00)$.)

In \Cref{app:pw_expr} we give the explicit forms for the projected $G$ in terms of the $Q_i(z)$.

\subsection{Projection of $\cK_3$}
\label{app:projectK3}

We consider the threshold expansion for $\cK_3$ given in \Cref{eq:K3thr}. The first step is to express $\cK_3$ in the $\{k s \lambda\}$ basis. This has been carried out in ref.~\cite{\implement}, and we can carry over the results
with only two adjustments. The first is that here we use the helicity basis, as discussed for $G$ above. The second adjustment arises from our use of standard complex spherical harmonics 
(as opposed to the real versions used in ref.~\cite{\implement}).
Thus we must specify which harmonics are complex conjugated.
With our choice, the final state spherical harmonic that is pulled out of $\cK_3$ (which depends on the direction of the primary member of the final state pair in the pair c.m. frame)
is not conjugated, while the corresponding initial-state harmonic is conjugated. This implies that the harmonic functions that appear in the resulting helicity-basis form of $\cK_3$
have the same conjugation structure as for $G$, 
namely the final-state harmonic is conjugated while the initial-state one is not. One result of this structure is that the phase cancellation described above for $G$ between \Cref{eq:final_Wigner,eq:harmonic_phases}
applies to $\cK_3$ as well.
This can be shown by an extension of the methods used above.

Thus, again, all angles except $\Theta$ can be trivially integrated, leading to the master formula
\begin{multline}
\cK^{(ij)J}_{3; \ell' s'; \ell s} = 
\frac{\sqrt{(2\ell'+1)(2\ell+1)}}{2J+1} \sum_{\lambda' \lambda}
\braket{J,-\!\lambda' | \ell', 0; s', -\!\lambda'} \braket{J,-\!\lambda | \ell, 0; s, -\!\lambda}
\\
\times \int \frac{dc_\Theta}2\
d_{\lambda, \lambda'}^J(\Theta)
\cK^{(ij)}_{3;s' \lambda'; s \lambda}(\bm p, \bm k)\bigg|_{\alpha=\beta=\gamma=0}
\,.
\label{eq:masterK3}
\end{multline}
We recall that the superscript $(ij)$ refers to the flavors of the final- and initial-state spectators, respectively.

In \Cref{app:pw_K3} we give the explicit forms for the projections of $\cK_3$ onto the channels of interest.
To our knowledge, these results have not been presented previously.
The integrals involved are all elementary.
The only subtlety in the computation is
the need to write the Cartesian Kronecker delta in the helicity basis. We do so using
\begin{equation}
\delta_{ij} = (\hat x_{-p})_i (\hat x_{-p})_j + (\hat y_{-p})_i (\hat y_{-p})_j + (\hat z_{-p})_i (\hat z_{-p})_j \,,
\end{equation}
where, in the final-state pair basis,
and for $\alpha=\beta=\gamma=0$,
\begin{equation}
\hat x_{-p} = (1,0,0)\,,\quad \hat y_{-p} = (0,1,0)\,, \quad \hat z_{-p}=(0,0,1)\,,
\end{equation}
while in the initial-state pair basis
\begin{equation}
\hat x_{-p} = (c_\Theta,0,-s_\Theta)\,,\quad \hat y_{-p} = (0,1,0)\,, \quad \hat z_{-p}=(s_\Theta,0,c_\Theta)\,.
\end{equation}

\subsection{Projection of $\cM_2$}
\label{app:projectM2}

We close this appendix with a derivation of the result \Cref{eq:M2new} for the projected form of $\cM_2$.
The starting point is the form for this quantity prior to projection, given in the description of integral equations in sec.~VII of ref.~\cite{\BSnondegen}, 
\begin{equation}
\cM_{2;s'\lambda';s\lambda}^{(ij)}(\bm p, \bm k) = \delta_{ij} \eta_i 2 \omega_k (2\pi)^3
\delta^3(\bm p-\bm k) \delta_{s's}\delta_{\lambda'\lambda}
\cM_{2,s}^{(i)}(p)\,.
\label{eq:M2orig}
\end{equation}
Notation is explained in the main text.

The projection onto definite $J,M$ is given,
as in \Cref{eq:Gt_hel}, by
\begin{equation}
    \cM_{2;s'\lambda';s\lambda}^{(ij)JM}(p,k)
= \frac{2J+1}{(4\pi)^2}
\int_{\Omega_p} \int_{\Omega_k}
D^J_{M,-\lambda'}(\Omega_p)
D^{J*}_{M,-\lambda}(\Omega_k)
\cM_{2;s'\lambda';s\lambda}^{(ij)}(\bm p, \bm k)\,.
\end{equation}
Writing the delta function as
\begin{equation}
    \delta^{(3)}(\bm p-\bm k) = \frac1{p^2}
    \delta(p-k) \delta^{(2)}(\Omega_p-\Omega_k)\,,
\end{equation}
performing the $\Omega_k$ integral to set $\Omega_k=\Omega_p$,
and using,
\begin{equation}
\int_{\Omega} D^{j_1*}_{\mu_1 m_1}(\Omega) D^{j_2}_{\mu_2 m_2}(\Omega)
= \frac{4\pi}{2 j_1+1} \delta_{j_1 j_2} \delta_{\mu_1 \mu_2} \delta_{m_1 m_2}\,,
\end{equation}
we find,
\begin{equation}
    \cM_{2;s'\lambda';s\lambda}^{(ij)JM}(p,k)
= \delta_{ij} \eta_i \tilde \delta(p-k) \delta_{s's} \delta_{\lambda'\lambda}
\cM_{2,s}^{(i)}(p)\,,
\end{equation}
where $\tilde \delta(p-k)$ is defined in \Cref{eq:delta_tilde}. The lack of dependence on $M$ is manifest.

Next we convert to the LS basis using \Cref{eq:toLS}, which yields
\begin{multline}
    \cM_{2; \ell' s';\ell s}^{(ij)J}(p,k)
    = \frac{2\ell+1}{2J+1} \sum_\lambda
    \braket{J,-\lambda|\ell', 0; s',-\lambda'}
    \braket{J,-\lambda|\ell, 0; s,-\lambda}
    \\
\times    \eta_i \delta_{ij} \delta_{s's}
    \delta_{\lambda'\lambda}
    \tilde \delta(p-k) \cM_{2,s}^{(i)}(p)\,.
\end{multline}
Using the unitarity of the transformation in
\Cref{eq:toLS},
one immediately finds the result given in \Cref{eq:M2new} of the main text.

A similar derivation holds for the phase-space matrix $\bm{\tilde \rho}$ that appears in the integral equations.

\section{Expressions for the partial-wave projection of OPE}
\label{app:pw_expr}

To keep the formulas compact, we define relativistic factors as
    %%%%%%
    \begin{align}
    \label{eq:relativistic-factors}
    \beta_p = \frac{p}{E - \omega_p} \, , ~~~ \gamma_p = \frac{1}{\sqrt{1-\beta_p^2}} \, , ~~~ g_{pk} = \frac{\beta_p \gamma_p \omega_k}{k} \, ,
    \end{align}
    %%%%%%
and analogous formulas with $k\leftrightarrow p$.
We also abbreviate $z_{ij}$ as simply $z$. Following ref.~\cite{Jackura:2023qtp}, we use the spectroscopic notation $(J,\ell,s) \equiv {}^{2s+1} \ell_J$ to denote transitioning pair-spectator states.

In the following, results in which
$\{\ell',s'\}$ and $\{\ell,s\}$ are interchanged are not given in off-diagonal cases. The missing results are simply obtained by switching momenta, $p \leftrightarrow k$, and switching the corresponding spectator masses.

We also stress that the following results only hold for allowed choices of $i$ and $j$.
In particular, if $i=2$ , so that the 
final-state pair consists of two $D$ mesons in a symmetric ($I=1$) state, then the pair spin $s'$ must be even. There is no such constraint for $i=1$. Similar comments apply to the initial state
and the corresponding index $j$.

Finally, in ref.~\cite{Jackura:2023qtp} it is noted that a check on the results is obtained by expanding the
expressions for $G^{(ij)}$ about $p=0$ and $k=0$ for fixed $\sigma_p$ and $\sigma_k$. One expects that the projected
$G$ will have a leading dependence
\begin{equation}
G^{(ij)J}_{\ell' s';\ell s}(p,k) \propto  p^{\ell'} k^\ell\,,
\label{eq:threxpect}
\end{equation} 
due to angular-momentum barrier factors.
This expectation holds for all the
following results.

\subsection{$J^P = 1^+$}

\paragraph{The $\bm{^1 P_1 \to {}^1 P_1}$ amplitude}     
    %%%%%%
    \begin{align}
    \label{eq:OPE-1}
    G^{(ij)1}_{10;10}(p,k) = \frac{H_{ij}(p,k)}{2 p k} \, Q_1(z) \, .
    \end{align}
    %%%%%%

\paragraph{The $\bm{^1 P_1 \rightarrow {}^3 S_1}$ amplitude}
    %%%%%%
    \begin{align}
    \label{eq:S-wave-G}
    G^{(ij)1}_{01;10}(p,k) = -\frac{H_{ij}(p,k)}{2 p q_p^\star} \left[ \frac{1}{3} \Big(\gamma_p + 2 \Big) \, Q_0(z)
    %%%
    + g_{pk} \, Q_1(z)
    %%%
    + \frac{2}{3} ( \gamma_p - 1 ) \, Q_2(z) \right] \, .
    \end{align}
    %%%%%%

\paragraph{The $\bm{{}^1 P_1 \rightarrow {}^3 D_1 }$ amplitude}
    %%%%%
    \begin{align}
    G^{(ij)1}_{21;10}(p,k) = \frac{H_{ij}(p,k)}{\sqrt{2} \, p q_p^\star} \left[ \frac{1}{3} (\gamma_p - 1 ) \, Q_0(z) 
    %%%
    + g_{pk} \, Q_1(z)
    %%%
    + \frac{1}{3} (2\gamma_p + 1) \, Q_2(z) \right] \, .
    \end{align}
    %%%%%

\paragraph{The $\bm{^3 S_1 \rightarrow {}^3 S_1}$ amplitude} 
    %%%%%
    \begin{align}
    G^{(ij)1}_{01;01}(p,k) &= \frac{H_{ij}(p,k)}{2 q_p^\star q_k^\star} \Bigg[ \frac{1}{3} 
    \left\{ g_{kp} (\gamma_p + 2) 
    + g_{pk} (\gamma_k + 2) \right\} \, Q_0(z) \\ \nonumber
    %%%
    & + \left\{ g_{pk} g_{kp} + \frac{1}{5} \left( -2 + 2 \gamma_p + 2 \gamma_k + 3 \gamma_p \gamma_k \right) \right\}  \, Q_1(z) \\ \nonumber
    %%%
    & + \frac{2}{3} \left\{ g_{kp} (\gamma_p - 1) + g_{pk} (\gamma_k - 1) \right\} \, Q_2(z) 
    %%%
    + \frac{2}{5} (\gamma_k - 1)(\gamma_p - 1) \, Q_3(z) \Bigg] \, .
    \end{align}
    %%%%%

\paragraph{The $\bm{{}^3 D_1 \rightarrow {}^3 S_1}$ amplitude} 
    %%%%%
    \begin{align}
    G^{(ij)1}_{01;21}(p,k) &= - \frac{H_{ij}(p,k)}{\sqrt{2} q_p^\star q_k^\star} \Bigg[ \frac{1}{3} \left\{g_{kp}  (\gamma_p + 2) + g_{pk} (\gamma_k - 1) \right\} \, Q_0(z) \\ \nonumber
    %%%
    & + \left\{ g_{pk} g_{kp} + \frac{1}{5} ( 1 - \gamma_p + 2 \gamma_k + 3 \gamma_p \gamma_k ) \right\}  \, Q_1(z) \\ \nonumber
    %%%
    & + \frac{1}{3} \left\{ g_{kp} (2\gamma_p - 2) + g_{pk} (2\gamma_k + 1) \right\} \, Q_2(z) 
    %%%
    + \frac{1}{5} (\gamma_p-1)(2 \gamma_k + 1) \, Q_3(z)  \Bigg] \, .
    \end{align}
    %%%%%

\paragraph{The $\bm{^3 D_1 \rightarrow {}^3 D_1}$ amplitude} 
    %%%%%
    \begin{align}
    \label{eq:OPE-2}
    G_{21;21}^{(ij) 1}(p,k) &= \frac{H_{ij}(p,k)}{2 q_p^\star q_k^\star} \Bigg[ \frac{2}{3} \left\{ g_{kp} (\gamma_p - 1) + g_{pk} (\gamma_k - 1) \right\} \, Q_0(z) \\ \nonumber
    %%%
    & + \left\{ 2 \, g_{kp} g_{pk} + \frac{1}{5} \left( -1 - 2\gamma_p - 2 \gamma_k + 6 \gamma_p \gamma_k \right) \right\}  \, Q_1(z) \\ \nonumber
    %%%
    & + \frac{2}{3} \left\{ g_{kp} (2 \gamma_p + 1) + g_{pk} (2  \gamma_k + 1) \right\} \, Q_2(z) 
    %%%
    + \frac{1}{5} ( 1 + 2 \gamma_k)(1 + 2 \gamma_p) \, Q_3(z) \Bigg] \, .
    \end{align}
    %%%%%

\subsection{$J^P = 0^-$}
\label{sec:pw-g-j0}

\paragraph{The $\bm{{}^1 S_0 \to {}^1 S_0}$ amplitude}     
    %%%%%%
    \begin{align}
    \label{eq:OPE-0}
    G^{(ij)0}_{00;00}(p,k) = \frac{H_{ij}(p,k)}{2 p k} \, Q_0(z) \, .
    \end{align}
    %%%%%%

\paragraph{The $\bm{^1 S_0 \rightarrow {}^3 P_0}$ amplitude}
    %%%%%%
    \begin{align}
    G^{(ij)0}_{11;00}(p,k) = \frac{\sqrt{3} H_{ij}(p,k)}{2 p q_p^\star} \, \left[ g_{pk} \, Q_0(z) + \gamma_p Q_1(z) \right] \, .
    \end{align}
    %%%%%%

\paragraph{The $\bm{{}^3 P_0 \rightarrow {}^3 P_0 }$ amplitude}
    %%%%%
    \begin{align}
    G^{(ij)0}_{11;11} = \frac{H_{ij}(p,k)}{2 \, q_k^\star q_p^\star} \, \left[ \Big( 3 g_{pk} g_{kp} +  \gamma_k \gamma_p \Big) Q_0(z) 
    + 3 \left( g_{pk} \gamma_k 
    + g_{kp} \gamma_p \right) Q_1(z)  + 2  \gamma_k \gamma_p Q_2(z) \right] \, .
    \end{align}
    %%%%%

\section{Expressions for the partial-wave projection of $\cK_3$}
\label{app:pw_K3}

This appendix provides explicit expressions for the partial wave projection of $\cK_3$,
based on the master formula given in \Cref{eq:masterK3}.
To write the results compactly, 
we introduce some notation in addition to that used in the previous appendix,
    %%%%%
    \begin{align}
    \omega_{q_p^\star}^- & = \sqrt{(q_p^\star)^2 + m_D^2} - \sqrt{(q_p^\star)^2 + m_\pi^2} \, , \\
    p_p^\star & = \gamma_p (p + \omega_p \beta_p ) \, , \\
    \omega_{p_p^\star} & = \gamma_p (\omega_p + \beta_p p) \, ,   
    \end{align}
    %%%%%
as well as the analogous expressions with
$p\to k$. Since the $\cK_3$ matrix is symmetric under interchanges of final and initial states, we provide only the minimal number of non-zero elements necessary to reproduce all partial-wave and flavor compositions.

The expected threshold behavior is the same as for $G$, \Cref{eq:threxpect}, and this expectation holds for all the following results.

We stress that the results in this appendix do not include the symmetry factors discussed at the end of \Cref{sec:threebodyforces} in the main text.

%%%%%%%%%%%%%%%%%%%%%%%%%%%%%%%%%%%
\subsection{${J^P = 1^+}$}
%%%%%%%%%%%%%%%%%%%%%%%%%%%%%%%%%%%
The isotropic and $\cK_3^B$ terms
do not contribute. For the former this result
is trivial since an internal $p$-wave
is required to create $J=1$; 
for the latter this is a nontrivial result.

\paragraph{The $\bm{^1 P_1 \to {}^1 P_1}$ amplitude}     

    %%%%%
    \begin{align}
    \cK_{3;10;10}^{(11)1}(p,k) &= \frac{1}{6}\, \cK_3^E \, \frac{1}{M^2}
    \Big( p - \gamma_p \beta_p \omega_{q_p^\star}^- \Big)
    \Big( k - \gamma_k \beta_k  \omega_{q_k^\star}^- \Big) \, , \\
    %%%
    \cK_{3;10;10}^{(12)1}(p,k) &= - \frac{1}{3} \, \cK_3^E \, \frac{1}{M^2} \big( p - \gamma_p \beta_p \omega_{q_p^\star}^- \big) k \, , \\
    %%%
    \cK_{3;10;10}^{(22)1}(p,k) &= \frac{2}{3} \, \cK_3^E \, \frac{p k}{M^2} \, .
    \end{align}
    %%%%%

\paragraph{The $\bm{^1 P_1 \to {}^3 S_1}$ amplitude}     

    %%%%%
    \begin{align}
    \cK_{3;01;10}^{(11)1}(p,k) &= - \frac{1}{9} \, \cK_3^E \, \frac{1}{M^2} q_p^\star \, \left(\gamma_p + 2 \right) \left(k - \gamma_k \beta_k  \omega_{q_k^\star}^- \right) \, , \\
    %%%
    \cK_{3;01;10}^{(12)1}(p,k) &= \frac{2}{9} \, \cK_3^E \, \frac{1}{M^2} \, q_p^\star \big( \gamma_p + 2 \big) k , \\
    %%%
    \cK_{3;01;10}^{(22)1}(p,k) &= 0 \, .
    \end{align}
    %%%%%

\paragraph{The $\bm{^1 P_1 \rightarrow {}^3 D_1}$ amplitude} 

    %%%%%
    \begin{align}
    \cK_{3;21;10}^{(11)1}(p,k) &= \frac{\sqrt{2}}{9} \, \cK_3^E \, \frac{1}{M^2} q_p^\star \, \left(\gamma_p - 1 \right) \left(k - \gamma_k \beta_k  \omega_{q_k^\star}^- \right) \, , \\
    %%%
    \cK_{3;21;10}^{(12)1}(p,k) &= - \frac{2\sqrt{2}}{9}  \, \cK_3^E \,  \frac{1}{M^2} q_p^\star \left( \gamma_p - 1 \right) k \, , \\
    %%%
    \cK_{3;21;10}^{(22)1}(p,k) &= 0 \, .
    \end{align}
    %%%%%

\paragraph{The $\bm{^3 S_1 \rightarrow {}^3 S_1}$ amplitude}

    %%%%%
    \begin{align}
    \label{eq:K-matrix-s-wave}
    \cK_{3;01;01}^{(11)1}(p,k) &= \frac{2}{27} \, \cK_3^E \, \frac{1}{M^2} q_p^\star \, (\gamma_p + 2) q_k^\star ( \gamma_k + 2) \, , \\
    %%%
    \cK_{3;1;01;01}^{(12)}(p,k) &= 0 \, , \\
    %%%
    \cK_{3;1;01;01}^{(22)}(p,k) &= 0 \, .
    \end{align}
    %%%%%

\paragraph{The $\bm{^3 S_1 \rightarrow {}^3 D_1}$ amplitude} 

    %%%%%
    \begin{align}
    \cK_{3;21;01}^{(11)1}(p,k) &= -\frac{2\sqrt{2}}{27}  \, \cK_3^E \, \frac{1}{M^2} q_p^\star q_k^\star \left(\gamma_p - 1\right)  \left(\gamma_k+2\right) \, , \\
    %%%
    \cK_{3;1;21;01}^{(12)}(p,k) &= 0 \, , \\
    %%%
    \cK_{3;1;21;01}^{(22)}(p,k) &= 0 \, . \\ 
    \end{align}
    %%%%%
    
\paragraph{The $\bm{{}^3 D_1 \rightarrow {}^3 D_1}$ amplitude}

    %%%%%
    \begin{align}
    \cK_{3;21;21}^{(11)1}(p,k) &= \frac{4}{27}  \, \cK_3^E \, \frac{1}{M^2} q_p^\star \, (\gamma_p - 1) q_k^\star (\gamma_k - 1) \, , \\
    %%%
    \cK_{3;1;21;21}^{(12)}(p,k) &= 0 \, , \\
    %%%
    \cK_{3;1;21;21}^{(22)}(p,k) &= 0 \, .
    \end{align}
    %%%%%

\paragraph{Separated momentum dependence} Given the elements above, it is possible to cast the $J^P=1^+$ three-body $K$ matrix into the form of \Cref{eq:k3-separable} with $a_{\rm max}=1$. The corresponding non-zero ``left'' functions are,
    %%%%%%
    \begin{align}
    \cK_{L,10}^{(1)}(p) &= \sqrt{\frac{ \cK_3^E}{6} } \, \frac{p -  \gamma_p \beta_p \omega_{q_p^\star}^- }{M} \, , \\
    %%%
    \cK_{L,01}^{(1)}(p) &= - \sqrt{\frac{ 2 \cK_3^E}{27}} \, \frac{q_p^\star (\gamma_p + 2) }{ M } \, , \\
    %%%
    \cK_{L,21}^{(1)}(p) &= \sqrt{\frac{4 \cK_3^E}{27}} \, \frac{q_p^\star(\gamma_p - 1)}{M} \, , ~~~  \\
    %%%
    \cK_{L,10}^{(2)}(p) &= - \sqrt{\frac{2 \cK_3^E}{3} } \, \frac{p}{M} \, ,
    \end{align}
    %%%%%%
where we have now
dropped the trivial index $a$. The ``right'' functions are obtained from the above expressions by replacing labels $L\leftrightarrow R$, momenta $p \leftrightarrow k$, and the corresponding spectator's masses when appropriate. In this case, matrix $\cI$ of \Cref{eq:matrix-I} becomes a number.

%%%%%%%%%%%%%%%%%%%%%%%%%%%%%%%%%%%
\subsection{${J^P = 0^-}$}
\label{sec:pw-k3-j0}
%%%%%%%%%%%%%%%%%%%%%%%%%%%%%%%%%%%

In the following, we use the abbreviation $\cK_3^{\rm iso} = \cK_{3}^{\rm iso, 0} + \cK_{3}^{\rm iso, 1} \, \Delta$.

\paragraph{$\bm{{}^1 S_0 \to {}^1 S_0}$ elements}     

    %%%%%
    \begin{align}
    \cK_{3;00;00}^{(11)0}(p,k) & = \cK_3^{\rm iso} + \cK_3^B \frac{1}{M^2} \Big[ E(\omega_p + \omega_k) - 6 m_D^2 + \omega_{k_k^\star} \omega_{q_k^\star}^- + \omega_{p_p^\star} \omega_{q_p^\star}^- \Big] \\ \nonumber
    %%%
    & + \cK_3^E \frac{1}{M^2} \Big[ 2 m_\pi^2 - \frac{1}{2} (E - \omega_{q_p^\star}^- \gamma_p - \omega_p) (E - \omega_{q_k^\star}^- \gamma_k - \omega_k) \Big]  \, , \\
    %%%
    \cK_{3;00;00}^{(12)0}(p,k) & = \cK_3^{\rm iso} + \cK_3^B \frac{1}{M^2} \Big[ E \omega_p - 3m_D^2 + \omega_{p_p^\star} \omega_{q_p^\star}^- + (\sigma_k - 4 m_D^2) \Big] \\ \nonumber
    & + \cK_3^E \frac{1}{M^2} \Big[2 m_\pi^2 - (E - \omega_{q_p^\star}^- \gamma_p- \omega_p ) \omega_k \Big] \, , \\
    %%%
    \cK_{3;00;00}^{(22)0}(p,k) &= \cK_3^{\rm iso} + \cK_3^B \, \frac{1}{M^2} \Big[ (\sigma_k - 4 m_D^2) + (\sigma_p - 4 m_D^2) \Big] \\ \nonumber
    & + \cK_3^E \frac{2}{M^2} \big( m_\pi^2 - \omega_p \omega_k \big)  \, .
    \end{align}
    %%%%%

\paragraph{$\bm{^1 S_0 \rightarrow {}^3 P_0}$ elements}

    %%%%%
    \begin{align}
     \cK_{3;11;00}^{(11)0}(p,k) & =  - \cK_3^B \, \frac{2}{\sqrt{3}} \frac{1}{M^2} \, p_p^\star q_p^\star 
     - \cK_3^E \frac{1}{\sqrt{3}} \frac{1}{M^2} \, q_p^\star \beta_p \gamma_p \Big( E -  \omega_{q_k^\star}^- \gamma_k - \omega_k\Big) \, , \\
     %%%
     \cK_{3;11;00}^{(12)0}(p,k) & = - \frac{2}{\sqrt{3}} \cK_3^B \frac{1}{M^2} \, p_p^\star q_p^\star - \frac{2}{\sqrt{3}} \cK_3^E \frac{1}{M^2} \, q_p^\star \beta_p \gamma_p \omega_k \, , \\
     %%%
     \cK_{3;11;00}^{(12)0}(p,k) & = 0 \, .
    \end{align}
    %%%%%
    
\paragraph{The $\bm{{}^3 P_0 \rightarrow {}^3 P_0 }$ elements}

    %%%%%
    \begin{align}
    \cK_{3;11;11}^{(11)0}(p,k) & = - \frac{2}{3} \cK_3^E \frac{1}{M^2} \, q_p^\star \beta_p \gamma_p q_k^\star \beta_k \gamma_k \, , \\
    %%%
    \cK_{3;11;11}^{(12)0}(p,k) & = 0 \, \\
    %%%
    \cK_{3;11;11}^{(22)0}(p,k) & = 0 \, .
    \end{align}
    %%%%%
    
\paragraph{Separated momentum dependence} Below we provide the ``left'' and ``right" functions reproducing the above partial-wave projected $J^P=0^-$ $\cK_3$ matrix. In the most general case, when all coefficients are non-zero, it requires $a_{\rm max} = 4$. The $a \leq 2$ ``left'' functions involve only $\cK^{\rm iso}$ and $\cK^E$ terms,
    %%%%%%
    \begin{align}
    %%%
    \cK_{L,00}^{1,(1)}(p) &= i\sqrt{\frac{\cK_3^E}{2}} \, \frac{E - \omega_{q_p^\star}^- \gamma_p - \omega_p}{M} \, , 
    ~~~ \cK_{L,00}^{2,(1)}(p) = \sqrt{ \cK_3^{\rm iso} + 2 \cK_3^E \, \frac{m_\pi^2}{M^2} } \, , \\
    %%%
    \cK_{L,11}^{1,(1)}(p) &= i \sqrt{\frac{2 \cK_3^E}{3}} \, \frac{q_p^\star \beta_p \gamma_p}{M} \,, 
    ~~~  \cK_{L,11}^{2,(1)}(p) = 0 \, , \\
    %%%
    \cK_{L,00}^{1,(2)}(p) &= i \sqrt{2 \cK_3^E} \, \frac{\omega_p}{M} \, , 
    ~~~ \cK_{L,00}^{2,(2)}(p) = \sqrt{ \cK_3^{\rm iso} + 2 \cK_3^E \, \frac{m_\pi^2}{M^2} } \, .
    \end{align}
    %%%%%%
The corresponding ``right'' functions are obtained by replacing momenta $p \leftrightarrow k$ and the corresponding masses. To include $\cK_3^B$ terms we need to add new non-symmetric terms,
    %%%%%%
    \begin{align}
    \cK_{L,00}^{3,(1)}(p) &= {\cK_3^B} \, \frac{E \omega_p - 3 m_D^2 + \omega_{p_p^\star} \omega_{q_p^\star}^-}{M^2} \, ,
    ~~~ 
    \cK_{R,00}^{3,(1)}(k) = 1 \, , \\
    %%%
    \cK_{L,00}^{4,(1)}(p) &= 1 \, , 
    ~~~ 
    \cK_{R,00}^{4,(1)}(k) = {\cK_3^B} \, \frac{E \omega_k - 3 m_D^2 + \omega_{k_k^\star} \omega_{q_k^\star}^- }{M^2} \, , \\
    %%%
    \cK_{L,11}^{3,(1)}(p) &= -\cK_3^B \, \frac{2}{\sqrt{3}} \, \frac{q_p^\star p_p^\star}{M^2} \, , ~~~ 
    \cK_{R,11}^{3,(1)}(k) = 0 \, , \\
    %%%
    \cK_{L,11}^{4,(1)}(p) &= 0 \, , ~~~ 
    \cK_{R,11}^{4,(1)}(k) = -\cK_3^B \, \frac{2}{\sqrt{3}} \, \frac{q_k^\star k_k^\star}{M^2} \, , \\
    %%%
    \cK_{L,00}^{3,(2)}(p) &= {\cK_3^B} \, \frac{\sigma_p - 4 m_D^2}{M^2} \, , 
    ~~~ \cK_{R,00}^{3,(2)}(k) = 1 \, , \\
    %%%
    \cK_{L,00}^{4,(2)}(p) &= 1 \, , ~~~ \cK_{R,00}^{4,(2)}(k) = {\cK_3^B} \, \frac{\sigma_k - 4 m_D^2}{M^2} \, .
    %%%
    \end{align}
    %%%%%%

%%%%%%%%%%%%%%%%%%%%%%%%%%%%%%%%%%%%%%%
%%%%%%%%%%%%%%%%%%%%%%%%%%%%%%%%%%%%%%%
\section{Residue of $D^*$ pole}
\label{app:chpt}

In this appendix we derive the result \Cref{eq:Dpi-r} used in the main text, which provides the relation between the effective range in $p$-wave $D\pi$ scattering amplitude to the $DD^*\pi$ coupling. We simplify the notation by writing this amplitude as $\cM_{2,1}$, using $s$ for the square of the two-particle c.m.~frame energy, and denoting the magnitude of the momentum of each particle in this frame as $q$ and the phase shift simply as $\delta(q)$.

We first observe that it follows from the physical crossing condition described after \Cref{eq:pole-condition} that the $p$-wave amplitude,
    %%%%%
    \begin{align}
    \cM_{2,1}^{-1} &= \frac1{8\pi \sqrt{s}} \frac1{q^2}
    \left(q^3 \cot \delta(q) - i q^3 \right)
    \,,
    \label{eq:appD1}
    \end{align}
    %%%%%
must have a positive residue at a physical bound-state pole
(as already incorporated into the form \Cref{eq:LSZ-1} used in the LSZ reduction of $\cM_3$).
We thus introduce a (real, positive) $D^*D\pi$ coupling constant, $g_{D^* D \pi}$, by taking the following form for the inverse amplitude close to the pole,
    %%%%%
    \begin{align}
    \cM_{2,1}^{-1} &= - \frac{2}{q_0^2}\frac{s-m_{D^*}^2}{ g_{D^*D\pi}^2}
    \left[ 1+ \cO(s-m_{D^*}^2) \right]\,,
    \label{eq:appD2}
    \end{align}
    %%%%%
The overall factors are chosen so that the coupling constant agrees with that used in the chiral perturbation theory (ChPT) literature~\cite{Belyaev:1994zk}, as discussed further below. Note that the residue is positive because the momentum at the bound state, $q_0$ is imaginary.
We will use the two-term ERE expansion,
$q^3 \cot \delta = -1/a_0 + r q^2/2$
(again with a simplified notation),
in which case $q_0$ is given by
\Cref{eq:D-star-pole-spec}.
Matching the derivatives of \Cref{eq:appD1,eq:appD2} at the pole position yields
    %%%%%
    \begin{align}
    g_{D^* D\pi}^2 &= -\frac{128\pi m_{D^*}}
    {1- (m_D^2-m_\pi^2)^2/m_{D^*}^4} 
    \frac{1}{r + 3 |q_0|}\,.
    \label{eq:appD3}
    \end{align}
    %%%%%
In order for the right-hand side to be positive, we must have $r < - 3 |q_0|$. The result \Cref{eq:appD3} leads immediately to \Cref{eq:Dpi-r}.

We now explain why the form \Cref{eq:appD2} agrees with conventions used in ChPT. The coupling constant is defined by~\cite{Belyaev:1994zk},
    %%%%%
    \begin{equation}
    \braket{D^{0}(p_1) \pi^+(p_2)| D^{*+} (p_1+p_2)} = - (\epsilon \cdot p_2)  g_{D^* D \pi}
    = - \sqrt2 \braket{D^{+}(p_1) \pi^0(p_2)| D^{*+} (p_1+p_2)}\,,
    \end{equation}
    %%%%%
where $\epsilon$ is the $D^*$ polarization vector,
and the second equality follows from isospin symmetry. The decay widths are then,
    %%%%%
    \begin{equation}
    \Gamma(D^{*+} \to D^0 \pi^+) =\frac{g_{D^* D\pi}^2 q^{3}}{24 \pi m_{D^*}^2} =  2 \, \Gamma(D^{*+}\to D^+ \pi^0) \, .
    \end{equation}
    %%%%%
To match the normalization of $\cM_{2,1}$ we need to project onto $I=1/2$ and $\ell=1$.
For the former, we use,
    %%%%%
    \begin{equation}
    \bra{D\pi, I=1/2,m_I=1/2} = \sqrt{\tfrac23} \bra{D^0 \pi^+} - \sqrt{\tfrac13} \bra{D^+ \pi^0}
    \end{equation}
    %%%%%
to obtain,
    %%%%%
    \begin{equation}
    \braket{D(p_1)\pi(p_2),I=1/2 | D^{*+}(p_1+p_2)} = - \sqrt{\tfrac32}(\epsilon \cdot p_2) g_{D^*D\pi}\,.
    \end{equation}
    %%%%%
Using this, and summing over polarizations, the pole contribution to $\cM_2$ is,
    %%%%%
    \begin{equation}
    \cM_2(I=1/2) = -\frac32 \, g_{D^* D\pi}^2 \frac{\bm q'_\pi \cdot \bm q_\pi}{s - m_{D^*}^2}\,,
    \end{equation}
    %%%%%
with $\bm q'_\pi$ and $\bm q_\pi$ being the final and initial pion momenta in the c.m. frame. An equivalent result can be obtained by using the effective Lagrangian and Feynman rules in ref.~\cite{Wise:1993wa}. The projection onto $\ell=1$ can be done using
    %%%%%
    \begin{equation}
    \bm a\cdot \bm b = \frac{4\pi}3 a b \sum_m Y_{1m}(\hat a) Y^*_{1m} (\hat b)\,,
    \end{equation}
    %%%%%
and leads to
    %%%%%
    \begin{equation}
    \cM_{2,1}(I=1/2) = -\frac12 g_{D^* D\pi}^2 \frac{q^2}{s-m_{D^*}^2}\,,
    \end{equation}
    %%%%%
which agrees with \Cref{eq:appD2}.

%%%%%%%%%%%%%%%%%%%%%%%%%%%%%%%%%%%%%%%%
\bibliographystyle{JHEP} 
\bibliography{ref}
%%%%%%%%%%%%%%%%%%%%%%%%%%%%%%%%%%%%%%%%

\end{document}